# Intrinsic spin-momentum dynamics of surface electromagnetic waves in complex dispersive system

Peng Shi[†], Xinrui Lei, Qiang Zhang, Heng Li, Luping Du, Xiaocong Yuan[‡]

*Nanophotonics Research Centre, Shenzhen Key Laboratory of Micro-Scale Optical Information Technology, Shenzhen University, 518060, China*

Spin-momentum locking is an intrinsic property of surface electromagnetic fields and its study has led to the discovery of photonic spin lattices and diverse applications. Previously, dispersion was ignored in the spin-momentum locking, giving rise to abnormal phenomena contradictory to the physical realities. Here, we formulate four dispersive spin-momentum equations for surface waves, revealing universally that the transverse spin vector is locked with the momentum. The locking property obeys the right-hand rule in the dielectric but the left-hand rule in the dispersive metal/magnetic materials. In addition to the dispersion, the structural features can affect the spin-momentum locking significantly. Remarkably, an extraordinary longitudinal spin originating from the coupling polarization ellipticity is uncovered even for the purely polarized state. We further demonstrate the spin-momentum locking properties with diverse photonic topological lattices by engineering the rotating symmetry. The findings open up opportunities for designing robust nanodevices with practical importance in chiral quantum optics.

Momentum and angular momentum (AM) are the two fundamental dynamical characters of matter and waves [1-7] and are important in understanding and predicting the behaviors in wave-matter interactions. Through the active manipulation of electron spins in solid-state systems, a multidisciplinary field that is referred to as spintronics and has potential applications in the field of information technology has grown [7,8]. Almost simultaneously, optical scientists have developed similar concepts for electromagnetic (EM) systems, giving rise to the discovery of the spin-dependent position or momentum of light, including the spin-Hall effect [9-17] and optical magnus effect [18], the spin-dependent optical vortex [19-21] and the spin-dependent unidirectional propagation of light [22-27]. Therein, by engineering the extrinsic spin-orbit coupling in artificial structures [22-24], the photonic analogy of unidirectional topological spin states has been demonstrated with the pseudo-spin. Meanwhile, intrinsic spin-momentum locking (*i*SML) originating from spin–orbit coupling in Maxwell's equations has been demonstrated for surface EM systems [25-27]. This *i*SML describes the photonic spin dynamics in dispersionless system and has been exploited in diverse applications based on the spin AM (SAM) and momentum degrees of freedom for optical manipulation [28-30], nanometrology [31,32], spin-based robust optical devices [33-36] and data processing with photonic topological solitons [37-43].

Previous research on *i*SML adopted structureless and dispersionless EM surface modes [25-41], which limited further application. To exploit the applications, one can manipulate the *i*SML in artificial materials, including the metamaterials [44-47], metasurface [48-50], photonic crystal [22-24], artificial anisotropic materials [51] and chiral structures [52,53], which contain dispersion inevitably. If the dispersive and structural properties are not considered in *i*SML, it would give rise to abnormal phenomena contradictory to physical reality [54]. The energy density in a dispersive isotropic medium is described by the Brillouin formula [55]. However, there are challenges in characterizing the momentum and AM in dispersive media owing to the long-standing Abraham–Minkowski debate [56]. Although the kinetic Abraham-Poynting momentum can be used to describe the classical current feature of photons [57], it does not relate to dispersion. It is thus difficult to carry out the spin-orbit decomposition and to evaluate the *i*SML properties of a complex dispersive system.

Here, we reexamine the spin-orbit decomposition in a complex dispersive system and utilize a dispersion-related momentum to formulate four Maxwell-like spin-momentum equations (SMEs) for the surface waves in multilayered structure. The SMEs unveil that the transverse spin is locked with the dispersive momentum: the *i*SML satisfies the right-hand rule in the dielectric but the left-hand rule in the dispersive metals/magnetic materials. Moreover, the SMEs reveal that the structural and material dispersions can affect the *i*SML appreciably, which provides guidance for tuning the *i*SML by designing the structure and dispersion. Additionally, we uncover an extraordinary longitudinal spin component that does not possess *i*SML but depends on the symmetry of the EM mode. To verify the *i*SML properties, we investigate the spin-momentum properties of photonic topological lattices under diverse rotating symmetry. The present theoretical framework is important to the development of field theory with the spin and momentum of photons and is expected to have application in physical and integrated optics.

We consider the purely polarized monochromatic surface mode (*p*- or *s*- polarized mode) propagating in a multilayered structure as shown in Fig. 1 with the complex electric field $\mathbf{E}$ and magnetic field $\mathbf{H}$ and angular frequency $\omega$. Considering the dispersive effect [58-61], the Minkowski-type canonical momentum is $\tilde{\mathbf{p}}_o = \langle\tilde{\psi}|\hat{\mathbf{p}}|\psi\rangle/\hbar\omega$, where $\hat{\mathbf{p}}$ is the momentum operator, and the SAM is $\tilde{\mathbf{S}} = \langle\tilde{\psi}|\hat{\mathbf{S}}|\psi\rangle/\hbar\omega$, where $\hat{\mathbf{S}}$ is the spin-1 matrix in SO(3). Here, the bra vector $\langle\tilde{\psi}| = (\tilde{\varepsilon}\mathbf{E}^*, -i\tilde{\mu}\mathbf{H}^*)/2$ depends on the group permittivity $\tilde{\varepsilon} = \partial[\omega\varepsilon]/\partial\omega$ and the group permeability $\tilde{\mu} = \partial[\omega\mu]/\partial\omega$ whereas the ket vector $|\psi\rangle = (\mathbf{E}, i\mathbf{H})^{\mathrm{T}}/2$.

In the case of dispersive media, by performing inverse processing with respect to spin-orbit decomposition [57], a dispersive momentum $\tilde{\mathbf{p}} = [\tilde{\varepsilon}\mu + \varepsilon\tilde{\mu}]\,\mathbf{p}/2\varepsilon_0\mu_0$ can be obtained, where $\mathbf{p}=\varepsilon_0\mu_0\mathrm{Re}\{\mathbf{E}^*\times\mathbf{H}\}/2$ the kinetic momentum, $\varepsilon_0$ is the permittivity and $\mu_0$ is the permeability in vacuum [54]. Obviously, the dispersive momentum is consistent with the kinetic momentum in the free space and dielectric. In dispersive metals with the negative permittivity $\varepsilon=\varepsilon_0(1-\omega_{ep}^2/\omega^2)$, where $\omega_{ep}$ is the electric plasma frequency, or magnetic materials with the negative permeability $\mu=\mu_0(1-\omega_{mp}^2/\omega^2)$, where $\omega_{mp}$ is the magnetic plasma frequency [62], we have $[\tilde{\varepsilon}\mu + \varepsilon\tilde{\mu}]/2\varepsilon_0\mu_0 = 1$, and the dispersive momentum is converted into the kinetic momentum of photons. Notably, only the metal/magnetic materials with negative real permittivity/permeability are considered here. Thus, irrespective of there being a dispersionless dielectric or dispersive media, the dispersive momentum is proportional to the kinetic momentum and includes the dispersive effect, which is beneficial in evaluating the *i*SML of light. With the dispersive momentum, the Maxwell-like dispersive SMEs can be summarized as [54]

$$\nabla\cdot\tilde{\mathbf{p}} = 0\,, \tag{1}$$

$$\nabla\cdot\tilde{\mathbf{S}} = \left(\varepsilon\tilde{\mu} - \tilde{\varepsilon}\mu\right)\left(\mathbf{E}^*\cdot\mathbf{H} + \mathbf{E}\cdot\mathbf{H}^*\right)/4\,, \tag{2}$$

$$\nabla\times\tilde{\mathbf{S}} = 2\left(\tilde{\mathbf{p}} - \tilde{\mathbf{p}}_o\right), \tag{3}$$

$$\nabla\times\tilde{\mathbf{p}} = \frac{\chi}{2k^2}\tilde{\mathbf{S}}_t = \frac{\chi}{2k^2}\left(\tilde{\mathbf{S}} - \tilde{\mathbf{S}}_l\right). \tag{4}$$

The parameter $\chi$ in Eq. (4) is

$$\chi = \begin{cases} 2/\left[1+\tilde{\eta}/\eta\right] & \text{horozontal} \\ \left[2-2\left(1-\tilde{\eta}/\eta\right)/n_{eff}^2\right]/\left[1+\tilde{\eta}/\eta\right] & \text{normal} \end{cases}, \tag{5}$$

where $n_{eff}^2=\beta^2/\omega^2\varepsilon\mu$ is the relative effective index, $\beta$ is the propagating constant, $k^2=\omega^2\varepsilon\mu$, $\tilde{\eta} = \tilde{\mu}/\tilde{\varepsilon}$ and $\eta=\mu/\varepsilon$.

Equation (1) shows that the normal component of the dispersive momentum is continuous through the interface because the dispersive momentum only has horizontal components and thus $\nabla\cdot\tilde{\mathbf{p}} = \nabla\frac{\tilde{\varepsilon}\mu+\varepsilon\tilde{\mu}}{2\varepsilon_0\mu_0}\cdot\mathbf{P} + \frac{\tilde{\varepsilon}\mu+\varepsilon\tilde{\mu}}{2\varepsilon_0\mu_0}\nabla\cdot\mathbf{P} = 0$. Equation (2) shows that the normal component of SAM is active or dissipating in the dispersive medium owing to the dispersion-induced breaking of the dual symmetry between the electric and magnetic properties, $\tilde{\varepsilon}\mu - \varepsilon\tilde{\mu} \neq 0$ [63], which is dramatically different from the dispersionless case. While the tendencies of electric and magnetic dispersions are identical (i.e., $\omega_{ep}=\omega_{mp}$), the dual symmetry is protected and the normal SAM component is passive. Anyhow, as the material's dispersion is present, the normal SAM component is discontinuous through the interface owing to the present of the additional dispersion-related terms in the group permittivity/permeability. The normal component of SAM depends on the electric ellipticity $(\mathbf{E}^*\times\mathbf{E})_z$ and magnetic ellipticity $(\mathbf{H}^*\times\mathbf{H})_z$, which are determined by the continuous horizontal electric/magnetic field. Thus, the normal SAM components at the two sides of an interface are parallel because $\tilde{\varepsilon} > 0$ and $\tilde{\mu} > 0$. Subsequently, Eq. (3) represents the spin-orbit decomposition of the dispersive momentum $\tilde{\mathbf{p}} = \tilde{\mathbf{p}}_o + \tilde{\mathbf{p}}_s$, in which the canonical momentum $\tilde{\mathbf{p}}_o$ also represents the orbital momentum (the orbital angular momentum $\tilde{\mathbf{L}} = \mathbf{r}\times\tilde{\mathbf{p}}_o$) and $\tilde{\mathbf{p}}_s = \nabla\times\tilde{\mathbf{S}}/2$ is the dispersive Belinfante spin momentum [64]. Equation (3) shows that the horizontal SAM component is discontinuous through the interface owing to the additional dispersive terms in the group permittivity $(\omega\partial\varepsilon/\partial\omega)$ for the *p*-polarized surface wave and in the group permeability $(\omega\partial\mu/\partial\omega)$ for the *s*-polarized surface wave, respectively. Moreover, the directions of the horizontal SAM components are opposite one another at the two sides of an interface owing to the opposing signs of $\varepsilon$ and $\tilde{\varepsilon}$ for the *p*-polarized surface wave or $\mu$ and $\tilde{\mu}$ for the *s*-polarized surface wave. Finally, Eq. (4) reveals that the horizontal dispersive momentum is discontinuous through the interface. In particular, for the purely polarized surface modes at the interface between the dielectric and metals/magnetic materials, the horizontal dispersive momenta have opposing signs at the two sides because the normal electric/magnetic field is discontinuous for the *p*/*s*-polarized surface wave.

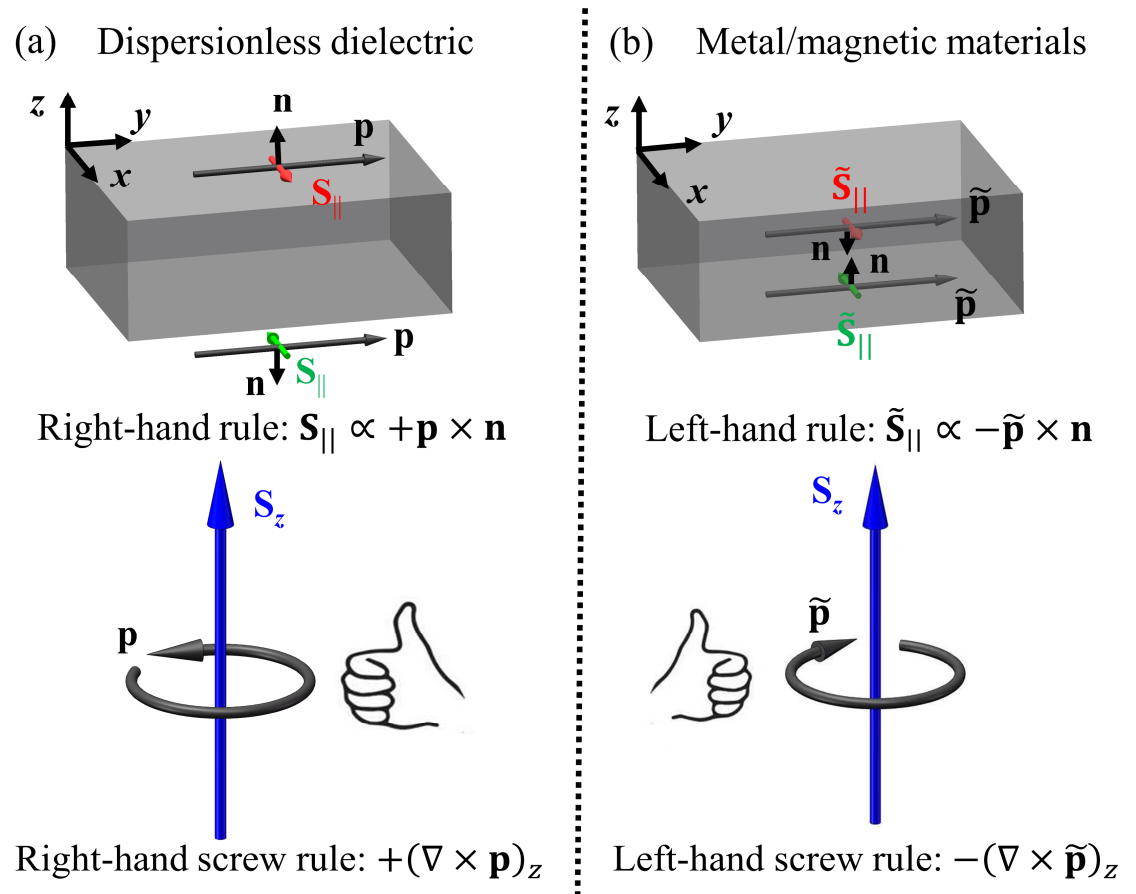

**FIG. 1.** Schematic diagram of *i*SML in the multilayered system. (a) in the dispersionless dielectric, the horizontal SAM are locked with the $\mathbf{p}$ and the locking property satisfies the right-hand rule $\mathbf{S}_{\|}\propto+\mathbf{p}\times\mathbf{n}$ whereas the normal SAM is locked with $\mathbf{p}$ and the locking property satisfies the right-hand screw rule $\mathbf{S}_z\propto+(\nabla\times\mathbf{p})_z$. (b) in the dispersive medium, the horizontal SAM are locked with the dispersive momentum $\tilde{\mathbf{p}}$ and the locking property satisfies the left-hand rule $\tilde{\mathbf{S}}_{\|} \propto -\tilde{\mathbf{p}}\times\mathbf{n}$ whereas the normal SAM is locked with $\tilde{\mathbf{p}}$ and the locking property satisfies the left-hand screw rule $\tilde{\mathbf{S}}_z \propto -(\nabla\times\tilde{\mathbf{p}})_z$. $\mathbf{n}$ denotes the outer normal direction.

Remarkably, Eq. (4) also expresses the *i*SML between the momentum of photons and transverse spin [2,25,27]. The transverse spin originates from the transverse inhomogeneities of the EM field [64,65]. In a dispersionless medium, such as the upper and lower space in Fig. 1, $\tilde{\eta}=\eta$ and the $\chi$ is equal to 2 uniformly. The transverse spin is locked with the kinetic momentum and the locking property satisfies the right-hand rule as shown in Fig. 1(a). However, in the dispersive metal/magnetic materials, the *i*SML properties are dramatically different. As an example, in the air-metal-air structure, the permittivity of the metal takes a negative real value at optical frequency $\omega < \omega_p$, and the group permittivity $\tilde{\varepsilon} = \varepsilon_0\left(1 + \omega_{ep}^2/\omega^2\right)$ is positive such that the electric energy density has physical meaning. One can obtain that $\tilde{\eta}/\eta=(1-\omega_{ep}^2/\omega^2)/(1+\omega_{ep}^2/\omega^2)\in(-1,0)$, $1+\tilde{\eta}/\eta\in(0,1)$ and the $\chi$ for the horizontal components is positive. Because $k^2<0$ for metal materials, the horizontal SAM is locked with the dispersive momentum and the locking property satisfies the left-hand rule. Meanwhile, because $n_{eff}^2$ is negative in the metal and $1-\tilde{\eta}/\eta\in(1,2)$, the $\chi$ for the normal component is positive as well.

Similarly, because $k^2<0$ for metal materials, the transverse spin in the normal direction is also locked with the dispersive momentum and the locking property satisfies the left-hand screw rule. Notably, the difference between the horizontal $\chi$ and normal $\chi$ originates from the breaking of dual symmetry between the group permittivity and group permeability in the dispersive medium. Assuming the specific case that the permittivity and permeability have identical dispersive properties, the dispersion-induced dual symmetry is protected, and horizontal $\chi$ and normal $\chi$ are found to be equal to 1 simultaneously, which is consistent with the dispersionless SMEs [27].

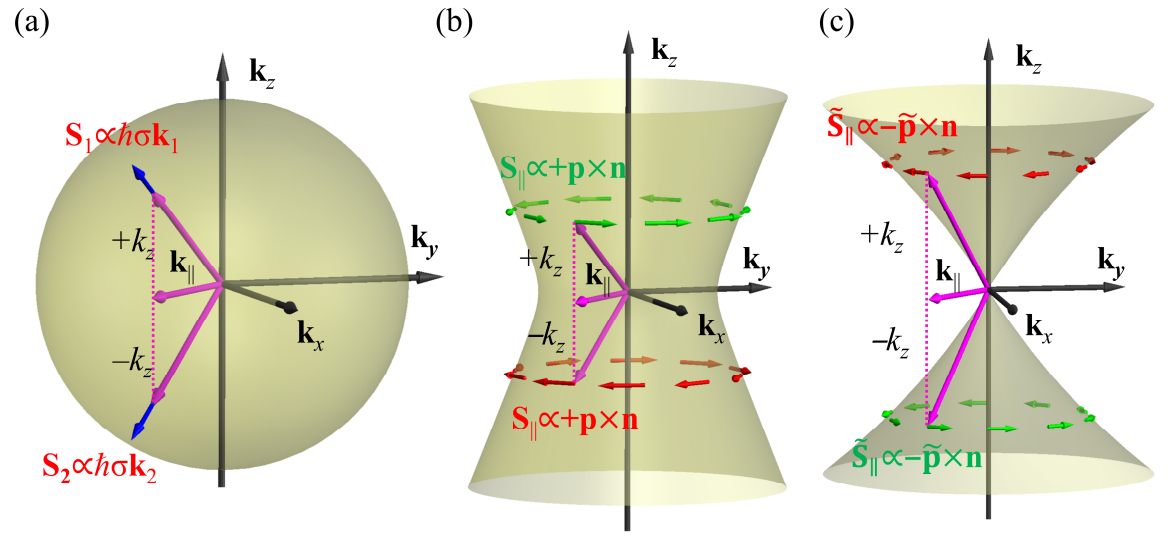

**FIG. 2.** Origins of *i*SML in the *k*-space. The wavevector and transverse spin for (a) the propagating waves, (b) the surface waves in dispersionless dielectric and (c) the surface waves in dispersive metal/magnetic materials. For the propagating plane wave, the spin $\mathbf{S}\propto\hbar\sigma\mathbf{k}$ is purely longitudinal and the wavevector satisfies $k_x^2+k_y^2+k_z^2=k^2$, corresponding to a sphere. Thus, for a transverse wavevector $k_x^2+k_y^2=k_{||}^2<k^2$, there are two solutions of $z$-component wavevector ($\pm k_z$) representing two propagating directions. Together with the two circularly polarized basises, the propagating wave possesses $\mathbb{Z}_4$ topological invariant [25]. For the purely polarized surface plane wave in dispersionless dielectric, the wavevector satisfies the relation $k_x^2+k_y^2-k_z^2=k^2$, corresponding to an uniparted hyperboloid. For the purely polarized surface plane wave in dispersive metal/magnetic materials ($k^2<0$), the wavevector satisfies $-k_x^2-k_y^2+k_z^2=-k^2$, corresponding to a parted hyperboloid. In these two cases, the dual symmetry is broken, and thus the two solutions $\mp k_z$ are corresponding to surface waves at the upper and lower sides of interface. The $\mathbb{Z}_4$ index is degraded into a pair of $\mathbb{Z}_2$ indices, which indicates the *i*SML [66]. Notably, the local wavevector is normally proportional to the canonical momentum. However, for the purely polarized surface waves, the dispersive momentum is proportional to the canonical momentum. The topology of the dispersive momentum is thus consistent with that of the wavevector.

Additionally, dispersion includes the spatial dispersion and the material dispersion, which both can affect the optical spin-orbit interaction between the photon's spin and position/momentum, leading to the spin-dependent momentum or propagation of light [2,9-18]. This spin-dependent effect can be also observed in the Eq. (4). Moreover, Eq. (5) reveals that the *i*SML property is relative to the effective index $n_{eff}^2$. Thus, by well-designing the dispersive property of artificial materials, such as the electric and magnetic plasma frequencies in photonic metamaterials, the chirality of *i*SML can be engineered flexibly. Meanwhile, because the dispersive momentum of surface waves can be re-expressed as $\tilde{\mathbf{p}} \propto \langle\psi|i\nabla|\psi\rangle$, where $|\psi\rangle$ is the Hertz potential. The transverse spin $\tilde{\mathbf{S}}_t \propto \chi\langle\nabla\psi| \times i|\nabla\psi\rangle$ has a similar form with the Berry curvature in the representation of the Hertz potential [1,2], which is beneficial for the analysis of spin-orbit interactions and geometric phases of surface waves in dispersive systems.

The topological origins of *i*SML can be understood in the momentum space in Fig. 2. For the circularly polarized propagating plane waves in the free space, the longitudinal spin possesses $\mathbb{Z}_4$ topological invariant [25]. However, in the presence of the interface, the $\mathbb{Z}_4$ index is degraded into a pair of $\mathbb{Z}_2$ indices [66]. Moreover, owing to the breaking of dual symmetry between electric and magnetic properties of the dielectric and metal/magnetic materials, only one polarized state survives and thus the surface wave possesses *i*SML.

Subsequently, we exhibit the spin-momentum properties using Bessel-type surface modes for the air–metal–air layered structure in Fig. 3. In Figs. 3(a-b) and 3(e-f), the dispersive momenta and horizontal SAMs of the symmetric and anti-symmetric modes are simultaneously inverted through the interface owing to the discontinuity of normal electric field $E_z$. Thus, in the upper and lower space, the directional vectors of horizontal SAMs can be recognized by the right-hand rule expressed by $\mathbf{S}_{||}\propto+\mathbf{p}\times\mathbf{n}$, whereas in the layer, the horizontal SAM is also locked with the dispersive momenta but the locking properties satisfy the left-hand rule $\tilde{\mathbf{S}}_{\parallel} \propto -\tilde{\mathbf{p}} \times \mathbf{n}$. Meanwhile, the normal components of SAMs are parallel through the interface as shown in Figs. (c) and (g). Together with the reversal of dispersive momenta on the two sides of the interface, the *i*SMLs between the dispersive momenta and the normal SAM components satisfy the right-hand screw rule in the upper and lower space but the left-hand screw rule in the layer. These conclusions are totally different from the evaluation of *i*SML properties made by ignoring the dispersive effect, where the *i*SML satisfies the right-hand rule in the dispersive medium.

Interestingly, in addition to the transverse spins, there is an additional spin component (in Figs. 3(d) and (h)) owing to the coupling between the individual waves at the upper and lower interfaces in the layer. By ignoring the dispersion, this coupling spin will lead to abnormity in judging the chirality of *i*SML for the symmetric and anti-symmetric modes. The coupling spin can be decomposed into two contributions: (1) the interference spin between the upper and the lower waves; and (2) the coupling polarization ellipticity between the *x*/*y*-component of upper wave and the *y*/*x*-component of lower wave. For contribution (1), the interference between the upper and lower waves introduces inhomogeneities into the field, and these inhomogeneities result in the transverse spin. Thus, the contribution (1) is consistent with the unified property of the transverse spin [65], whereas the contribution (2) should be regarded as the longitudinal spin. Figures 3(d) and (h) show that the longitudinal spins do not possess the *i*SML, because the directional vectors of the longitudinal spins depend on the symmetry and propagating direction of the modes simultaneously. This extraordinary longitudinal spin generated from the purely polarized field can also be found for dipole radiations [67].

The *i*SML properties are further demonstrated with chiral spin textures. The formation of photonic chiral spin texture originates from the conservation of total angular momentum and subluminal transportation of photons, and the stability is assured by the system's symmetry [68]. In the C4 symmetry, the photonic meron spin lattice can be obtained in the presence of optical spin–orbit coupling (where the photonic skyrmion lattices are present in the C6 rotating symmetry [54]). The vector diagrams of dispersive momenta at different values of $z$ of the material layer are similar and thus only one diagram is presented in Fig. 4(a), which contains multiple positive and negative vortexes. The SAMs at $z$ = +12.5, 0 and –12.5 nm are respectively shown in Figs. 4(b–d). From the vector diagrams in Figs. 4(b) and 4(d), the spin vectors in the two planes can be regarded as photonic meron lattices whose skyrmion numbers are ±1/2. It is observed that the normal spin vector is locked with the dispersive momentum and satisfies the left-hand rule. Moreover, the directions of chiral whirling for the photonic meron lattices are opposite in Figs. 4(b) and 4(d). This is because the horizontal SAM components are in opposite directions owing to the reversal of the outer normal direction in the two planes. In the center of the material layer, the horizontal SAM disappears and only the normal SAM component exists. This lattice can be regarded as photons with alternating positive and negative spins, which has potential application in data storage [69].

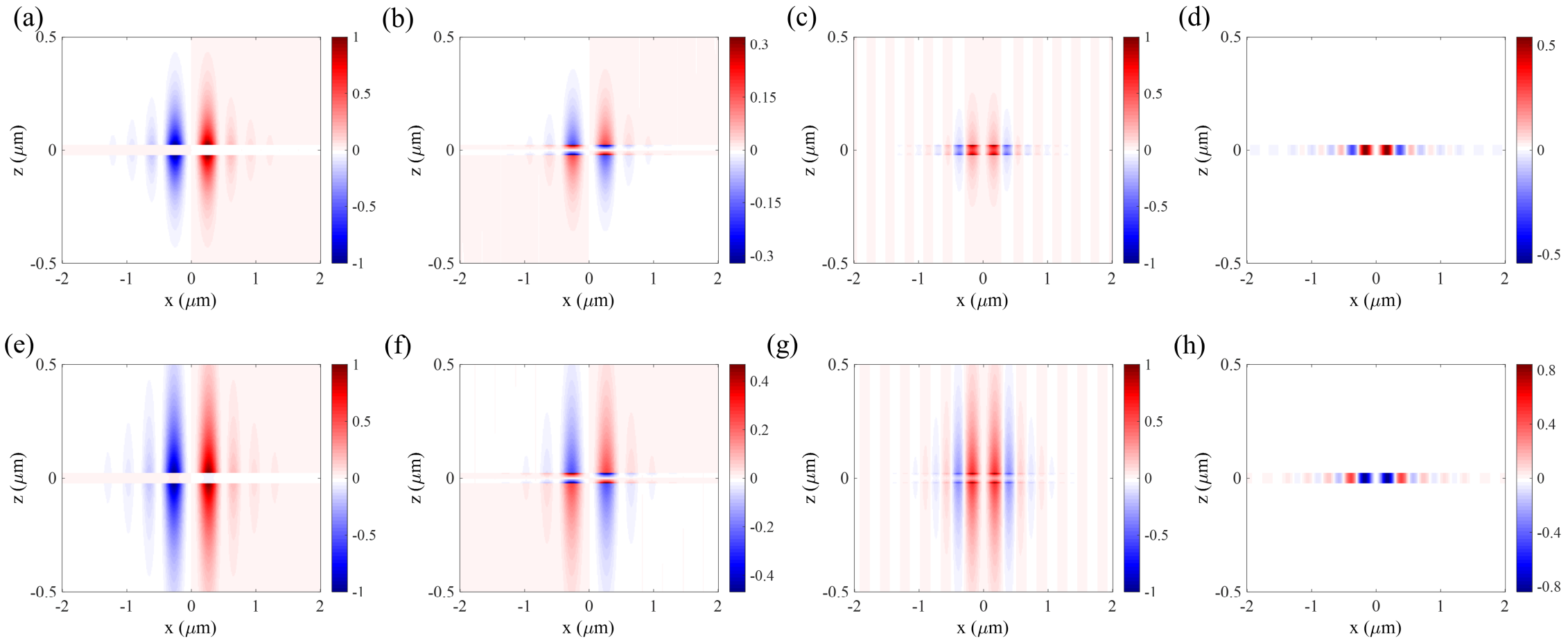

**FIG. 3.** *i*SML for +2-order Bessel-type surface EM modes. In the *xz*-plane, the figure presents the (a) momentum $\tilde{p}_y$, (b) SAM $\tilde{S}_x$, (c) $\tilde{S}_z$, and (d) longitudinal spin $\tilde{S}_l$ for the symmetric mode and the (e) momentum $\tilde{p}_y$, (f) SAM $\tilde{S}_x$, (g) $\tilde{S}_z$, and (h) longitudinal spin $\tilde{S}_l$ for the anti-symmetric mode. In the layer, the transverse spins are locked with the dispersive momentum and the locking property satisfies the left-hand rule. The longitudinal spins originated from the coupling between the surface waves only exist in the layer. The longitudinal spins are opposite for the two modes, which indicates the longitudinal spin does not possess *i*SML. The thickness of metal (Au [70]) is 50nm.

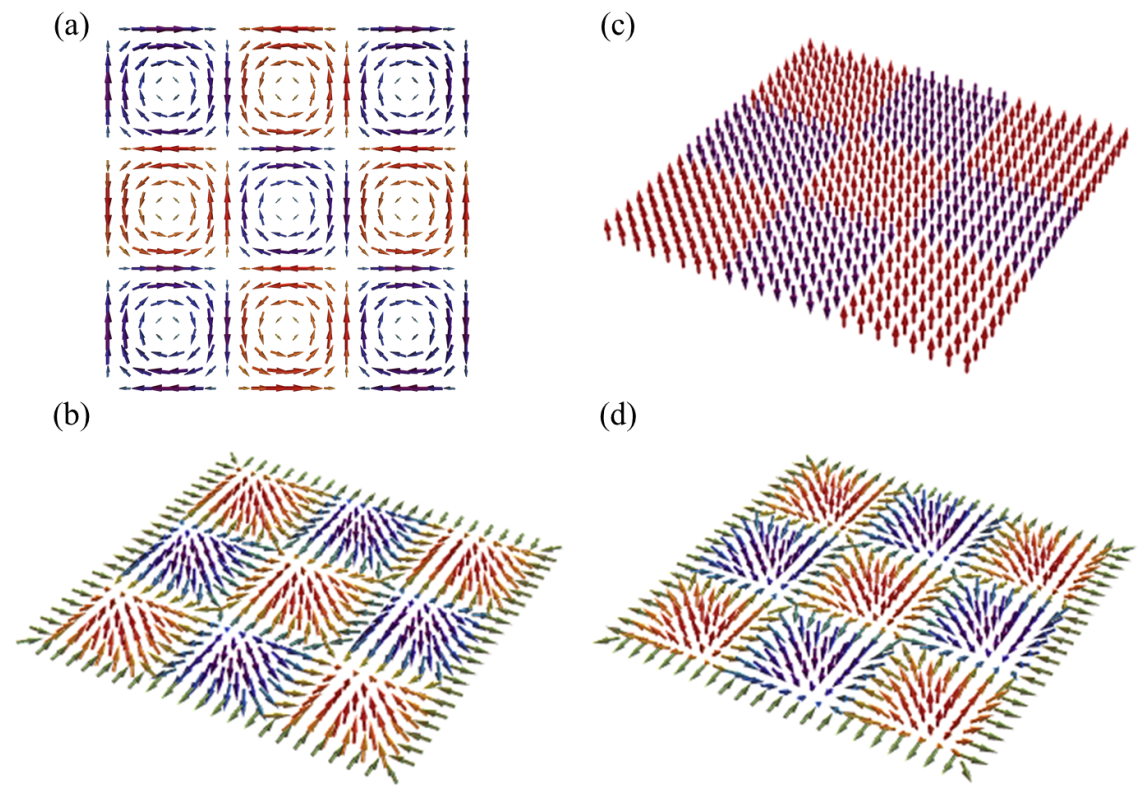

**FIG. 4.** Vector diagrams of (a) the dispersive momentum and spin textures at the planes (b) $z$=+12.5nm, (c) $z$=0 and (d) $z$=–12.5nm for the meron lattices in the layer of air-metal-air structure constructed by the symmetric mode. The vector properties of dispersive momentum are the same for the three spin textures. The spin textures are locked with the dispersive momentum and the locking properties satisfy the left-hand rule universally. The skyrmion number of the central photonic meron unit is –1/2 for (b) and (d). The horizontal components in (b) and (d) are enhanced by 5 times.

In summary, we demonstrated dispersive SMEs and associated *i*SML properties of surface EM waves in dispersive system. In the metal/magnetic materials, the transverse spin is locked with the dispersive momentum and satisfies the left-hand rule universally. Remarkably, the dispersive SMEs show that the *i*SML is affected by the structural and material properties, which provides guidance for tuning the *i*SML by designing the structure and dispersion. Moreover, in addition to the transverse spin, there is longitudinal spin due to coupling polarization ellipticity between the orthometric polarized components of the surface waves. The longitudinal spin is determined by the mode's symmetry and does not possess *i*SML. This extraordinary longitudinal spin generated by the purely polarized field is fascinating and was barely known previously. Finally, we exhibited diverse photonic topological lattices under varying rotating symmetry to demonstrate the *i*SML property. Our theory provides an efficient toolbox for the description of *i*SML of light in both dispersive and nondispersive systems and is expected to have widespread application in spin-based nanodevices.

This work was supported, in part, by Guangdong Major Project of Basic Research No. 2020B0301030009, National Natural Science Foundation of China grants U1701661, 61935013, 62075139, 61427819, 61622504, 12174266, and 12047540. L.D. acknowledges the support given by the Guangdong Special Support Program.

## REFERENCES

†pittshiustc@gmail.com
‡xcyuan@szu.edu.cn

# Supplemental materials for
# Intrinsic spin-momentum dynamics of surface electromagnetic waves in complex dispersive system

Peng Shi†, Xinrui Lei, Qiang Zhang, Heng Li, Luping Du, Xiaocong Yuan‡

*Nanophotonics Research Center, Shenzhen Key Laboratory of Micro-Scale Optical Information Technology & Institute of Microscale Optoelectronics, Shenzhen University, Shenzhen 518060, China*

*Corresponding author: †pittshiustc@gmail.com*; *‡xcyuan@szu.edu.cn*

## Contents:

# I. Mode properties for the surface EM waves in multilayered configuration

Multilayered configuration is beneficial for designing and fabricating the dispersion-engineered artificial materials, such as: photonic crystals [S1], plasmonics [S2], negative-index metamaterials [S3-S4], the hyperbolic metamaterials [S5-S6] and the metasurface [S7], etc. This is the main motivation that we use the multilayered system to approach the complex dispersive system in our work.

The *p*-polarized (transverse magnetic, TM) or *s*-polarized (transverse electric, TE) surface electromagnetic (EM) mode can be excited at the optical interfaces in multilayered systems. In the section, we will deduce and summarize the mode properties of *p*-polarized and *s*-polarized surface EM modes for the three types of multilayered structures as shown in Fig. S1. We mainly aim to show that there will be coupling EM terms while the layer is introduced into the structure, which lead to the different spin-momentum properties comparing to those of single interfacial system in Fig. S1(a). Noteworthily, the calculated methods to obtain the field distributions and the dispersion relations can be extended into arbitrary multilayered configurations.

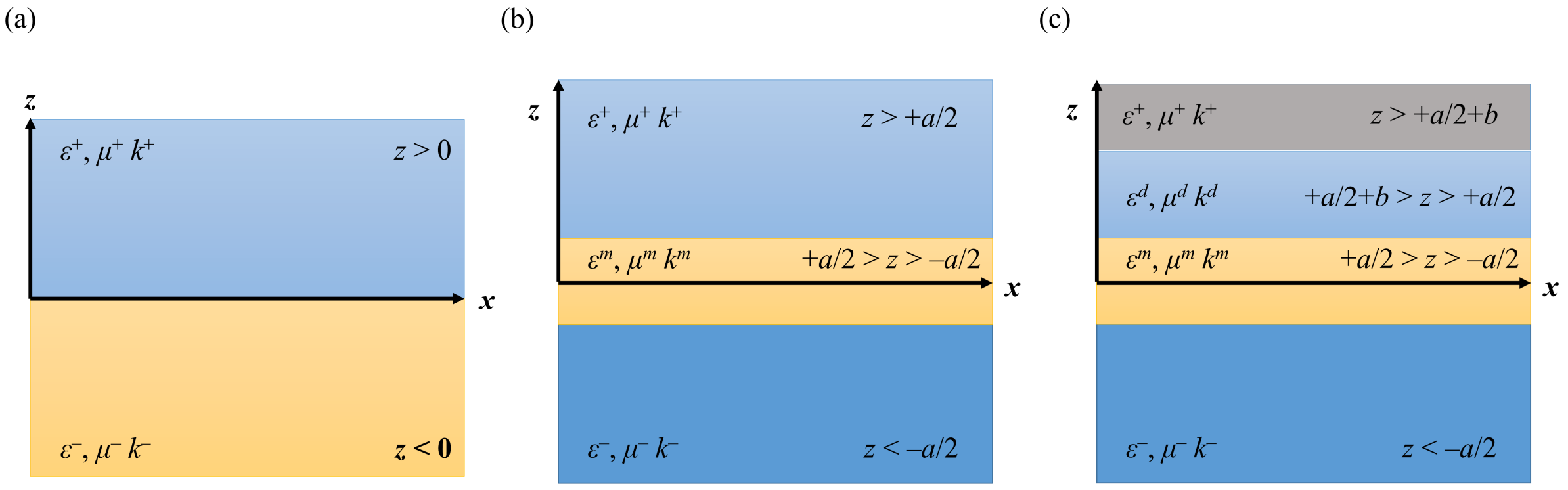

**Fig. S1.** (a) Schematic diagram of singe interface configuration to excite the *p*/*s*-polarized surface modes. The interface is localized at the plane $z$ = 0. (b) Schematic diagram of one-layer configuration to excite the *p*/*s*-polarized surface modes. The interfaces are localized at the planes $z = +a/2$ and $z = -a/2$. (c) Schematic diagram of two-layers configuration (such as: the dielectric waveguide structure) to excite the *p*/*s*-polarized surface modes. The interfaces are localized at the planes $z = -a/2$, $z = +a/2$ and $z = +a/2+b$, respectively.

As shown in Fig. S1(a), at an interface between two materials (permittivity $\varepsilon^{\pm}$, permeability $\mu^{\pm}$ and total wavevector $k^{\pm}$), the *p*-polarized surface modes should satisfy the relations: $H_z = 0$ and $\partial/\partial z = \mp k_z^{\pm}$, where $ik_z^{\pm}$ is the normal wavevector. From the Maxwell's equations, the electric/magnetic field components in Cartesian coordinates ($x$, $y$, $z$) have the expressions:

$$\begin{array}{lll} E_x^{\pm} = \mp\dfrac{k_z^{\pm}}{\beta^2}\dfrac{\partial E_z^{\pm}}{\partial x} & E_y^{\pm} = \mp\dfrac{k_z^{\pm}}{\beta^2}\dfrac{\partial E_z^{\pm}}{\partial y} & E_z^{\pm} = \dfrac{A_{\pm}}{\varepsilon^{\pm}}\xi(x,y)e^{\mp k_z^{\pm} z} \\ H_x^{\pm} = -\dfrac{i\omega\varepsilon^{\pm}}{\beta^2}\dfrac{\partial E_z^{\pm}}{\partial y} & H_y^{\pm} = +\dfrac{i\omega\varepsilon^{\pm}}{\beta^2}\dfrac{\partial E_z^{\pm}}{\partial x} & H_z^{\pm} = 0 \end{array} . \quad \text{(S1)}$$

Here, $\beta = \sqrt{k^{\pm 2} + k_z^{\pm 2}}$ is the horizontal wavevector (propagating constant) and the $z$-component electric field fulfills the transverse Helmholtz equation $\nabla_\perp^2 \xi + \beta^2 \xi = 0$ with $\nabla_\perp^2 = \partial^2/\partial x^2 + \partial^2/\partial y^2$. By considering the EM boundary conditions, the dispersion relation can be expressed as

$$A_+ = A_- , \ \frac{k_z^+}{k_z^-} = -\frac{\varepsilon^+}{\varepsilon^-} \text{ and } \beta = \omega\sqrt{\frac{\varepsilon^+\varepsilon^-\left(\varepsilon^-\mu^+ - \varepsilon^+\mu^-\right)}{\left(\varepsilon^-\right)^2 - \left(\varepsilon^+\right)^2}} . \quad \text{(S2)}$$

On the other hand, the $s$-polarized surface modes should satisfy the relations: $E_z = 0$ and $\partial/\partial z = \mp k_z^{\pm}$. From the Maxwell's equations, the electric/magnetic field components of $s$-polarized surface modes have the expressions:

$$\begin{array}{lll} E_x^{\pm} = +\dfrac{i\omega\mu^{\pm}}{\beta^2}\dfrac{\partial H_z^{\pm}}{\partial y} & E_y^{\pm} = -\dfrac{i\omega\mu^{\pm}}{\beta^2}\dfrac{\partial H_z^{\pm}}{\partial x} & E_z^{\pm} = 0 \\ H_x^{\pm} = \mp\dfrac{k_z^{\pm}}{\beta^2}\dfrac{\partial H_z^{\pm}}{\partial x} & H_y^{\pm} = \mp\dfrac{k_z^{\pm}}{\beta^2}\dfrac{\partial H_z^{\pm}}{\partial y} & H_z^{\pm} = \dfrac{A_{\pm}}{\mu^{\pm}}\zeta(x,y)e^{\mp k_z^{\pm} z} \end{array}. \tag{S3}$$

Here, $\beta = \sqrt{k^{\pm 2} + k_z^{\pm 2}}$ is also the horizontal wavevector of the $s$-polarized surface modes and the $z$-component magnetic field fulfills the transverse Helmholtz equation $\nabla_{\perp}^2\zeta + \beta^2\zeta = 0$. By considering the EM boundary conditions, the dispersion relation can be expressed as

$$A_+ = A_-,\ \frac{k_z^+}{k_z^-} = -\frac{\mu^+}{\mu^-} \text{ and } \beta = \omega\sqrt{\frac{\mu^+\mu^-\left(\varepsilon^+\mu^- - \varepsilon^-\mu^+\right)}{\left(\mu^-\right)^2 - \left(\mu^+\right)^2}}. \tag{S4}$$

To transfer the Eq. (S1) and Eq. (S3) from the Cartesian coordinates into the cylindrical coordinates $(r,\varphi,z)$, one can employ the matrices:

$$\begin{pmatrix} E_r \\ E_\varphi \\ E_z \end{pmatrix} = \begin{pmatrix} \cos\varphi & \sin\varphi & 0 \\ -\sin\varphi & \cos\varphi & 0 \\ 0 & 0 & 1 \end{pmatrix}\begin{pmatrix} E_x \\ E_y \\ E_z \end{pmatrix}, \tag{S5}$$

$$\begin{pmatrix} E_x \\ E_y \\ E_z \end{pmatrix} = \begin{pmatrix} \cos\varphi & -\sin\varphi & 0 \\ \sin\varphi & \cos\varphi & 0 \\ 0 & 0 & 1 \end{pmatrix}\begin{pmatrix} E_r \\ E_\varphi \\ E_z \end{pmatrix}, \tag{S6}$$

$$\begin{pmatrix} \partial/\partial x \\ \partial/\partial y \\ \partial/\partial z \end{pmatrix} = \begin{pmatrix} \cos\varphi & -\sin\varphi/r & 0 \\ \sin\varphi & \cos\varphi/r & 0 \\ 0 & 0 & 1 \end{pmatrix}\begin{pmatrix} \partial/\partial r \\ \partial/\partial \varphi \\ \partial/\partial z \end{pmatrix}. \tag{S7}$$

The Eqs. (S1-S7) are consistent with the Ref. [27].

Then, to derivate the dispersion relations and mode distributions in the layered system, we need to employ the results given in Eqs. (S1-S7). For the one-layer system as shown in Fig. S1(b), the electric/magnetic field components of the $p$-polarized surface modes have the expressions:

$$\begin{array}{ccc} z > +\dfrac{a}{2} & -\dfrac{a}{2} < z < +\dfrac{a}{2} & z < -\dfrac{a}{2} \\ E_x^+ = -\dfrac{k_z^+}{\beta^2}\dfrac{\partial E_z^+}{\partial x} & E_x^m = E_x^{m+} + E_x^{m-} = +\dfrac{k_z^m}{\beta^2}\dfrac{\partial E_z^{m+}}{\partial x} - \dfrac{k_z^m}{\beta^2}\dfrac{\partial E_z^{m-}}{\partial x} & E_x^- = +\dfrac{k_z^-}{\beta^2}\dfrac{\partial E_z^-}{\partial x} \\ E_y^+ = -\dfrac{k_z^+}{\beta^2}\dfrac{\partial E_z^+}{\partial y} & E_y^m = E_y^{m+} + E_y^{m-} = +\dfrac{k_z^m}{\beta^2}\dfrac{\partial E_z^{m+}}{\partial y} - \dfrac{k_z^m}{\beta^2}\dfrac{\partial E_z^{m-}}{\partial y} & E_y^- = +\dfrac{k_z^-}{\beta^2}\dfrac{\partial E_z^-}{\partial y} \\ E_z^+ = \dfrac{A_+}{\varepsilon^+}\xi e^{-k_z^+(z-a/2)} & E_z^m = E_z^{m+} + E_z^{m-} = \dfrac{B_+}{\varepsilon^m}\xi e^{+k_z^m(z-a/2)} + \dfrac{B_-}{\varepsilon^m}\xi e^{-k_z^m(z+a/2)} & E_z^- = \dfrac{A_-}{\varepsilon^-}\xi e^{+k_z^-(z+a/2)} \\ H_x^+ = -\dfrac{i\omega\varepsilon^+}{\beta^2}\dfrac{\partial E_z^+}{\partial y} & H_x^m = H_x^{m+} + H_x^{m-} = -\dfrac{i\omega\varepsilon^m}{\beta^2}\dfrac{\partial E_z^{m+}}{\partial y} - \dfrac{i\omega\varepsilon^m}{\beta^2}\dfrac{\partial E_z^{m-}}{\partial y} & H_x^- = -\dfrac{i\omega\varepsilon^-}{\beta^2}\dfrac{\partial E_z^-}{\partial y} \\ H_y^+ = +\dfrac{i\omega\varepsilon^+}{\beta^2}\dfrac{\partial E_z^+}{\partial x} & H_y^m = H_y^{m+} + H_y^{m-} = +\dfrac{i\omega\varepsilon^m}{\beta^2}\dfrac{\partial E_z^{m+}}{\partial x} + \dfrac{i\omega\varepsilon^m}{\beta^2}\dfrac{\partial E_z^{m-}}{\partial x} & H_y^- = +\dfrac{i\omega\varepsilon^-}{\beta^2}\dfrac{\partial E_z^-}{\partial x} \\ H_z^+ = 0 & H_z^m = 0 & H_z^- = 0 \end{array}. \tag{S8}$$

Here, $\xi(x,y)$ is the function of horizontal coordinates $(x,y)$ and we ignore the $(x,y)$-dependent in the expressions for convenience. The $z$-component electric field fulfills the transverse Helmholtz equation $\nabla_\perp^2\xi+\beta^2\xi=0$. By considering the EM boundary conditions, the dispersion relation can be expressed as

$$\frac{B_+}{B_-}=\frac{\dfrac{k_z^m}{\varepsilon^m}-\dfrac{k_z^+}{\varepsilon^+}}{\dfrac{k_z^m}{\varepsilon^m}+\dfrac{k_z^+}{\varepsilon^+}}e^{-k_z^m a}\quad \frac{B_-}{B_+}=\frac{\dfrac{k_z^m}{\varepsilon^m}-\dfrac{k_z^-}{\varepsilon^-}}{\dfrac{k_z^m}{\varepsilon^m}+\dfrac{k_z^-}{\varepsilon^-}}e^{-k_z^m a}\quad \begin{array}{l}+A_+=+B_++B_-e^{-k_z^m a}\\ +B_+e^{-k_z^m a}+B_-=+A_-\end{array}, \tag{S9a}$$

$$e^{-2k_z^m a}=\frac{\left(k_z^m/\varepsilon^m+k_z^+/\varepsilon^+\right)\left(k_z^m/\varepsilon^m+k_z^-/\varepsilon^-\right)}{\left(k_z^m/\varepsilon^m-k_z^+/\varepsilon^+\right)\left(k_z^m/\varepsilon^m-k_z^-/\varepsilon^-\right)}, \tag{S9b}$$

and

$$\beta^2=\omega^2\varepsilon^i\mu^i+\left(k_z^i\right)^2, \tag{S9c}$$

where $i$ = +, – and $m$ are corresponding to the materials in the regions $z>+a/2$, $z<-a/2$ and $-a/2<z<+a/2$, respectively. Since only the relative amplitude makes physical sense, one can set $B_+$ or $B_-$ to be 1 and the other amplitude coefficients can be calculated properly.

On the other hand, for the $s$-polarized surface modes in the three-layers system, the electric/magnetic field components of the $s$-polarized surface modes have the expressions:

$$\begin{array}{ccc}
z>+\dfrac{a}{2} & -\dfrac{a}{2}<z<+\dfrac{a}{2} & z<-\dfrac{a}{2}\\
E_x^+=+\dfrac{i\omega\mu^+}{\beta^2}\dfrac{\partial H_z^+}{\partial y} & E_x^m=E_x^{m+}+E_x^{m-}=\dfrac{i\omega\mu^m}{\beta^2}\dfrac{\partial H_z^{m+}}{\partial y}+\dfrac{i\omega\mu^m}{\beta^2}\dfrac{\partial H_z^{m-}}{\partial y} & E_x^-=+\dfrac{i\omega\mu^-}{\beta^2}\dfrac{\partial H_z^-}{\partial y}\\
E_y^+=-\dfrac{i\omega\mu^+}{\beta^2}\dfrac{\partial H_z^+}{\partial x} & E_y^m=E_y^{m+}+E_y^{m-}=-\dfrac{i\omega\mu^m}{\beta^2}\dfrac{\partial H_z^{m+}}{\partial x}-\dfrac{i\omega\mu^m}{\beta^2}\dfrac{\partial H_z^{m-}}{\partial x} & E_y^-=-\dfrac{i\omega\mu^-}{\beta^2}\dfrac{\partial H_z^-}{\partial x}\\
E_z^+=0 & E_z^m=0 & E_z^-=0\\
H_x^+=-\dfrac{k_z^+}{\beta^2}\dfrac{\partial H_z^+}{\partial x} & H_x^m=H_x^{m+}+H_x^{m-}=+\dfrac{k_z^m}{\beta^2}\dfrac{\partial H_z^{m+}}{\partial x}-\dfrac{k_z^m}{\beta^2}\dfrac{\partial H_z^{m-}}{\partial x} & H_x^-=+\dfrac{k_z^-}{\beta^2}\dfrac{\partial H_z^-}{\partial x}\\
H_y^+=-\dfrac{k_z^+}{\beta^2}\dfrac{\partial H_z^+}{\partial y} & H_y^m=H_y^{m+}+H_y^{m-}=+\dfrac{k_z^m}{\beta^2}\dfrac{\partial H_z^{m+}}{\partial y}-\dfrac{k_z^m}{\beta^2}\dfrac{\partial H_z^{m-}}{\partial y} & H_y^-=+\dfrac{k_z^-}{\beta^2}\dfrac{\partial H_z^-}{\partial y}\\
H_z^+=\dfrac{A_+}{\mu^+}\zeta e^{-k_z^+(z-a/2)} & H_z^m=H_z^{m+}+H_z^{m-}=\dfrac{B_+}{\mu^m}\zeta e^{+k_z^m(z-a/2)}+\dfrac{B_-}{\mu^m}\zeta e^{-k_z^m(z+a/2)} & H_z^-=\dfrac{A_-}{\mu^-}\zeta e^{+k_z^-(z+a/2)}
\end{array}. \tag{S10}$$

Here, $\zeta(x,y)$ is the function of horizontal coordinates $(x,y)$ and we ignore the $(x,y)$-dependent in the expressions for convenience. The $z$-component magnetic field fulfills the transverse Helmholtz equation $\nabla_\perp^2\zeta+\beta^2\zeta=0$. By considering the boundary conditions, the dispersion relation can be expressed as

$$\frac{B_+}{B_-}=\frac{\dfrac{k_z^m}{\mu^m}-\dfrac{k_z^+}{\mu^+}}{\dfrac{k_z^m}{\mu^m}+\dfrac{k_z^+}{\mu^+}}e^{-k_z^m a}\quad \frac{B_-}{B_+}=\frac{\dfrac{k_z^m}{\mu^m}-\dfrac{k_z^-}{\mu^-}}{\dfrac{k_z^m}{\mu^m}+\dfrac{k_z^-}{\mu^-}}e^{-k_z^m a}\quad \begin{array}{l}+A_+=+B_++B_-e^{-k_z^m a}\\ +B_+e^{-k_z^m a}+B_-=+A_-\end{array}, \tag{S11a}$$

$$e^{-2k_z^m a}=\frac{\left(k_z^m/\mu^m+k_z^+/\mu^+\right)\left(k_z^m/\mu^m+k_z^-/\mu^-\right)}{\left(k_z^m/\mu^m-k_z^+/\mu^+\right)\left(k_z^m/\mu^m-k_z^-/\mu^-\right)}, \tag{S11b}$$

and

$$\beta^2=\omega^2\varepsilon^i\mu^i+\left(k_z^i\right)^2, \tag{S11c}$$

where $i$ = +, – and $m$ are corresponding to the materials in the regions $z>+a/2$, $z<-a/2$ and $-a/2<z<+a/2$, respectively. In the same way, one can set $B_+$ or $B_-$ to be 1 to calculate the other amplitude coefficients.

Subsequently, more intricately, we consider the field distributions and dispersion relations for the multilayered structure containing layers more than one layer as shown in Fig. 1(c). The electric/magnetic field components of the $p$-polarized surface modes have the expressions:

$$z>+\frac{a}{2}+b:\quad
\begin{aligned}
E_x^{+} &= -\frac{k_z^{+}}{\beta^2}\frac{\partial E_z^{+}}{\partial x}\\
E_y^{+} &= -\frac{k_z^{+}}{\beta^2}\frac{\partial E_z^{+}}{\partial y}\\
E_z^{+} &= \frac{A_+}{\varepsilon^{+}}\xi e^{-k_z^{+}(z-a/2-b)}\\
H_x^{+} &= -\frac{i\omega\varepsilon^{+}}{\beta^2}\frac{\partial E_z^{+}}{\partial y}\\
H_y^{+} &= +\frac{i\omega\varepsilon^{+}}{\beta^2}\frac{\partial E_z^{+}}{\partial x}\\
H_z^{+} &= 0
\end{aligned}$$

$$+\frac{a}{2}<z<+\frac{a}{2}+b:\quad
\begin{aligned}
E_x^{d} &= E_x^{d+}+E_x^{d-} = +\frac{k_z^{d}}{\beta^2}\frac{\partial E_z^{d+}}{\partial x}-\frac{k_z^{d}}{\beta^2}\frac{\partial E_z^{d-}}{\partial x}\\
E_y^{d} &= E_y^{d+}+E_y^{d-} = +\frac{k_z^{d}}{\beta^2}\frac{\partial E_z^{d+}}{\partial y}-\frac{k_z^{d}}{\beta^2}\frac{\partial E_z^{d-}}{\partial y}\\
E_z^{d} &= E_z^{d+}+E_z^{d-} = \frac{C_+}{\varepsilon^{d}}\xi e^{+k_z^{d}(z-a/2-b)}+\frac{C_-}{\varepsilon^{d}}\xi e^{-k_z^{d}(z-a/2)}\\
H_x^{d} &= H_x^{d+}+H_x^{d-} = -\frac{i\omega\varepsilon^{d}}{\beta^2}\frac{\partial E_z^{d+}}{\partial y}-\frac{i\omega\varepsilon^{d}}{\beta^2}\frac{\partial E_z^{d-}}{\partial y}\\
H_y^{d} &= H_y^{d+}+H_y^{d-} = \frac{i\omega\varepsilon^{d}}{\beta^2}\frac{\partial E_z^{d+}}{\partial x}+\frac{i\omega\varepsilon^{d}}{\beta^2}\frac{\partial E_z^{d-}}{\partial x}\\
H_z^{d} &= 0
\end{aligned}$$

$$-\frac{a}{2}<z<+\frac{a}{2}:\quad
\begin{aligned}
E_x^{m} &= E_x^{m+}+E_x^{m-} = +\frac{k_z^{m}}{\beta^2}\frac{\partial E_z^{m+}}{\partial x}-\frac{k_z^{m}}{\beta^2}\frac{\partial E_z^{m-}}{\partial x}\\
E_y^{m} &= E_y^{m+}+E_y^{m-} = +\frac{k_z^{m}}{\beta^2}\frac{\partial E_z^{m+}}{\partial y}-\frac{k_z^{m}}{\beta^2}\frac{\partial E_z^{m-}}{\partial y}\\
E_z^{m} &= E_z^{m+}+E_z^{m-} = \frac{B_+}{\varepsilon^{m}}\xi e^{+k_z^{m}(z-a/2)}+\frac{B_-}{\varepsilon^{m}}\xi e^{-k_z^{m}(z+a/2)}\\
H_x^{m} &= H_x^{m+}+H_x^{m-} = -\frac{i\omega\varepsilon^{m}}{\beta^2}\frac{\partial E_z^{m+}}{\partial y}-\frac{i\omega\varepsilon^{m}}{\beta^2}\frac{\partial E_z^{m-}}{\partial y}\\
H_y^{m} &= H_y^{m+}+H_y^{m-} = \frac{i\omega\varepsilon^{m}}{\beta^2}\frac{\partial E_z^{m+}}{\partial x}+\frac{i\omega\varepsilon^{m}}{\beta^2}\frac{\partial E_z^{m-}}{\partial x}\\
H_z^{m} &= 0
\end{aligned}$$

$$z<-\frac{a}{2}:\quad
\begin{aligned}
E_x^{-} &= +\frac{k_z^{-}}{\beta^2}\frac{\partial E_z^{-}}{\partial x}\\
E_y^{-} &= +\frac{k_z^{-}}{\beta^2}\frac{\partial E_z^{-}}{\partial y}\\
E_z^{-} &= \frac{A_-}{\varepsilon^{-}}\xi e^{+k_z^{-}(z+a/2)}\\
H_x^{-} &= -\frac{i\omega\varepsilon^{-}}{\beta^2}\frac{\partial E_z^{-}}{\partial y}\\
H_y^{-} &= +\frac{i\omega\varepsilon^{-}}{\beta^2}\frac{\partial E_z^{-}}{\partial x}\\
H_z^{-} &= 0
\end{aligned}
\quad . \tag{S12}$$

The $z$-component electric field fulfills the transverse Helmholtz equation $\nabla_\perp^2\xi+\beta^2\xi=0$. By considering the boundary conditions, the dispersion relation can be expressed as

$$\begin{aligned}
&A_- = 1\\
&\frac{B_+k_z^m}{\varepsilon^m}e^{-k_z^m a}-\frac{B_-k_z^m}{\varepsilon^m}=\frac{k_z^-}{\varepsilon^-} \qquad -B_+e^{-k_z^m a}-B_-=-1\\
&\frac{C_+k_z^d}{\varepsilon^d}e^{-k_z^d b}-\frac{C_-k_z^d}{\varepsilon^d}=\frac{B_+k_z^m}{\varepsilon^m}-\frac{B_-k_z^m}{\varepsilon^m}e^{-k_z^m a} \qquad -C_+e^{-k_z^d b}-C_-=-B_+-B_-e^{-k_z^m a}\\
&-\frac{A_+k_z^+}{\varepsilon^+}=\frac{C_+k_z^d}{\varepsilon^d}-\frac{C_-k_z^d}{\varepsilon^d}e^{-k_z^d b} \qquad -A_+=-C_+-C_-e^{-k_z^d b}
\end{aligned}, \tag{S13a}$$

$$-\frac{\varepsilon^d k_z^+}{\varepsilon^+ k_z^d}\frac{C_+ + C_-e^{-k_z^d b}}{C_+ - C_-e^{-k_z^d b}}=1, \tag{S13b}$$

and

$$\beta^2=\omega^2\varepsilon^i\mu^i+\left(k_z^i\right)^2, \tag{S13c}$$

where $i$ = +, –, $m$, and $d$ are corresponding to the materials in the regions $z>+a/2$, $z<-a/2$, $-a/2<z<+a/2$ and $+a/2<z<+a/2+b$, respectively. The parameters $C_+$ and $C_-$ can be solved by Eq. (S13a).

On the other hand, for the $s$-polarized surface modes in the four-layers system, the electric/magnetic field components of the $s$-polarized surface modes have the expressions:

$$
\begin{array}{llll}
z>+\frac{a}{2}+b & +\frac{a}{2}<z<+\frac{a}{2}+b & -\frac{a}{2}<z<+\frac{a}{2} & z<-\frac{a}{2} \\
E_x^{+}=+\frac{i\omega\mu^{+}}{\beta^{2}}\frac{\partial H_z^{+}}{\partial y} & E_x^{d}=E_x^{d+}+E_x^{d-}=\frac{i\omega\mu^{d}}{\beta^{2}}\frac{\partial H_z^{d+}}{\partial y}+\frac{i\omega\mu^{d}}{\beta^{2}}\frac{\partial H_z^{d-}}{\partial y} & E_x^{m}=E_x^{m+}+E_x^{m-}=\frac{i\omega\mu^{m}}{\beta^{2}}\frac{\partial H_z^{m+}}{\partial y}+\frac{i\omega\mu^{m}}{\beta^{2}}\frac{\partial H_z^{m-}}{\partial y} & E_x^{-}=+\frac{i\omega\mu^{-}}{\beta^{2}}\frac{\partial H_z^{-}}{\partial y} \\
E_y^{+}=-\frac{i\omega\mu^{+}}{\beta^{2}}\frac{\partial H_z^{+}}{\partial x} & E_y^{d}=E_y^{d+}+E_y^{d-}=-\frac{i\omega\mu^{d}}{\beta^{2}}\frac{\partial H_z^{d+}}{\partial x}-\frac{i\omega\mu^{d}}{\beta^{2}}\frac{\partial H_z^{d-}}{\partial x} & E_y^{m}=E_y^{m+}+E_y^{m-}=-\frac{i\omega\mu^{m}}{\beta^{2}}\frac{\partial H_z^{m+}}{\partial x}-\frac{i\omega\mu^{m}}{\beta^{2}}\frac{\partial H_z^{m-}}{\partial x} & E_y^{-}=-\frac{i\omega\mu^{-}}{\beta^{2}}\frac{\partial H_z^{-}}{\partial x} \\
E_z^{+}=0 & E_z^{d}=0 & E_z^{m}=0 & E_z^{-}=0 \\
H_x^{+}=-\frac{k_z^{+}}{\beta^{2}}\frac{\partial H_z^{+}}{\partial x} & H_x^{d}=H_x^{d+}+H_x^{d-}=+\frac{k_z^{d}}{\beta^{2}}\frac{\partial H_z^{d+}}{\partial x}-\frac{k_z^{d}}{\beta^{2}}\frac{\partial H_z^{d-}}{\partial x} & H_x^{m}=H_x^{m+}+H_x^{m-}=+\frac{k_z^{m}}{\beta^{2}}\frac{\partial H_z^{m+}}{\partial x}-\frac{k_z^{m}}{\beta^{2}}\frac{\partial H_z^{m-}}{\partial x} & H_x^{-}=+\frac{k_z^{-}}{\beta^{2}}\frac{\partial H_z^{-}}{\partial x} \\
H_y^{+}=-\frac{k_z^{+}}{\beta^{2}}\frac{\partial H_z^{+}}{\partial y} & H_y^{d}=H_y^{d+}+H_y^{d-}=+\frac{k_z^{d}}{\beta^{2}}\frac{\partial H_z^{d+}}{\partial y}-\frac{k_z^{d}}{\beta^{2}}\frac{\partial H_z^{d-}}{\partial y} & H_y^{m}=H_y^{m+}+H_y^{m-}=+\frac{k_z^{m}}{\beta^{2}}\frac{\partial H_z^{m+}}{\partial y}-\frac{k_z^{m}}{\beta^{2}}\frac{\partial H_z^{m-}}{\partial y} & H_y^{-}=+\frac{k_z^{-}}{\beta^{2}}\frac{\partial H_z^{-}}{\partial y} \\
H_z^{+}=\frac{A_{+}}{\mu^{+}}\zeta e^{-k_z^{+}(z-a/2-b)} & H_z^{d}=H_z^{d+}+H_z^{d-}=\frac{C_{+}}{\mu^{d}}\zeta e^{+k_z^{d}(z-a/2-b)}+\frac{C_{-}}{\mu^{d}}\zeta e^{-k_z^{d}(z-a/2)} & H_z^{m}=H_z^{m+}+H_z^{m-}=\frac{B_{+}}{\mu^{m}}\zeta e^{+k_z^{m}(z-a/2)}+\frac{B_{-}}{\mu^{m}}\zeta e^{-k_z^{m}(z+a/2)} & H_z^{-}=\frac{A_{-}}{\mu^{-}}\zeta e^{+k_z^{-}(z+a/2)}
\end{array}
$$

. (S14)

The $z$-component magnetic field fulfills the transverse Helmholtz equation $\nabla_{\perp}^{2}\zeta+\beta^{2}\zeta=0$. By considering the boundary conditions, the dispersion relation can be expressed as

$$
\begin{array}{ll}
A_{-}=1 & \\
B_{+}e^{-k_z^{m}a}+B_{-}=1 & \frac{B_{+}k_z^{m}}{\mu^{m}}e^{-k_z^{m}a}-\frac{B_{-}k_z^{m}}{\mu^{m}}=+\frac{k_z^{-}}{\mu^{-}} \\
C_{+}e^{-k_z^{d}b}+C_{-}=B_{+}+B_{-}e^{-k_z^{m}a} & \frac{C_{+}k_z^{d}}{\mu^{d}}e^{-k_z^{d}b}-\frac{C_{-}k_z^{d}}{\mu^{d}}=\frac{B_{+}k_z^{m}}{\mu^{m}}-\frac{B_{-}k_z^{m}}{\mu^{m}}e^{-k_z^{m}a} \\
+A_{+}=C_{+}+C_{-}e^{-k_z^{d}b} & -\frac{A_{+}k_z^{+}}{\mu^{+}}=\frac{C_{+}k_z^{d}}{\mu^{d}}-\frac{C_{-}k_z^{d}}{\mu^{d}}e^{-k_z^{d}b}
\end{array},
\tag{S15a}
$$

$$
-\frac{\mu^{d}k_z^{+}}{\mu^{+}k_z^{d}}\frac{C_{+}+C_{-}e^{-k_z^{d}b}}{C_{+}-C_{-}e^{-k_z^{d}b}}=1,
\tag{S15b}
$$

and

$$
\beta^{2}=\omega^{2}\varepsilon^{i}\mu^{i}+\left(k_z^{i}\right)^{2},
\tag{S15c}
$$

where $i$ = +, –, $m$, and $d$ are corresponding to the materials in the regions $z>+a/2$, $z<-a/2$, $-a/2<z<+a/2$ and $+a/2<z<+a/2+b$, respectively. In the same way, the parameters $C_+$ and $C_-$ can be solved by Eq. (S15a).

From the derivations, one can found that the expressions of electric/magnetic fields for the one-layer system as shown in Fig. S1(b) are similar with those of two-layers system in Fig. S1(c). Thus, we can conclude that the theoretical results of one-layer system can be generalized to arbitrary multilayered systems.

In the following, to exhibit the mode properties visually, we will show the dispersion relations and the field distributions of plane wave solution and Bessel function solution for the one-layer configuration (air-metal-air). In the air-metal-air structure, there are symmetric and anti-symmetric modes exist. The symmetric mode has a larger propagating wavevector ($\beta$) comparing to that of the anti-symmetric mode as shown in Fig. S2. Noteworthily, the symmetry of mode is evaluated by the horizontal electric field component for the $p$-polarized mode.

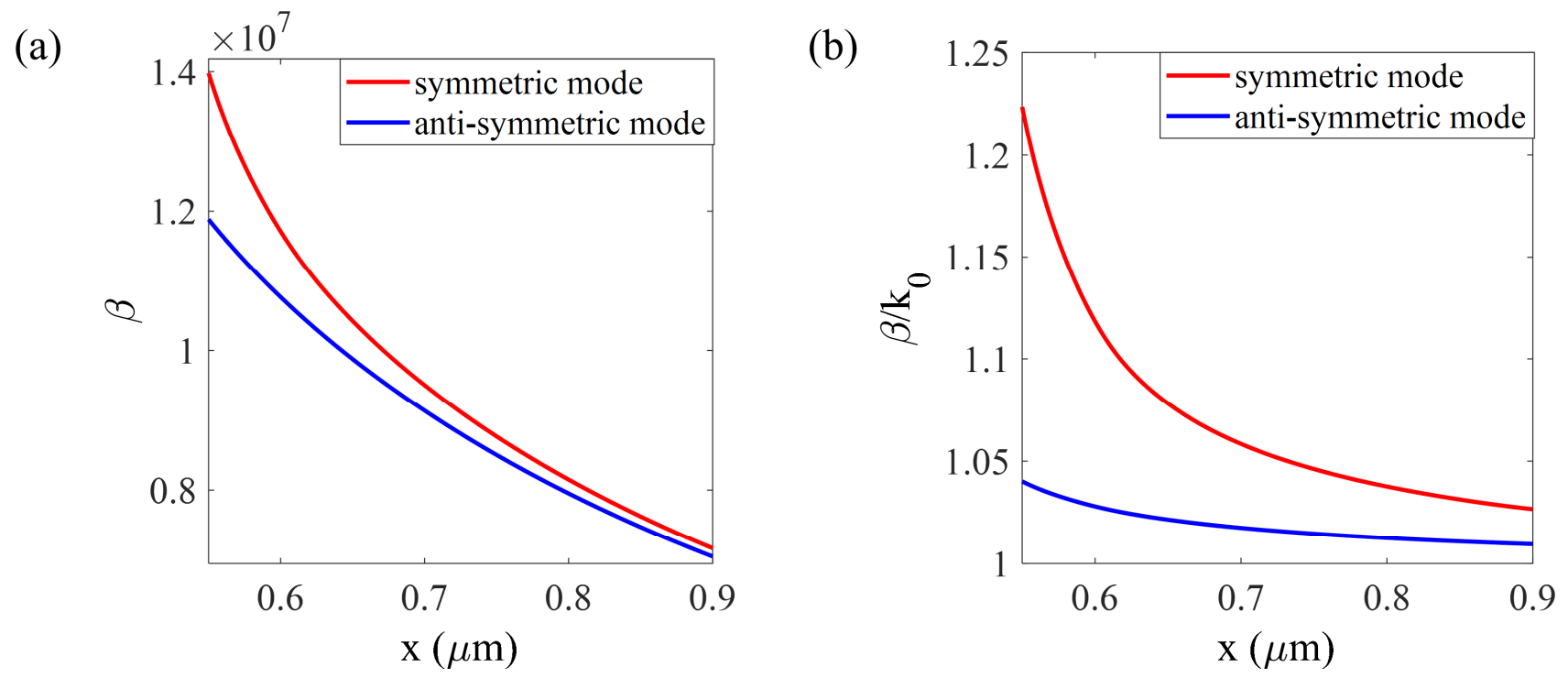

**Fig. S2.** (a) Propagating constant $\beta$ and (b) effective index $\beta/k_0$ via the wavelength ($\lambda$). The red line denotes the symmetric mode while the blue line indicates the anti-symmetric mode. The structure is air-metal-air and the material of metal layer is Au [70]. $k_0$ is the wavevector in vacuum.

For the surface plane wave solution, the expressions of electric Hertz potential can be expressed as

$$\xi = e^{i\beta x}, \tag{S16}$$

and the field distributions are:

$$\begin{array}{ccc}
z > +\dfrac{a}{2} & -\dfrac{a}{2} < z < +\dfrac{a}{2} & z < -\dfrac{a}{2} \\
E_x^{+} = -i\dfrac{A_+ k_z^{+}}{\varepsilon^{+}\beta} e^{i\beta x - k_z^{+}(z-a/2)} & \begin{aligned} E_x^{m} &= E_x^{m+} + E_x^{m-} \\ &= +i\dfrac{B_+ k_z^{m}}{\varepsilon^{m}\beta} e^{i\beta x + k_z^{m}(z-a/2)} - i\dfrac{B_- k_z^{m}}{\varepsilon^{m}\beta} e^{i\beta x - k_z^{m}(z+a/2)} \end{aligned} & E_x^{-} = +i\dfrac{A_- k_z^{-}}{\varepsilon^{-}\beta} e^{i\beta x + k_z^{-}(z+a/2)} \\
E_y^{+} = 0 & E_y^{m} = E_y^{m+} + E_y^{m-} = 0 & E_y^{-} = 0 \\
E_z^{+} = \dfrac{A_+}{\varepsilon^{+}} e^{i\beta x} e^{-k_z^{+}(z-a/2)} & \begin{aligned} E_z^{m} &= E_z^{m+} + E_z^{m-} \\ &= \dfrac{B_+}{\varepsilon^{m}} e^{i\beta x} e^{+k_z^{m}(z-a/2)} + \dfrac{B_-}{\varepsilon^{m}} e^{i\beta x} e^{-k_z^{m}(z+a/2)} \end{aligned} & E_z^{-} = \dfrac{A_-}{\varepsilon^{-}} e^{i\beta x} e^{+k_z^{-}(z+a/2)} \\
H_x^{+} = 0 & H_x^{m} = H_x^{m+} + H_x^{m-} = 0 & H_x^{-} = 0 \\
H_y^{+} = -\dfrac{A_+\omega}{\beta} e^{i\beta x - k_z^{+}(z-a/2)} & \begin{aligned} H_y^{m} &= H_y^{m+} + H_y^{m-} \\ &= -\dfrac{B_+\omega}{\beta} e^{i\beta x + k_z^{m}(z-a/2)} - \dfrac{B_-\omega}{\beta} e^{i\beta x - k_z^{m}(z+a/2)} \end{aligned} & H_y^{-} = -\dfrac{A_-\omega}{\beta} e^{i\beta x + k_z^{-}(z+a/2)} \\
H_z^{+} = 0 & H_z^{m} = 0 & H_z^{-} = 0
\end{array} . \tag{S17}$$

The field distributions are shown in Fig. S3. Particularly, we show the kinetic momentum and dispersionless spin angular momentum (SAM) for the symmetric and anti-symmetric modes (the kinetic momentum $\mathbf{p} = \varepsilon_0\mu_0 \,\mathrm{Re}\{\mathbf{E}^* \times \mathbf{H}\}/2$ in vacuum and the dispersionless SAM $\mathbf{S} = \mathrm{Im}\{\varepsilon\mathbf{E}^* \times \mathbf{E} + \mu\mathbf{H}^* \times \mathbf{H}\}/4\omega$ ). In the case, one can observe that the SAM is locked with the kinetic momentum obeying the right-hand rule, no matter whether in the dielectric or in the metal. Since the plane wave solution does not contain $S_z$, we consider a Bessel function solution [S8]

$$\xi = J_m(\beta r) e^{im\varphi}. \tag{S18}$$

Here, $J_m$ is the $m$-order Bessel function of first type. We also show the kinetic momentum and dispersionless SAM for the symmetric and anti-symmetric modes of Bessel function solution. Since the locking properties between the kinetic momentum and horizontal spin ($S_r$) are similar with those of plane wave solution, we do not show those SAM components here. In Fig. S4 (e, f, k, l), one can find that the spin-momentum locking properties are opposite for the symmetric and anti-symmetric modes. This phenomenon is abnormal, which is one of main motivation of our work. We will explain that this is originated from two reasons: 1. the dispersion is ignored in the investigation of spin-momentum properties of EM field; 2. There is a longitudinal spin component owing to the coupling between the orthogonal components of the horizontal EM fields in the upper and lower interfaces and this longitudinal spin does not possess the property of spin-momentum locking.

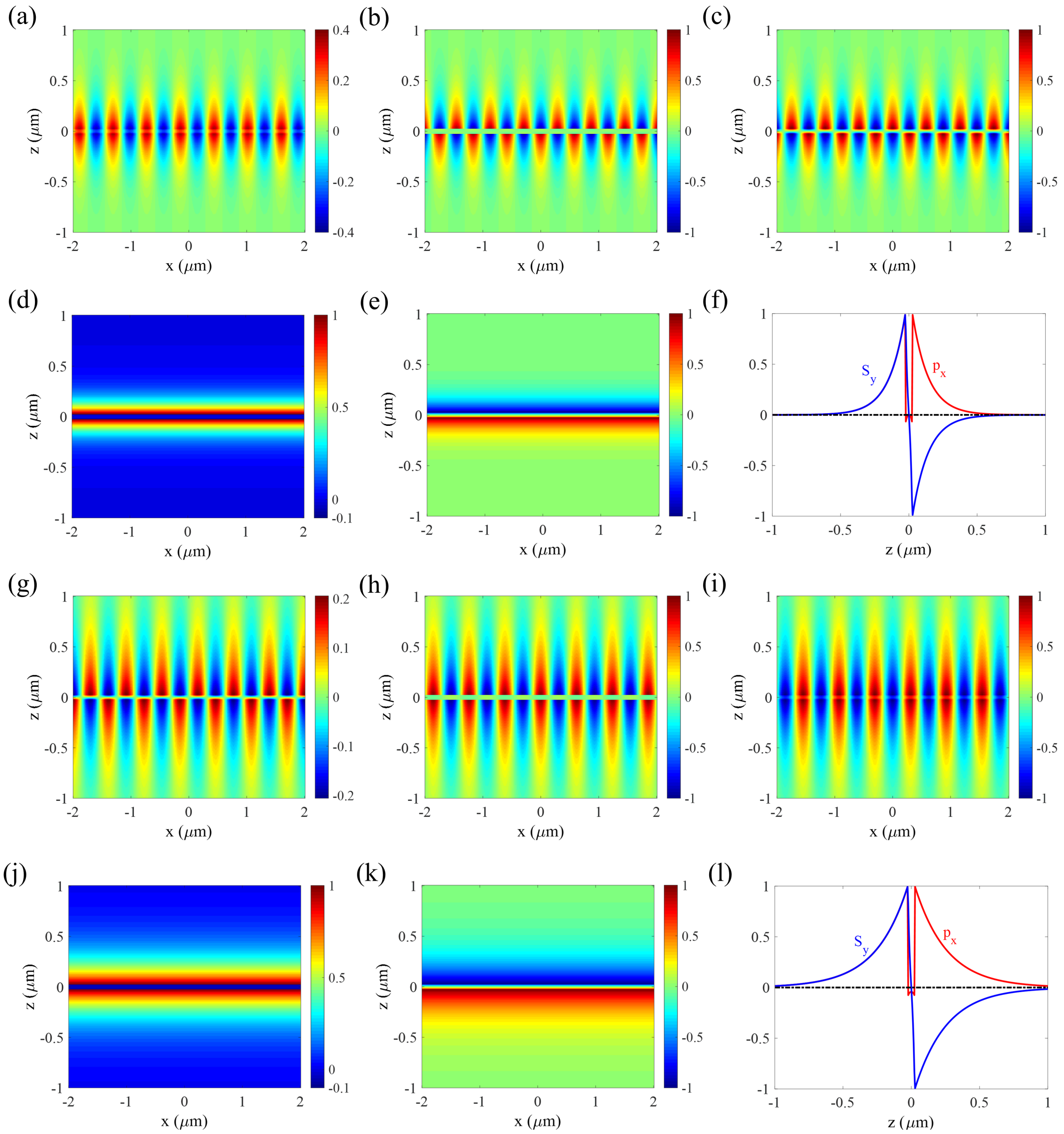

**Fig. S3.** The *xz*-plane distributions of the real part of (a) $E_x$, (b) $E_z$, (c) $H_y$, and the (d) kinetic momentum $p_x$, (e) dispersionless SAM $S_y$ for the symmetric plane wave mode at the *xz*-plane. The 1D contour of $p_x$ and $S_y$ at $x = 0$ of *xz*-plane is shown in (f). The *xz*-plane distributions of the real part of (g) $E_x$, (h) $E_z$, (i) $H_y$, and (j) kinetic momentum $p_x$, (k) dispersionless SAM $S_y$ for the anti-symmetric plane wave mode at the *xz*-plane. The 1D contour of $p_x$ and $S_y$ at $x = 0$ of *xz*-plane is shown in (l). It is worth noting that the symmetry or anti-symmetry of a mode is evaluated by the horizontal electric field component for the transverse magnetic modes. The structure is air-metal-air and the material of metal layer is Au [70]. The wavelength is 632.8nm.

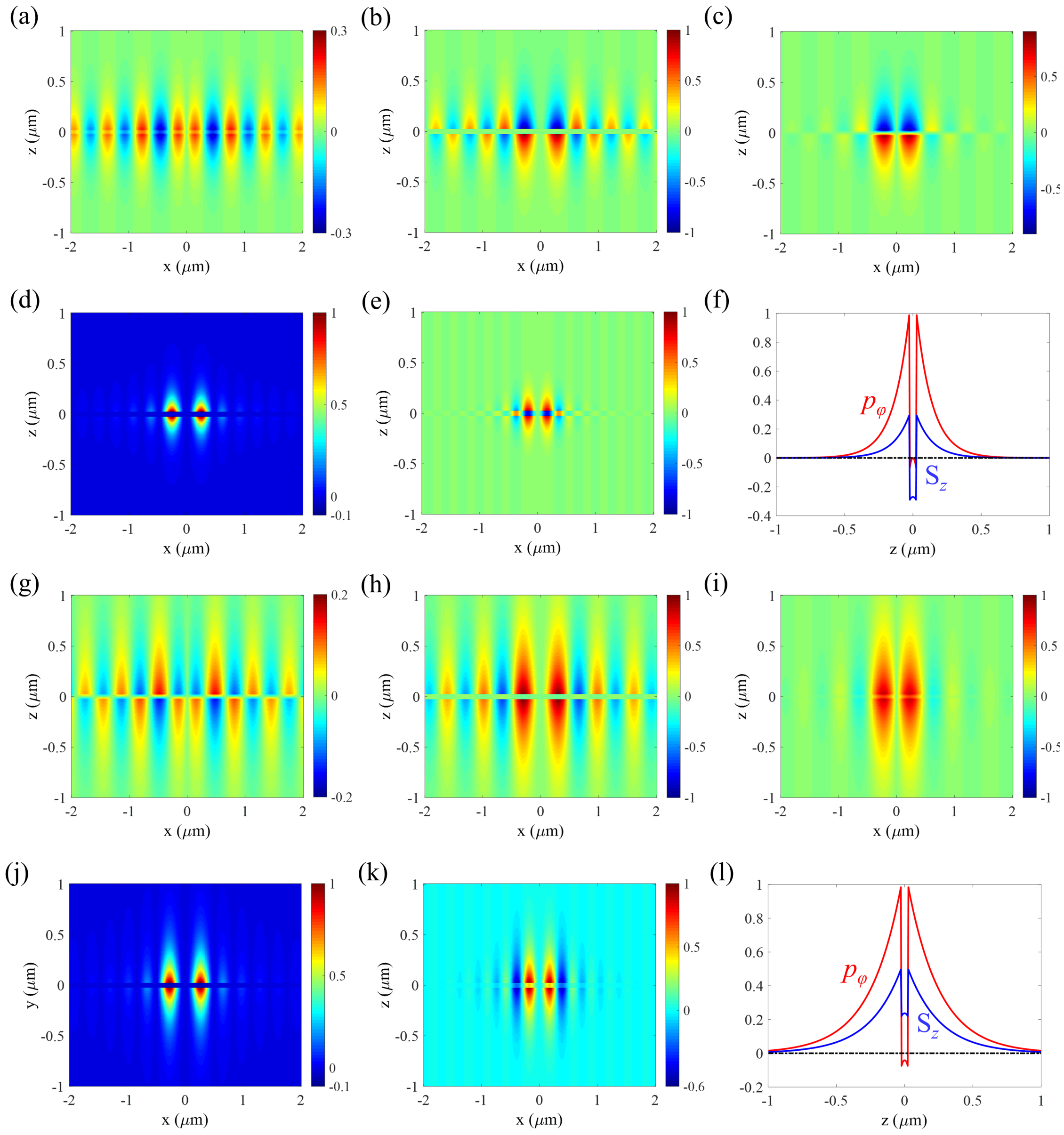

**Fig. S4.** The *xz*-plane distributions of the real part of (a) $E_r$, (b) $E_z$, (c) $H_r$, and the (d) kinetic momentum $p_\varphi$, (e) dispersionless SAM $S_z$ for the symmetric plane wave mode at the *xz*-plane. The 1D contour of $p_\varphi$ and $S_z$ at $x$ = 0.15μm of *xz*-plane is shown in (f). The *xz*-plane distributions of the real part of (g) $E_r$, (h) $E_z$, (i) $H_r$, and (j) kinetic momentum $p_\varphi$, (k) dispersionless SAM $S_z$ for the anti-symmetric plane wave mode at the *xz*-plane. The 1D contour of $p_\varphi$ and $S_z$ at $x$ = 0.15μm of *xz*-plane is shown in (l). As mentioned above, the symmetry or anti-symmetry of a mode is evaluated by the horizontal electric field component ($E_r$ here) for the transverse magnetic modes. It is worth noting that, there are longitudinal spin components due to the coupling between horizontal EM field components in the $S_z$ (the details can be found in the following section), which makes the signs of $S_z$ opposite for the two modes in (e) and (k). Here, the order of Bessel function is +2. The structure is air-metal-air and the material of metal layer is Au [70]. The wavelength is 632.8nm.

## II. Dispersionless SMEs for the *p*-polarized surface EM modes

The intrinsic spin-momentum locking property of surface EM waves by ignoring the dispersion:

$$\mathbf{S}=\frac{1}{2\omega^2}\nabla\times\mathbf{P}=\frac{1}{2\omega^2\varepsilon_0\mu_0}\nabla\times\mathbf{p}=\frac{1}{2k_0^2}\nabla\times\mathbf{p}\,, \tag{S19}$$

where $\mathbf{p}=\varepsilon_0\mu_0\mathrm{Re}\{\mathbf{E}^*\times\mathbf{H}\}/2$ is the kinetic momentum of photons and proportional to the Poynting vector $\mathbf{P}$ due to the request of relativity [S9], was demonstrated in the single interface configurations [27]. The equation (S19) definitely reveals that [27]: 1. the transversal feature of SAM with respect to kinetic momentum (thus, the total three-dimensional spin vector can be regarded as the transverse spin universally); 2. the spin-momentum locking property between the SAM and the kinetic momentum (the locking property evaluating by the kinetic momentum obeys the right-hand rule, no matter in dielectric or metal/magnetic materials); 3. the derivative feature of the transverse spin (the transverse spin is originated form the transverse inhomogeneities of EM field).

In the section, we will demonstrate theoretically that, for the multilayered structures, the spin-momentum locking will still be satisfied. However, unlike spin-momentum properties demonstrated in the single interface system [27] in Fig. S1(a), there is a hidden longitudinal spin component due to the local coupling between the horizontal EM components exist, no matter whether for the transverse magnetic modes or the transverse electric modes. Here, we only exhibit derivations of spin-momentum locking for the *p*-polarized surface EM waves in the one-layer system following the process in reference [27], but the results can be flexibly extended to the *p*-polarized or *s*-polarized surface mode in arbitrary multilayered systems from the expressions (S10), (S12) and (S14). The kinetic momentum of photons $\mathbf{p}=\varepsilon_0\mu_0\,\mathrm{Re}\left(\mathbf{E}^*\times\mathbf{H}\right)/2$ of the *p*-polarized surface mode in Fig. S1(b) is:

$$p_x=\varepsilon_0\mu_0\begin{bmatrix}\dfrac{\omega}{2}\dfrac{A_+^*A_+}{\varepsilon^+\varepsilon^+}\mathrm{Im}\left\{\dfrac{\varepsilon^+}{\beta^2}\xi^*\dfrac{\partial\xi}{\partial x}\right\}e^{-2k_z^+(z-a/2)}=\rho_x^+ & z>+\dfrac{a}{2}\\ \left.\begin{array}{l}+\dfrac{\omega}{2}\dfrac{B_+^*B_+}{\varepsilon^m\varepsilon^m}\mathrm{Im}\left\{\dfrac{\varepsilon^m}{\beta^2}\xi^*\dfrac{\partial\xi}{\partial x}\right\}e^{+2k_z^m(z-a/2)}\\ +\dfrac{\omega}{2}\dfrac{B_+^*B_-+B_-^*B_+}{\varepsilon^m\varepsilon^m}\mathrm{Im}\left\{\dfrac{\varepsilon^m}{\beta^2}\xi^*\dfrac{\partial\xi}{\partial x}\right\}e^{-k_z^m a}\\ +\dfrac{\omega}{2}\dfrac{B_-^*B_-}{\varepsilon^m\varepsilon^m}\mathrm{Im}\left\{\dfrac{\varepsilon^m}{\beta^2}\xi^*\dfrac{\partial\xi}{\partial x}\right\}e^{-2k_z^m(z+a/2)}\end{array}\right\}=\rho_x^{m+}+\rho_x^{mc}+\rho_x^{m-} & -\dfrac{a}{2}<z<+\dfrac{a}{2}\\ \dfrac{\omega}{2}\dfrac{A_-^*A_-}{\varepsilon^-\varepsilon^-}\mathrm{Im}\left\{\dfrac{\varepsilon^-}{\beta^2}\xi^*\dfrac{\partial\xi}{\partial x}\right\}e^{+2k_z^-(z+a/2)}=\rho_x^- & z<-\dfrac{a}{2}\end{bmatrix}, \tag{S20a}$$

$$p_y=\varepsilon_0\mu_0\begin{bmatrix}\dfrac{\omega}{2}\dfrac{A_+^*A_+}{\varepsilon^+\varepsilon^+}\mathrm{Im}\left\{\dfrac{\varepsilon^+}{\beta^2}\xi^*\dfrac{\partial\xi}{\partial y}\right\}e^{-2k_z^+(z-a/2)}=\rho_y^+ & z>+\dfrac{a}{2}\\ \left.\begin{array}{l}+\dfrac{\omega}{2}\dfrac{B_+^*B_+}{\varepsilon^m\varepsilon^m}\mathrm{Im}\left\{\dfrac{\varepsilon^m}{\beta^2}\xi^*\dfrac{\partial\xi}{\partial y}\right\}e^{+2k_z^m(z-a/2)}\\ +\dfrac{\omega}{2}\dfrac{B_+^*B_-+B_-^*B_+}{\varepsilon^m\varepsilon^m}\mathrm{Im}\left\{\dfrac{\varepsilon^m}{\beta^2}\xi^*\dfrac{\partial\xi}{\partial y}\right\}e^{-2k_z^m a}\\ +\dfrac{\omega}{2}\dfrac{B_-^*B_-}{\varepsilon^m\varepsilon^m}\mathrm{Im}\left\{\dfrac{\varepsilon^m}{\beta^2}\xi^*\dfrac{\partial\xi}{\partial y}\right\}e^{-2k_z^m(z+a/2)}\end{array}\right\}=\rho_y^{m+}+\rho_y^{mc}+\rho_y^{m-} & -\dfrac{a}{2}<z<+\dfrac{a}{2}\\ \dfrac{\omega}{2}\dfrac{A_-^*A_-}{\varepsilon^-\varepsilon^-}\mathrm{Im}\left\{\dfrac{\varepsilon^-}{\beta^2}\xi^*\dfrac{\partial\xi}{\partial y}\right\}e^{+2k_z^-(z+a/2)}=\rho_y^- & z<-\dfrac{a}{2}\end{bmatrix}, \tag{S20b}$$

and

$$p_z = \begin{bmatrix} 0 = \rho_z^+ & z > +\frac{a}{2} \\ 0 = \rho_z^{m+} + \rho_z^{mc} + \rho_z^{m-} & -\frac{a}{2} < z < +\frac{a}{2} \\ 0 = \rho_z^- & z < -\frac{a}{2} \end{bmatrix}. \tag{S20c}$$

On the other hand, the dispersionless SAM $\mathbf{S} = \mathrm{Im}\left(\varepsilon \mathbf{E}^* \times \mathbf{E} + \mu \mathbf{H}^* \times \mathbf{H}\right)/4\omega$ can be expressed as:

$$S_x = \begin{bmatrix} +\frac{\varepsilon^+}{4\omega}\frac{A_+^* A_+}{\varepsilon^+\varepsilon^+}\frac{k_z^+}{\beta^2}\mathrm{Im}\left\{\xi^*\frac{\partial \xi}{\partial y} - \xi\frac{\partial \xi^*}{\partial y}\right\}e^{-2k_z^+(z-a/2)} = \sigma_x^+ & z > +\frac{a}{2} \\ \left\{\begin{aligned} &-\frac{\varepsilon^m}{4\omega}\frac{B_+^* B_+}{\varepsilon^m\varepsilon^m}\frac{k_z^m}{\beta^2}\mathrm{Im}\left\{\xi^*\frac{\partial \xi}{\partial y} - \xi\frac{\partial \xi^*}{\partial y}\right\}e^{+2k_z^m(z-a/2)} \\ &+\frac{\varepsilon^m}{4\omega}\frac{\mathrm{Im}\left\{B_+^* B_- - B_+ B_-^*\right\}}{\varepsilon^m\varepsilon^m}\frac{k_z^m}{\beta^2}\left(\xi^*\frac{\partial \xi}{\partial y} + \xi\frac{\partial \xi^*}{\partial y}\right)e^{-2k_z^m a} \\ &+\frac{\varepsilon^m}{4\omega}\frac{B_-^* B_-}{\varepsilon^m\varepsilon^m}\frac{k_z^m}{\beta^2}\mathrm{Im}\left\{\xi^*\frac{\partial \xi}{\partial y} - \xi\frac{\partial \xi^*}{\partial y}\right\}e^{-2k_z^m(z+a/2)} \end{aligned}\right\} = \sigma_x^{m+} + \sigma_x^{mc} + \sigma_x^{m-} & -\frac{a}{2} < z < +\frac{a}{2} \\ -\frac{\varepsilon^-}{4\omega}\frac{A_-^* A_-}{\varepsilon^-\varepsilon^-}\frac{k_z^-}{\beta^2}\mathrm{Im}\left\{\xi^*\frac{\partial \xi}{\partial y} - \xi\frac{\partial \xi^*}{\partial y}\right\}e^{+2k_z^-(z+a/2)} = \sigma_x^- & z < -\frac{a}{2} \end{bmatrix}, \tag{S21a}$$

$$S_y = \begin{bmatrix} -\frac{\varepsilon^+}{4\omega}\frac{A_+^* A_+}{\varepsilon^+\varepsilon^+}\frac{k_z^+}{\beta^2}\mathrm{Im}\left\{\xi^*\frac{\partial \xi}{\partial x} - \xi\frac{\partial \xi^*}{\partial x}\right\}e^{-2k_z^+(z-a/2)} = \sigma_y^+ & z > +\frac{a}{2} \\ \left\{\begin{aligned} &+\frac{\varepsilon^m}{4\omega}\frac{B_+^* B_+}{\varepsilon^m\varepsilon^m}\frac{k_z^m}{\beta^2}\mathrm{Im}\left\{\xi^*\frac{\partial \xi}{\partial x} - \xi\frac{\partial \xi^*}{\partial x}\right\}e^{+2k_z^m(z-a/2)} \\ &-\frac{\varepsilon^m}{4\omega}\frac{\mathrm{Im}\left\{B_+^* B_- - B_-^* B_+\right\}}{\varepsilon^m\varepsilon^m}\frac{k_z^m}{\beta^2}\left(\xi^*\frac{\partial \xi}{\partial x} + \xi\frac{\partial \xi^*}{\partial x}\right)e^{-k_z^m a} \\ &-\frac{\varepsilon^m}{4\omega}\frac{B_-^* B_-}{\varepsilon^m\varepsilon^m}\frac{k_z^m}{\beta^2}\mathrm{Im}\left\{\xi^*\frac{\partial \xi}{\partial x} - \xi\frac{\partial \xi^*}{\partial x}\right\}e^{-2k_z^m(z+a/2)} \end{aligned}\right\} = \sigma_y^{m+} + \sigma_y^{mc} + \sigma_y^{m-} & -\frac{a}{2} < z < +\frac{a}{2} \\ +\frac{\varepsilon^-}{4\omega}\frac{A_-^* A_-}{\varepsilon^-\varepsilon^-}\frac{k_z^-}{\beta^2}\mathrm{Im}\left\{\xi^*\frac{\partial \xi}{\partial x} - \xi\frac{\partial \xi^*}{\partial x}\right\}e^{+2k_z^-(z+a/2)} = \sigma_y^- & z < -\frac{a}{2} \end{bmatrix}, \tag{S21b}$$

and

$$S_z = \begin{bmatrix} +\frac{\varepsilon^+}{4\omega}\frac{A_+^* A_+}{\varepsilon^+\varepsilon^+}\frac{1}{\beta^2}\mathrm{Im}\left\{\frac{\partial \xi^*}{\partial x}\frac{\partial \xi}{\partial y} - \frac{\partial \xi}{\partial x}\frac{\partial \xi^*}{\partial y}\right\}e^{-2k_z^+(z-a/2)} = \sigma_z^+ & z > +\frac{a}{2} \\ \left\{\begin{aligned} &+\frac{\varepsilon^m}{4\omega}\frac{B_+^* B_+}{\varepsilon^m\varepsilon^m}\frac{1}{\beta^2}\mathrm{Im}\left\{\frac{\partial \xi^*}{\partial x}\frac{\partial \xi}{\partial y} - \frac{\partial \xi}{\partial x}\frac{\partial \xi^*}{\partial y}\right\}e^{+2k_z^m(z-a/2)} \\ &-\frac{\varepsilon^m}{4\omega}\frac{+B_+^* B_- + B_-^* B_+}{\varepsilon^m\varepsilon^m}\frac{k_z^m k_z^m - \omega^2\varepsilon^m\mu^m}{\beta^4}\mathrm{Im}\left\{\frac{\partial \xi^*}{\partial x}\frac{\partial \xi}{\partial y} - \frac{\partial \xi}{\partial x}\frac{\partial \xi^*}{\partial y}\right\}e^{-k_z^m a} \\ &+\frac{\varepsilon^m}{4\omega}\frac{B_-^* B_-}{\varepsilon^m\varepsilon^m}\frac{1}{\beta^2}\mathrm{Im}\left\{\frac{\partial \xi^*}{\partial x}\frac{\partial \xi}{\partial y} - \frac{\partial \xi}{\partial x}\frac{\partial \xi^*}{\partial y}\right\}e^{-2k_z^m(z+a/2)} \end{aligned}\right\} = \sigma_z^{m+} + \sigma_z^{mc} + \sigma_z^{m-} & -\frac{a}{2} < z < +\frac{a}{2} \\ +\frac{\varepsilon^-}{4\omega}\frac{A_-^* A_-}{\varepsilon^-\varepsilon^-}\frac{1}{\beta^2}\mathrm{Im}\left\{\frac{\partial \xi^*}{\partial x}\frac{\partial \xi}{\partial y} - \frac{\partial \xi}{\partial x}\frac{\partial \xi^*}{\partial y}\right\}e^{+2k_z^-(z+a/2)} = \sigma_z^- & z < -\frac{a}{2} \end{bmatrix}. \tag{S21c}$$

From the relation (S9) and assuming that the medium is lossless, there is $\mathrm{Im}\left\{B_+^* B_- - B_-^* B_+\right\} = 0$ and the coupling terms of horizontal SAM vanish ($\sigma_x^{mc}=0$, $\sigma_y^{mc}=0$). Thus, only the coupling terms in the horizontal kinetic momentum ($\rho_x^{mc}\neq 0$, $\rho_y^{mc}\neq 0$) and in the normal component SAM ($\sigma_z^{mc}\neq 0$) are not zero and independent of $z$-axis.

From the expressions (S19-S21), if we ignore the coupling terms in kinetic momentum and dispersionless SAM, one can deduce that the individual waves definitely satisfy the spin-momentum locking relations:

$$\begin{aligned}\boldsymbol{\sigma}^{+}&=\frac{1}{2\omega^2\varepsilon_0\mu_0}\nabla\times\boldsymbol{\rho}^{+} \quad \boldsymbol{\sigma}^{-}=\frac{1}{2\omega^2\varepsilon_0\mu_0}\nabla\times\boldsymbol{\rho}^{-}\\ \boldsymbol{\sigma}^{m+}&=\frac{1}{2\omega^2\varepsilon_0\mu_0}\nabla\times\boldsymbol{\rho}^{m+} \quad \boldsymbol{\sigma}^{m-}=\frac{1}{2\omega^2\varepsilon_0\mu_0}\nabla\times\boldsymbol{\rho}^{m-}\end{aligned}. \tag{S22}$$

The remaining terms $\rho_x^{mc}$, $\rho_y^{mc}$, $\sigma_z^{mc}$ are all $z$-independent. Thus, there are $\left[\nabla\times\boldsymbol{\rho}^{mc}\right]_x=0$, $\left[\nabla\times\boldsymbol{\rho}^{mc}\right]_y=0$ and $\left[\nabla\times\boldsymbol{\rho}^{mc}\right]_z\neq 0$. The coupling SAM is

$$\boldsymbol{\sigma}^{mc}=0\hat{\mathbf{x}}+0\hat{\mathbf{y}}+\sigma_z^{mc}\hat{\mathbf{z}}=-\frac{\varepsilon^m}{4\omega}\frac{+B_+^*B_-+B_-^*B_+}{\varepsilon^m\varepsilon^m}\frac{k_z^m k_z^m-\omega^2\varepsilon^m\mu^m}{\beta^4}\mathrm{Im}\left\{\frac{\partial\xi^*}{\partial x}\frac{\partial\xi}{\partial y}-\frac{\partial\xi}{\partial x}\frac{\partial\xi^*}{\partial y}\right\}e^{-k_z^m a}\hat{\mathbf{z}}. \tag{S23}$$

Actually, the coupling term $\sigma_z^{mc}$ is originated from two contributions:

1. the interferential spin between the $B_+$ wave and the $B_-$ wave (named as the interferential spin term);

$$\boldsymbol{\sigma}_t^{m}=\frac{1}{2\omega^2}\nabla\times\boldsymbol{\rho}^{mc}=0\hat{\mathbf{x}}+0\hat{\mathbf{y}}+\frac{\varepsilon^m}{4\omega}\frac{+B_+^*B_-+B_-^*B_+}{\varepsilon^m\varepsilon^m}\frac{1}{\beta^2}\mathrm{Im}\left\{\frac{\partial\xi^*}{\partial x}\frac{\partial\xi}{\partial y}-\frac{\partial\xi}{\partial x}\frac{\partial\xi^*}{\partial y}\right\}e^{-k_z^m a}\hat{\mathbf{z}}; \tag{S24}$$

2. the directly coupling polarization ellipticities between the $x$/$y$-component of $B_+$ wave and the $y$/$x$-component of $B_-$ wave (named as the coupling spin term);

$$\boldsymbol{\sigma}_l^{m}=\boldsymbol{\sigma}^{mc}-\frac{1}{2\omega^2}\nabla\times\boldsymbol{\rho}^{mc}=0\hat{\mathbf{x}}+0\hat{\mathbf{y}}-\frac{\varepsilon^m}{4\omega}\frac{+B_+^*B_-+B_-^*B_+}{\varepsilon^m\varepsilon^m}\frac{2k_z^m k_z^m}{\beta^4}\mathrm{Im}\left\{\frac{\partial\xi^*}{\partial x}\frac{\partial\xi}{\partial y}-\frac{\partial\xi}{\partial x}\frac{\partial\xi^*}{\partial y}\right\}e^{-k_z^m a}\hat{\mathbf{z}}. \tag{S25}$$

Therein, the contribution 1 can be understood in two processes: 1. the interference between the $B_+$ wave and the $B_-$ wave causes the inhomogeneities of total EM field; 2. this inhomogeneities of EM field and associated inhomogeneous kinetic momentum results in the optical transverse spin. Thus, the contribution 1 is originated from the interferential inhomogeneities/structure properties of EM field that is consistent with the unified property of optical transverse spin [27, 64, 65]. Whereas for the contribution 2, we emphasize that it is local helix-dependent, and it should be regarded as the longitudinal spin (helix-dependent solely and irrelated to inhomogeneity/structure property of EM field [65]). Additionally, since $k_z^m k_z^m-\omega^2\varepsilon^m\mu^m>0$ for the noble metal materials, the $\boldsymbol{\sigma}^{mc}$ is parallel to the $\boldsymbol{\sigma}_l^m$ and antiparallel to the $\boldsymbol{\sigma}_t^m$. However, if the layer is dielectric, it is normally $k_z^m k_z^m-\omega^2\varepsilon^m\mu^m<0$. Thus, the $\boldsymbol{\sigma}^{mc}$ is parallel to the $\boldsymbol{\sigma}_t^m$ and antiparallel to the $\boldsymbol{\sigma}_l^m$.

Based on the former analysis, if the materials' dispersion is ignored here, we can reformulate the Maxwell-like spin-momentum equations of surface EM modes in a complex multilayered system to be

$$\nabla\cdot\mathbf{P}=\nabla\cdot\mathbf{p}=0, \tag{S26a}$$

$$\nabla\cdot\mathbf{S}=0, \tag{S26b}$$

$$\nabla\times\mathbf{S}=2\left(\varepsilon\mu\mathbf{P}-\mathbf{p}_o\right)=2\left(\varepsilon_r\mu_r\mathbf{p}-\mathbf{p}_o\right)=2\left(n^2\mathbf{p}-\mathbf{p}_o\right), \tag{S26c}$$

$$\nabla\times\mathbf{p}=2k_0^2\mathbf{S}_t=2k_0^2\left(\mathbf{S}-\mathbf{S}_l\right). \tag{S26d}$$

Here, $\varepsilon_r$ and $\mu_r$ are the relative permittivity and permeability, respectively; $n$ is the refractive index. Equation (S65c) denotes the spin-orbit decomposition by ignoring the dispersion, which the kinetic momentum $\varepsilon_r\mu_r\mathbf{p}$ can be decomposed into canonical/orbital momentum $\mathbf{p}_o$ and the Belinfante spin momentum $\mathbf{p}_s$ [S9]. In the equation (S26d), the longitudinal spin given by the difference between the total spin and the transverse spin $\mathbf{S}_l = \mathbf{S} - \mathbf{S}_t$ is

$$\mathbf{S}_l = \begin{bmatrix} 0 & z > +\frac{a}{2} \\ 0\hat{\mathbf{x}} + 0\hat{\mathbf{y}} - \frac{\varepsilon^m}{4\omega}\frac{+B_+^*B_- + B_-^*B_+}{\varepsilon^m\varepsilon^m}\frac{2k_z^m k_z^m}{\beta^4}\mathrm{Im}\left\{\frac{\partial\xi^*}{\partial x}\frac{\partial\xi}{\partial y} - \frac{\partial\xi}{\partial x}\frac{\partial\xi^*}{\partial y}\right\}e^{-k_z^m a}\hat{\mathbf{z}} & -\frac{a}{2} < z < +\frac{a}{2} \\ 0 & z < -\frac{a}{2} \end{bmatrix}. \quad \text{(S27)}$$

In addition, one can deduce the dispersionless Helmholtz-like spin momentum equation as

$$2\nabla\times\mathbf{p}_o = \nabla^2\mathbf{S} + 4k^2\left(\mathbf{S} - \mathbf{S}_l\right) = \left[\nabla^2\mathbf{S}_t + 4k^2\mathbf{S}_t\right] + \nabla^2\mathbf{S}_l. \quad \text{(S28)}$$

In the special case that the coupling longitudinal spin $\mathbf{S}_l$ can be ignored (in single interface systems or the layer is thick enough), the dispersionless Helmholtz-like spin momentum equation is downgraded into

$$2\nabla\times\mathbf{p}_o = \nabla^2\mathbf{S}_t + 4k^2\mathbf{S}_t, \quad \text{(S29)}$$

which is consistent with the Helmholtz-like spin momentum equation in the single interface system [27].

From equations (S26) to (S28), one can conclude that the transverse spin is still locked with the momentum in the complex multilayered structures while the longitudinal spin does not.

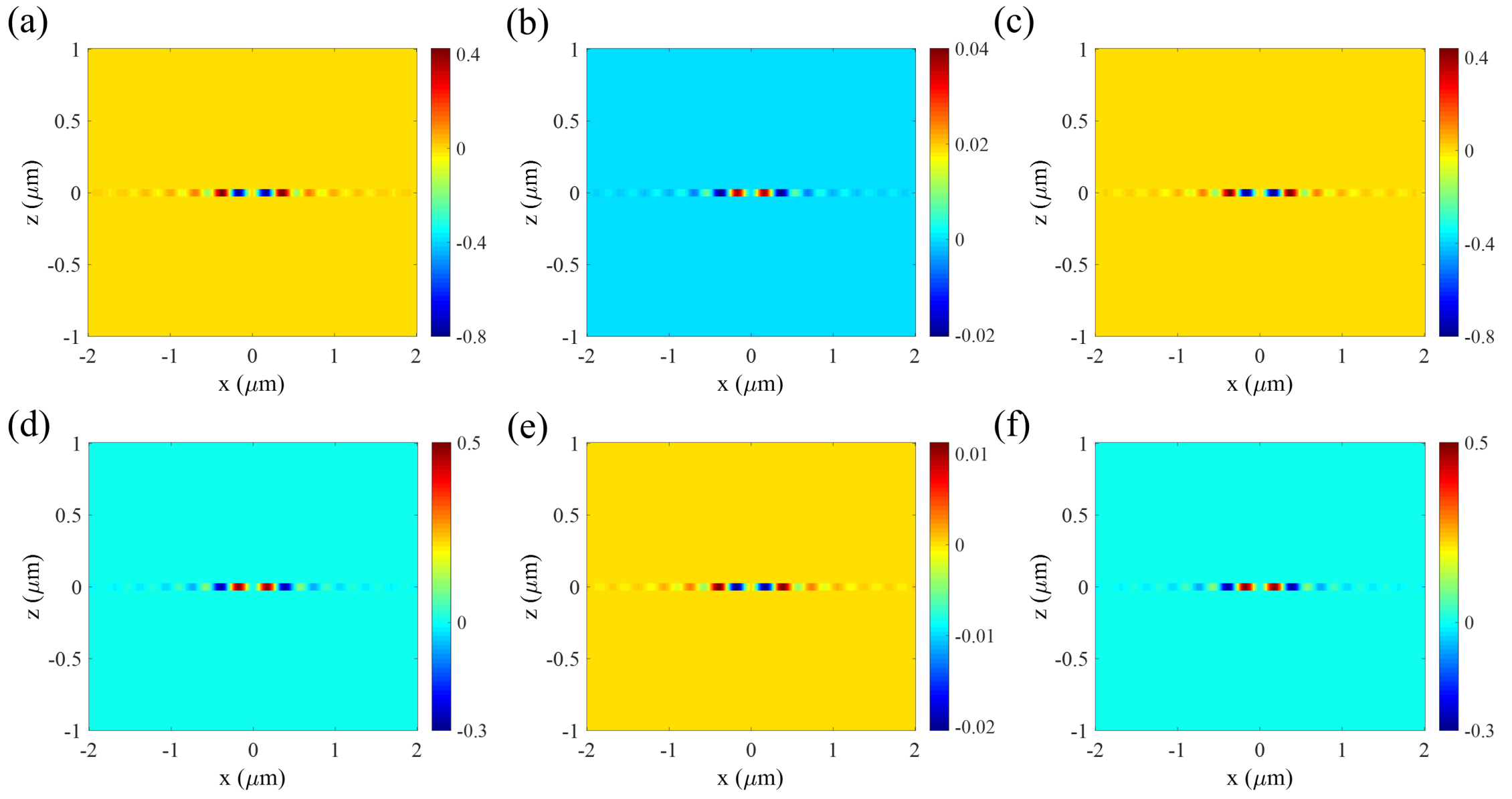

**Fig. S5.** (a) The $z$-component coupling SAM $\boldsymbol{\sigma}^{mc}$, (b) the $z$-component coupling transverse spin $\boldsymbol{\sigma}_t^m$, and (c) the $z$-component coupling longitudinal spin $\boldsymbol{\sigma}_l^m$ for the +2-order symmetric Bessel-type surface waves. (d) The $z$-component coupling SAM $\boldsymbol{\sigma}^{mc}$, (e) the $z$-component coupling transverse spin $\boldsymbol{\sigma}_t^m$, and (f) the $z$-component coupling longitudinal spin $\boldsymbol{\sigma}_l^m$ for the +2-order anti-symmetric Bessel-type surface waves. Obviously, since $k_z^m k_z^m - \omega^2\varepsilon^m\mu^m > 0$ for the noble metal materials, the $\boldsymbol{\sigma}^{mc}$ is parallel to the $\boldsymbol{\sigma}_l^m$ and antiparallel to the $\boldsymbol{\sigma}_t^m$. The thickness of gold layer is 50nm. The wavelength is 632.8nm.

To verify the spin-momentum properties mentioned above, we summarize the $z$ components of coupling SAMs for the symmetric and anti-symmetric Bessel surface modes of the air-metal-air structure in Fig. S5. Obviously, only in the layer, there is coupling SAM exist. Since $k_z^m k_z^m - \omega^2\varepsilon^m\mu^m > 0$ for the noble metal materials, the $\boldsymbol{\sigma}^{mc}$ is parallel to the $\boldsymbol{\sigma}_l^m$ and antiparallel to the $\boldsymbol{\sigma}_t^m$. Noteworthily, the three coupling SAMs are all $z$- independent as indicated in expressions (S23-S25). On the other hand, we show the $z$ components of diverse SAMs for the symmetric Bessel surface mode of the metal-air-metal structure in Fig. S6. Noteworthily, only symmetric mode exists in the metal-air-metal structure. It can be observed that the $\boldsymbol{\sigma}^{mc}$ is parallel to the $\boldsymbol{\sigma}_t^m$ and antiparallel to the $\boldsymbol{\sigma}_l^m$. This is because there is normally $k_z^m k_z^m - \omega^2 \varepsilon^m \mu^m < 0$ for the dielectric layer. Likewise, the three coupling SAMs are all $z$-independent.

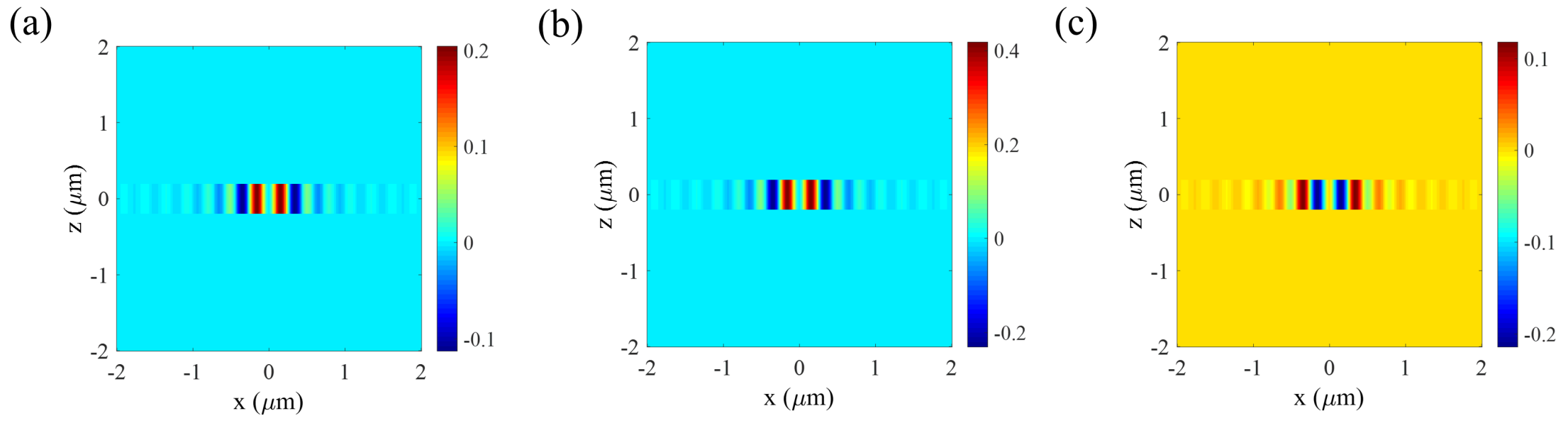

**Fig. S6.** (a) The $z$-component coupling SAM $\boldsymbol{\sigma}^{mc}$, (b) the $z$-component coupling transverse spin $\boldsymbol{\sigma}_t^m$ and (c) the $z$-component coupling longitudinal spin $\boldsymbol{\sigma}_l^m$ for the +2-order symmetric Bessel-type surface waves in the metal-air-metal structure. There is only symmetric mode exist here. As the layer is dielectric, it is normally $k_z^m k_z^m - \omega^2 \varepsilon^m \mu^m < 0$. Thus, the $\boldsymbol{\sigma}^{mc}$ is parallel to the $\boldsymbol{\sigma}_t^m$ and antiparallel to the $\boldsymbol{\sigma}_l^m$. The thickness of dielectric layer is 200nm. The wavelength is 632.8nm.

Finally, by decomposing the coupling $z$-component SAM into transverse spin and longitudinal spin components, the abnormal properties of spin-momentum locking in Fig. S4 are removed theoretically. Here, we draw the azimuthal kinetic momentum $p_\varphi$, the $z$-component SAMs $S_{t,z}$ and longitudinal spin $S_l$ for the symmetric and anti-symmetric modes of the +2 and −2 order surface Bessel waves at the air-metal-air structure in Fig. S7 for reference. It is observed that, for the surface EM waves in arbitrary multilayered systems, the symmetric and anti-symmetric modes satisfy an unique spin-momentum locking property: the horizontal spin components $\mathbf{S}_\parallel$ are purely transverse spin and thus there are locked with the kinetic momentum $\mathbf{p}$ and the locking property satisfies the right-hand rule $\mathbf{S}_\parallel \propto +\mathbf{p}\times\mathbf{n}$ with $\mathbf{n}$ the outer normal direction of interface, while the normal spin component $\mathbf{S}_\mathrm{n}$ contain the coupling helix (longitudinal spin) component and transverse spin component simultaneously and only the transverse spin component of $\mathbf{S}_\mathrm{n}$ is locked with the kinetic momentum $\mathbf{p}$ and the locking property satisfies the right-hand screw rule determined by $(\nabla\times\mathbf{p})_\mathrm{n}$. On the other hand, the coupling longitudinal spin appears in the multilayered system, and the coupling longitudinal spin only exists in the layers and does not possess the property of spin-momentum locking. The coupling longitudinal spin can be tuned by the mode's symmetry, which would destroy the right-hand rule in the spin-momentum locking between the kinetic momentum and normal component SAM if we do not decompose the total SAM into transverse spin component and longitudinal spin component in physics.

Till now, we remove the abnormal properties of spin-momentum locking caused by analyzing the mode's symmetry and coupling property. In the following section, we will focus on the abnormal properties of spin-momentum locking caused by the dispersions in Fig. S4.

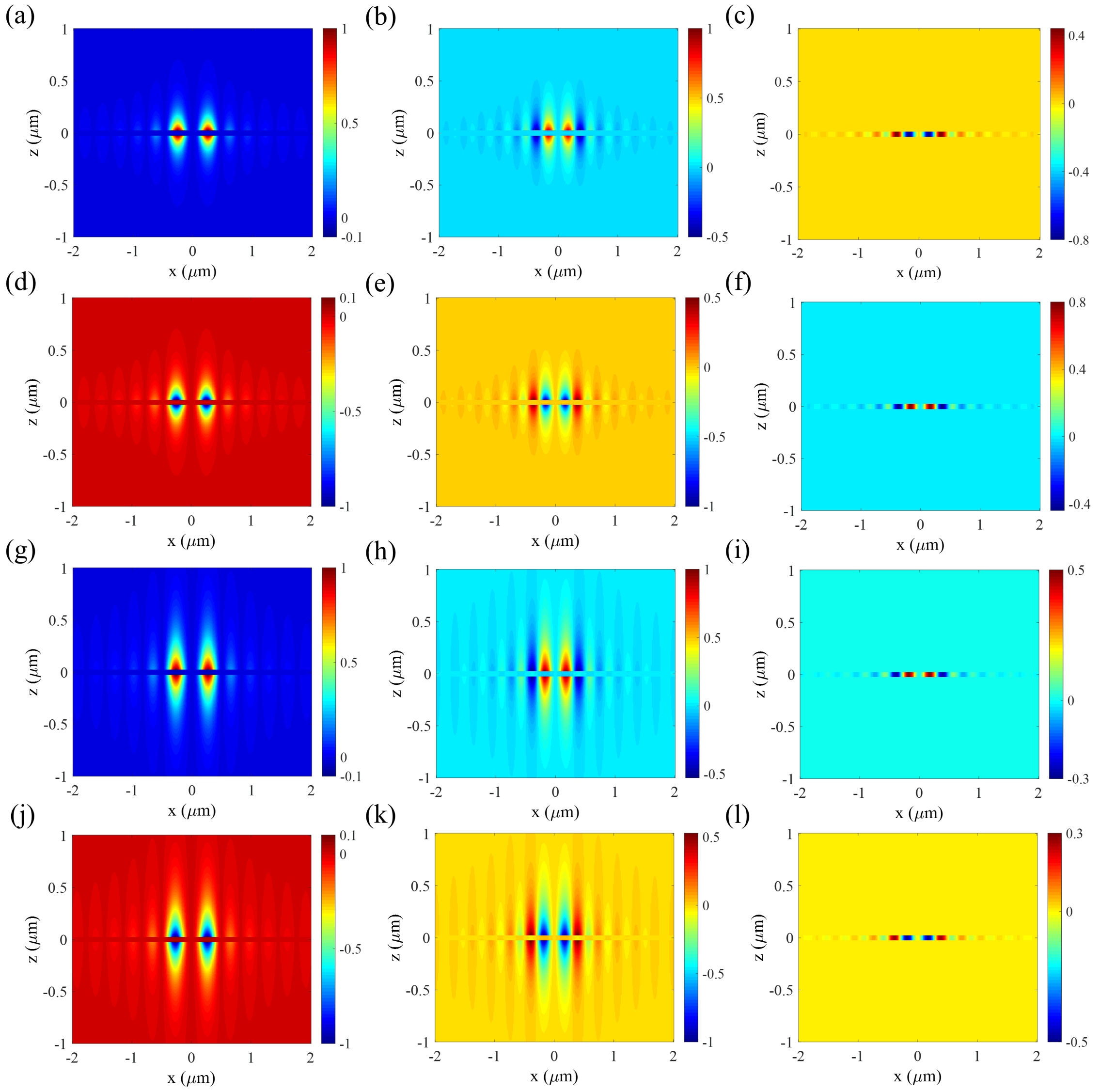

**Fig. S7.** (a) The kinetic momentum $p_\varphi$, (b) the *z*-component transverse spin $S_{t,z}$, (c) the *z*-component coupling longitudinal spin $S_l$ for the +2-order symmetric Bessel-type surface waves in the metal-air-metal structure; and (d) the kinetic momentum $p_\varphi$, (e) the *z*-component transverse spin $S_{t,z}$, (f) the *z*-component coupling longitudinal spin $S_l$ for the −2-order symmetric Bessel-type surface waves in the metal-air-metal structure. From (a), (b), (d) and (e), as the order changes from +2 to −2, the kinetic momentum and the *z*-component transverse spin $S_{t,z}$ are inverted correspondingly, which demonstrates the property of spin-momentum locking. (g) The kinetic momentum $p_\varphi$, (h) the *z*-component transverse spin $S_{t,z}$, (i) the *z*-component coupling longitudinal spin $S_l$ for the +2-order anti-symmetric Bessel-type surface waves in the metal-air-metal structure; and (j) the kinetic momentum $p_\varphi$, (k) the *z*-component transverse spin $S_{t,z}$, (l) the *z*-component coupling longitudinal $S_l$ for the −2-order anti-symmetric Bessel-type surface waves in the metal-air-metal structure. Likewise, as the order changes from +2 to −2, the kinetic momentum and the *z*-component transverse spin $S_{t,z}$ are inverted correspondingly, which demonstrates the property of spin-momentum locking. Noteworthily, the spin-momentum locking properties of symmetric and anti-symmetric modes are coincident here, which are totally different from those exhibited in Fig. S4. The thickness of Au layer is 50nm. The wavelength is 632.8nm.

## III. Dispersive SMEs for the *p*-polarized surface EM modes

In this section, we give the details of the derivations of dispersive SMEs in the multilayered structures and introduce the spin-momentum locking property for dispersive surface EM modes. In the dispersive medium, the kinetic momentum, canonical momentum density and spin angular momentum are [58]

$$\mathbf{p} = \varepsilon_0\mu_0\mathbf{P} = \frac{\varepsilon_0\mu_0}{2}\mathrm{Re}\left\{\mathbf{E}^*\times\mathbf{H}\right\}, \tag{S30}$$

$$\tilde{\mathbf{p}}_o = \frac{1}{4\omega}\mathrm{Im}\left\{\tilde{\varepsilon}\mathbf{E}^*\cdot(\nabla)\mathbf{E} + \tilde{\mu}\mathbf{H}^*\cdot(\nabla)\mathbf{H}\right\}, \tag{S31}$$

and

$$\tilde{\mathbf{S}} = \frac{1}{4\omega}\mathrm{Im}\left\{\tilde{\varepsilon}\mathbf{E}^*\times\mathbf{E} + \tilde{\mu}\mathbf{H}^*\times\mathbf{H}\right\}, \tag{S32}$$

respectively. It can be observed that an additional group term is introduced into the permittivity/permeability of the canonical momentum density and SAM to represent the dispersive effect:

$$\tilde{\varepsilon} = \varepsilon + \omega\frac{\partial\varepsilon}{\partial\omega} \text{ and } \tilde{\mu} = \mu + \omega\frac{\partial\mu}{\partial\omega} \tag{S33}$$

Remarkably, the dispersion will not change the expression of kinetic momentum/Poynting vector. From the section II, if we still use kinetic momentum/Poynting vector to evaluate the optical transverse spin, one can find that these group terms are not only originated from the inhomogeneities of EM field (they are originated from dispersion and inhomogeneities simultaneously). However, the group terms have the similar property with the dispersionless terms (The vector property of SAM depends on the electric ellipticity $\mathbf{E}^*\times\mathbf{E}$ and magnetic ellipticity $\mathbf{H}^*\times\mathbf{H}$, while the intensity of SAM is determined by the relative value of permittivity and permeability.). Therefore, it is meaningful to reformulate the spin-momentum equation by considering the dispersive terms.

We first consider the continuities of the kinetic momentum/Poynting vector

$$\nabla\cdot\mathbf{p} = \nabla\cdot\mathbf{P} = 0 \tag{S34}$$

since the kinetic momentum/Poynting vector does not contain the dispersive group terms. Then, the continuities of the dispersive canonical momentum and SAM densities are

$$\nabla\cdot\tilde{\mathbf{p}}_o = 0, \tag{S35}$$

$$\nabla\cdot\tilde{\mathbf{S}} = \tilde{C} = \frac{\omega}{4}\left(\varepsilon\frac{\partial\mu}{\partial\omega} - \mu\frac{\partial\varepsilon}{\partial\omega}\right)\left(\mathbf{E}^*\cdot\mathbf{H} + \mathbf{E}\cdot\mathbf{H}^*\right), \tag{S36}$$

respectively. Obviously, the dispersive SAM is active and the dispersive effect can be considered as the source of the SAM. However, if the dual symmetry between the dispersive permittivity and permeability is protected ($\varepsilon\,\partial\mu/\partial\omega = \mu\,\partial\varepsilon/\partial\omega$), the dispersive SAM would be conservative in the dispersive medium. On the other hand, the dispersive canonical momentum is continuous.

Since the kinetic momentum/Poynting vector is irrelative to the dispersion, the spin-orbit decomposition of kinetic momentum/Poynting vector cannot obtain the dispersive canonical momentum and SAM. By re-examining the process of the spin-orbit decomposition given by M. V. Berry [57], we define a dispersive momentum similar to the kinetic momentum:

$$\tilde{\mathbf{p}} = \frac{1}{4}\mathrm{Re}\left[\tilde{\varepsilon}\mu\mathbf{E}^*\times\mathbf{H} + \varepsilon\tilde{\mu}\mathbf{E}\times\mathbf{H}^*\right] = \frac{1}{4\omega}\mathrm{Im}\left[\tilde{\varepsilon}\mathbf{E}^*\times(\nabla\times\mathbf{E}) + \tilde{\mu}\mathbf{H}^*\times(\nabla\times\mathbf{H})\right]. \tag{S37}$$

In the free space, there are $\tilde{\varepsilon} = \varepsilon_0$ and $\tilde{\mu} = \mu_0$, and the dispersion momentum is consistent with the kinetic momentum. In particular, for the noble metals or magnetic materials, there are $\mu=\mu_0$, $\varepsilon = \varepsilon_0\left(1-\omega_{ep}^2/\omega^2\right)$ or $\varepsilon=\varepsilon_0$, $\mu = \mu_0\left(1-\omega_{mp}^2/\omega^2\right)$ [62], where $\mu_0$ and $\varepsilon_0$ are the permeability and permittivity in vacuum, respectively; $\omega_{ep}$ and $\omega_{mp}$ are the electric plasmon frequency and magnetic plasmon frequency, respectively. In the case, there is $\tilde{\varepsilon} = \varepsilon_0\left(1+\omega_{ep}^2/\omega^2\right)$ or $\tilde{\mu} = \mu_0\left(1+\omega_{mp}^2/\omega^2\right)$. Thus, the dispersion momentum is

$$\tilde{\mathbf{p}} = \frac{1}{4}\mathrm{Re}\left[\tilde{\varepsilon}\mu\mathbf{E}^*\times\mathbf{H} + \varepsilon\tilde{\mu}\mathbf{E}\times\mathbf{H}^*\right] = \frac{\varepsilon_0\mu_0}{2}\mathrm{Re}\left[\mathbf{E}^*\times\mathbf{H}\right] = \mathbf{p}\,, \tag{S38}$$

which is consistent with the kinetic momentum of photons. In addition, the dispersive momentum can be decomposed into:

$$\tilde{\mathbf{p}} = \tilde{\mathbf{p}}_o + \tilde{\mathbf{p}}_s\,, \tag{S39}$$

where the $\tilde{\mathbf{p}}_o$ is consistent with the dispersive canonical momentum and the dispersive spin momentum $\tilde{\mathbf{p}}_s$ is

$$\tilde{\mathbf{p}}_s = \frac{1}{2}\nabla\times\tilde{\mathbf{S}}\,. \tag{S40}$$

Through simple derivations, one can obtain the continuity of dispersive momentum and dispersive Belinfante spin momentum as:

$$\nabla\cdot\tilde{\mathbf{p}} = 0\,. \tag{S41}$$

and

$$\nabla\cdot\tilde{\mathbf{p}}_s = 0\,, \tag{S42}$$

respectively.

Here, we obtain three spin-momentum equations in equations (S41), (S36) and (S39). In the following, we deduce the fourth spin-momentum equation between the dispersive momentum and the dispersive SAM. In the three-layers system as shown in **Fig. S1(b)**, by calculating with the equations (S8) and (S37), the dispersive momentum density can be expressed as

$$\tilde{p}_x = \frac{\tilde{\varepsilon}(\omega)\mu + \varepsilon\tilde{\mu}(\omega)}{2}\begin{bmatrix} +\dfrac{\omega}{4}\dfrac{A_+^*A_+}{\varepsilon^+\varepsilon^+}\mathrm{Im}\left\{\dfrac{\varepsilon^+}{\beta^2}\xi^*\dfrac{\partial\xi}{\partial x} - \dfrac{\varepsilon^+}{\beta^2}\xi\dfrac{\partial\xi^*}{\partial x}\right\}e^{-2k_z^+(z-a/2)} & z > +\dfrac{a}{2} \\ \left\{\begin{array}{l} +\dfrac{\omega}{4}\dfrac{B_+^*B_+}{\varepsilon^m\varepsilon^m}\mathrm{Im}\left\{\dfrac{\varepsilon^m}{\beta^2}\xi^*\dfrac{\partial\xi}{\partial x} - \dfrac{\varepsilon^m}{\beta^2}\xi\dfrac{\partial\xi^*}{\partial x}\right\}e^{+2k_z^m(z-a/2)} \\ +\dfrac{\omega}{4}\dfrac{B_+^*B_- + B_-^*B_+}{\varepsilon^m\varepsilon^m}\mathrm{Im}\left\{\dfrac{\varepsilon^m}{\beta^2}\xi^*\dfrac{\partial\xi}{\partial x} - \dfrac{\varepsilon^m}{\beta^2}\xi\dfrac{\partial\xi^*}{\partial x}\right\}e^{-k_z^m a} \\ +\dfrac{\omega}{4}\dfrac{B_-^*B_-}{\varepsilon^m\varepsilon^m}\mathrm{Im}\left\{\dfrac{\varepsilon^m}{\beta^2}\xi^*\dfrac{\partial\xi}{\partial x} - \dfrac{\varepsilon^m}{\beta^2}\xi\dfrac{\partial\xi^*}{\partial x}\right\}e^{-2k_z^m(z+a/2)} \end{array}\right\} & -\dfrac{a}{2} < z < +\dfrac{a}{2} \\ +\dfrac{\omega}{4}\dfrac{A_-^*A_-}{\varepsilon^-\varepsilon^-}\mathrm{Im}\left\{\dfrac{\varepsilon^-}{\beta^2}\xi^*\dfrac{\partial\xi}{\partial x} - \dfrac{\varepsilon^-}{\beta^2}\xi\dfrac{\partial\xi^*}{\partial x}\right\}e^{+2k_z^-(z+a/2)} & z < -\dfrac{a}{2} \end{bmatrix}, \tag{S43a}$$

$$\tilde{p}_y = \frac{\tilde{\varepsilon}(\omega)\mu + \varepsilon\tilde{\mu}(\omega)}{2}\begin{bmatrix} +\dfrac{\omega}{4}\dfrac{A_+^*A_+}{\varepsilon^+\varepsilon^+}\mathrm{Im}\left\{\dfrac{\varepsilon^+}{\beta^2}\xi^*\dfrac{\partial\xi}{\partial y} - \dfrac{\varepsilon^+}{\beta^2}\xi\dfrac{\partial\xi^*}{\partial y}\right\}e^{-2k_z^+(z-a/2)} & z > +\dfrac{a}{2} \\ \left\{\begin{array}{l} +\dfrac{\omega}{4}\dfrac{B_+^*B_+}{\varepsilon^m\varepsilon^m}\mathrm{Im}\left\{\dfrac{\varepsilon^m}{\beta^2}\xi^*\dfrac{\partial\xi}{\partial y} - \dfrac{\varepsilon^m}{\beta^2}\xi\dfrac{\partial\xi^*}{\partial y}\right\}e^{+2k_z^m(z-a/2)} \\ +\dfrac{\omega}{4}\dfrac{B_+^*B_- + B_-^*B_+}{\varepsilon^m\varepsilon^m}\mathrm{Im}\left\{\dfrac{\varepsilon^m}{\beta^2}\xi^*\dfrac{\partial\xi}{\partial y} - \dfrac{\varepsilon^m}{\beta^2}\xi\dfrac{\partial\xi^*}{\partial y}\right\}e^{-k_z^m a} \\ +\dfrac{\omega}{4}\dfrac{B_-^*B_-}{\varepsilon^m\varepsilon^m}\mathrm{Im}\left\{\dfrac{\varepsilon^m}{\beta^2}\xi^*\dfrac{\partial\xi}{\partial y} - \dfrac{\varepsilon^m}{\beta^2}\xi\dfrac{\partial\xi^*}{\partial y}\right\}e^{-2k_z^m(z+a/2)} \end{array}\right\} & -\dfrac{a}{2} < z < +\dfrac{a}{2} \\ +\dfrac{\omega}{4}\dfrac{A_-^*A_-}{\varepsilon^-\varepsilon^-}\mathrm{Im}\left\{\dfrac{\varepsilon^-}{\beta^2}\xi^*\dfrac{\partial\xi}{\partial y} - \dfrac{\varepsilon^-}{\beta^2}\xi\dfrac{\partial\xi^*}{\partial y}\right\}e^{+2k_z^-(z+a/2)} & z < -\dfrac{a}{2} \end{bmatrix}, \tag{S43b}$$

$$\tilde{p}_z = \begin{bmatrix} 0 & z > +\frac{a}{2} \\ 0 & -\frac{a}{2} < z < +\frac{a}{2} \\ 0 & z < -\frac{a}{2} \end{bmatrix}. \qquad \text{(S43c)}$$

On the other hand, the three components of dispersive SAM are

$$\tilde{S}_x = \begin{bmatrix} +\frac{A_+^* A_+}{4\omega}\frac{\tilde{\varepsilon}^+}{\varepsilon^+\varepsilon^+}\frac{k_z^+}{\beta^2}\mathrm{Im}\left(\xi^*\frac{\partial \xi}{\partial y} - \xi\frac{\partial \xi^*}{\partial y}\right)e^{-2k_z^+(z-a/2)} & z > +\frac{a}{2} \\ \left\{\begin{array}{l} -\frac{B_+^* B_+}{4\omega}\frac{\tilde{\varepsilon}^m}{\varepsilon^m\varepsilon^m}\frac{k_z^m}{\beta^2}\mathrm{Im}\left(\xi^*\frac{\partial \xi}{\partial y} - \xi\frac{\partial \xi^*}{\partial y}\right)e^{+2k_z^m(z-a/2)} \\ +\frac{\tilde{\varepsilon}^m}{\varepsilon^m\varepsilon^m}\frac{k_z^m}{\beta^2}\mathrm{Im}\left[\frac{B_+^* B_- - B_+ B_-^*}{4\omega}\left(\xi^*\frac{\partial \xi}{\partial y} + \xi\frac{\partial \xi^*}{\partial y}\right)\right]e^{-k_z^m a} \\ +\frac{B_-^* B_-}{4\omega}\frac{\tilde{\varepsilon}^m}{\varepsilon^m\varepsilon^m}\frac{k_z^m}{\beta^2}\mathrm{Im}\left(\xi^*\frac{\partial \xi}{\partial y} - \xi\frac{\partial \xi^*}{\partial y}\right)e^{-2k_z^m(z+a/2)} \end{array}\right\} & -\frac{a}{2} < z < +\frac{a}{2} \\ -\frac{A_-^* A_-}{4\omega}\frac{\tilde{\varepsilon}^-}{\varepsilon^-\varepsilon^-}\frac{k_z^-}{\beta^2}\mathrm{Im}\left(\xi^*\frac{\partial \xi}{\partial y} - \xi\frac{\partial \xi^*}{\partial y}\right)e^{+2k_z^-(z+a/2)} & z < -\frac{a}{2} \end{bmatrix}, \qquad \text{(S44a)}$$

$$\tilde{S}_y = \begin{bmatrix} -\frac{A_+^* A_+}{4\omega}\frac{\tilde{\varepsilon}^+}{\varepsilon^+\varepsilon^+}\frac{k_z^+}{\beta^2}\mathrm{Im}\left(\xi^*\frac{\partial \xi}{\partial x} - \xi\frac{\partial \xi^*}{\partial x}\right)e^{-2k_z^+(z-a/2)} & z > +\frac{a}{2} \\ \left\{\begin{array}{l} +\frac{B_+^* B_+}{4\omega}\frac{\tilde{\varepsilon}^m}{\varepsilon^m\varepsilon^m}\frac{k_z^m}{\beta^2}\mathrm{Im}\left(\xi^*\frac{\partial \xi}{\partial x} - \xi\frac{\partial \xi^*}{\partial x}\right)e^{+2k_z^m(z-a/2)} \\ -\frac{\tilde{\varepsilon}^m}{\varepsilon^m\varepsilon^m}\frac{k_z^m}{\beta^2}\mathrm{Im}\left[\frac{B_+^* B_- - B_-^* B_+}{4\omega}\left(\xi^*\frac{\partial \xi}{\partial x} + \xi\frac{\partial \xi^*}{\partial x}\right)\right]e^{-k_z^m a} \\ -\frac{B_-^* B_-}{4\omega}\frac{\tilde{\varepsilon}^m}{\varepsilon^m\varepsilon^m}\frac{k_z^m}{\beta^2}\mathrm{Im}\left(\xi^*\frac{\partial \xi}{\partial x} - \xi\frac{\partial \xi^*}{\partial x}\right)e^{-2k_z^m(z+a/2)} \end{array}\right\} & -\frac{a}{2} < z < +\frac{a}{2} \\ +\frac{A_-^* A_-}{4\omega}\frac{\tilde{\varepsilon}^-}{\varepsilon^-\varepsilon^-}\frac{k_z^-}{\beta^2}\mathrm{Im}\left(\xi^*\frac{\partial \xi}{\partial x} - \xi\frac{\partial \xi^*}{\partial x}\right)e^{+2k_z^-(z+a/2)} & z < -\frac{a}{2} \end{bmatrix}, \qquad \text{(S44b)}$$

and

$$\tilde{S}_z = \begin{bmatrix} +\frac{A_+^* A_+}{4\omega}\left(\frac{\tilde{\varepsilon}^+}{\varepsilon^+\varepsilon^+}\frac{k_z^+ k_z^+}{\beta^4} + \frac{\omega^2\tilde{\mu}^+}{\beta^4}\right)\mathrm{Im}\left(\frac{\partial \xi^*}{\partial x}\frac{\partial \xi}{\partial y} - \frac{\partial \xi}{\partial x}\frac{\partial \xi^*}{\partial y}\right)e^{-2k_z^+(z-a/2)} & z > +\frac{a}{2} \\ \left\{\begin{array}{l} +\frac{B_+^* B_+}{4\omega}\left(\frac{\tilde{\varepsilon}^m}{\varepsilon^m\varepsilon^m}\frac{k_z^m k_z^m}{\beta^4} + \frac{\omega^2\tilde{\mu}^m}{\beta^4}\right)\mathrm{Im}\left(\frac{\partial \xi^*}{\partial x}\frac{\partial \xi}{\partial y} - \frac{\partial \xi}{\partial x}\frac{\partial \xi^*}{\partial y}\right)e^{+2k_z^m(z-a/2)} \\ -\frac{B_+^* B_- + B_-^* B_+}{4\omega}\left(\frac{\tilde{\varepsilon}^m}{\varepsilon^m\varepsilon^m}\frac{k_z^m k_z^m}{\beta^4} - \frac{\omega^2\tilde{\mu}^m}{\beta^4}\right)\mathrm{Im}\left(\frac{\partial \xi^*}{\partial x}\frac{\partial \xi}{\partial y} - \frac{\partial \xi}{\partial x}\frac{\partial \xi^*}{\partial y}\right)e^{-k_z^m a} \\ +\frac{B_-^* B_-}{4\omega}\left(\frac{\tilde{\varepsilon}^m}{\varepsilon^m\varepsilon^m}\frac{k_z^m k_z^m}{\beta^4} + \frac{\omega^2\tilde{\mu}^m}{\beta^4}\right)\mathrm{Im}\left(\frac{\partial \xi^*}{\partial x}\frac{\partial \xi}{\partial y} - \frac{\partial \xi}{\partial x}\frac{\partial \xi^*}{\partial y}\right)e^{-2k_z^m(z+a/2)} \end{array}\right\} & -\frac{a}{2} < z < +\frac{a}{2} \\ +\frac{A_-^* A_-}{4\omega}\left(\frac{\tilde{\varepsilon}^-}{\varepsilon^-\varepsilon^-}\frac{k_z^- k_z^-}{\beta^4} + \frac{\omega^2\tilde{\mu}^-}{\beta^4}\right)\mathrm{Im}\left(\frac{\partial \xi^*}{\partial x}\frac{\partial \xi}{\partial y} - \frac{\partial \xi}{\partial x}\frac{\partial \xi^*}{\partial y}\right)e^{+2k_z^-(z+a/2)} & z < -\frac{a}{2} \end{bmatrix}. \qquad \text{(S44c)}$$

By employing the expressions (S43) and (S44), the spin-momentum locking can be expressed as

$$\tilde{S}_x = \frac{1}{2\omega^2\varepsilon\mu}\frac{2}{1+\frac{\varepsilon}{\mu}\frac{\tilde{\mu}(\omega)}{\tilde{\varepsilon}(\omega)}}(\nabla\times\tilde{\mathbf{p}})_x, \tag{S45a}$$

$$\tilde{S}_y = \frac{1}{2\omega^2\varepsilon\mu}\frac{2}{1+\frac{\varepsilon}{\mu}\frac{\tilde{\mu}(\omega)}{\tilde{\varepsilon}(\omega)}}(\nabla\times\tilde{\mathbf{p}})_y, \tag{S45b}$$

and

$$\tilde{S}_z - \tilde{S}_l = \frac{1}{2\omega^2\varepsilon\mu}\frac{2-2\frac{\omega^2\varepsilon\mu}{\beta^2}\left(1-\frac{\varepsilon}{\mu}\frac{\tilde{\mu}(\omega)}{\tilde{\varepsilon}(\omega)}\right)}{1+\frac{\varepsilon}{\mu}\frac{\tilde{\mu}(\omega)}{\tilde{\varepsilon}(\omega)}}(\nabla\times\tilde{\mathbf{p}})_z. \tag{S45c}$$

In sum, the spin-momentum locking can be expressed as

$$\tilde{\mathbf{S}}_t = \tilde{\mathbf{S}} - \tilde{\mathbf{S}}_l, \tag{S46}$$

and

$$\tilde{\mathbf{S}}_t = \frac{1}{2\omega^2\varepsilon\mu}\chi\nabla\times\tilde{\mathbf{p}}. \tag{S47}$$

In the Eq. (S47), the parameter is

$$\chi = \begin{cases} 2/[1+\tilde{\eta}/\eta] & \text{horizontal}, \chi_{\parallel} \\ \left[2-\frac{2}{n_{eff}^2}(1-\tilde{\eta}/\eta)\right]\Big/[1+\tilde{\eta}/\eta] & \text{Normal}, \chi_n \end{cases}. \tag{S48}$$

Here, $\eta = \mu/\varepsilon$, $\tilde{\eta}(\omega) = \tilde{\mu}(\omega)/\tilde{\varepsilon}(\omega)$ and $n_{eff}^2 = \beta^2/k^2$ (Noteworthily, the propagation constant is identical in the structure, while the wavevector $k$ depends on the material's property in each layer). In the dispersionless limit, there is $\tilde{\mu}/\tilde{\varepsilon} = \mu/\varepsilon$, the parameter $\chi$ is equal to 1 universally. Thus, this asymmetry in the spin-momentum locking property between the horizontal and normal components is originated from the dispersion of materials. Moreover, in Eq. (S46), a coupling spin term, which is considered as the longitudinal spin as the section 2, is expressed as:

$$\tilde{\mathbf{S}}_l = 0\hat{\mathbf{x}} + 0\hat{\mathbf{y}} + \begin{bmatrix} 0\hat{\mathbf{z}} \\ -\frac{\tilde{\varepsilon}^m(\omega)}{4\omega}\frac{B_+^*B_- + B_-^*B_+}{\varepsilon^m\varepsilon^m}\frac{2k_z^m k_z^m}{\beta^4}\mathrm{Im}\left\{\frac{\partial\xi^*}{\partial x}\frac{\partial\xi}{\partial y} - \frac{\partial\xi}{\partial x}\frac{\partial\xi^*}{\partial y}\right\}e^{-k_z^m a}\hat{\mathbf{z}} \\ 0\hat{\mathbf{z}} \end{bmatrix}. \tag{S49}$$

The longitudinal spin only has the normal component in the layer. In the dispersionless limit, it has $\tilde{\varepsilon}^m/\varepsilon^m = 1$, $\tilde{\mu}^m/\mu^m = 1$ and $1-\tilde{\eta}/\eta = 0$ in the layer. Thus, one can reach that

$$\tilde{\mathbf{S}}_l = \mathbf{S}_l = \begin{bmatrix} 0 & z > +\frac{a}{2} \\ -\frac{\varepsilon^m}{4\omega}\frac{B_+^*B_- + B_-^*B_+}{\varepsilon^m\varepsilon^m}\frac{2k_z^m k_z^m}{\beta^4}\mathrm{Im}\left(\frac{\partial\xi^*}{\partial x}\frac{\partial\xi}{\partial y} - \frac{\partial\xi}{\partial x}\frac{\partial\xi^*}{\partial y}\right)e^{-k_z^m a}\hat{\mathbf{z}} & -\frac{a}{2} < z < +\frac{a}{2} \\ 0 & z < -\frac{a}{2} \end{bmatrix}, \tag{S50}$$

which matches well with the expression (S27). Thus, this term can be considered as the coupling spin term originated from the coupling ellipticities between the orthogonal polarized components in the horizontal plane uniformly.

In sum, the four Maxwell-like spin-momentum equations for the surface EM modes can be expressed as:

$$\nabla\cdot\tilde{\mathbf{p}}=0\,, \tag{S51a}$$

$$\nabla\cdot\tilde{\mathbf{S}}=\tilde{C}=\frac{1}{4}\left(\varepsilon\tilde{\mu}-\tilde{\varepsilon}\mu\right)\left(\mathbf{E}^{*}\cdot\mathbf{H}+\mathbf{E}\cdot\mathbf{H}^{*}\right), \tag{S51b}$$

$$\nabla\times\tilde{\mathbf{S}}=2\left(\tilde{\mathbf{p}}-\tilde{\mathbf{p}}_{o}\right), \tag{S51c}$$

$$\tilde{\mathbf{S}}_{t}=\tilde{\mathbf{S}}-\tilde{\mathbf{S}}_{l}=\frac{1}{2\omega^{2}\varepsilon\mu}\chi\nabla\times\tilde{\mathbf{p}}\,, \tag{S51d}$$

Here, the $\tilde{\mathbf{p}}$ is given in Eq. (S37); the $\tilde{\mathbf{p}}_{o}$ is given in Eq. (S31); $\tilde{C}$ indicates the dispersion induced EM spin. However, since $\nabla\cdot\tilde{\mathbf{S}}\neq 0$, the Helmholtz spin-momentum equation cannot be obtained directly.

To obtain the Helmholtz spin-momentum equation, we first calculate the dispersive canonical momentum density as

$$\tilde{p}_{ox}=\frac{1}{4\omega}\mathrm{Im}\left\{\begin{array}{l}
A_{+}^{*}A_{+}\left[\frac{\tilde{\varepsilon}^{+}}{\varepsilon^{+}\varepsilon^{+}}\xi^{*}\frac{\partial\xi}{\partial x}+\left(\frac{\tilde{\varepsilon}^{+}}{\varepsilon^{+}\varepsilon^{+}}\frac{k_{z}^{+}k_{z}^{+}}{\beta^{4}}+\frac{\omega^{2}\tilde{\mu}^{+}}{\beta^{4}}\right)\left(\frac{\partial\xi^{*}}{\partial x}\frac{\partial^{2}\xi}{\partial x^{2}}+\frac{\partial\xi^{*}}{\partial y}\frac{\partial^{2}\xi}{\partial x\partial y}\right)\right]e^{-2k_{z}^{+}(z-a/2)}\\
\left\{\begin{array}{l}
+B_{+}^{*}B_{+}\left[\frac{\tilde{\varepsilon}^{m}}{\varepsilon^{m}\varepsilon^{m}}\xi^{*}\frac{\partial\xi}{\partial x}+\left(\frac{\tilde{\varepsilon}^{m}}{\varepsilon^{m}\varepsilon^{m}}\frac{k_{z}^{m}k_{z}^{m}}{\beta^{4}}+\frac{\omega^{2}\tilde{\mu}^{m}}{\beta^{4}}\right)\left(\frac{\partial\xi^{*}}{\partial x}\frac{\partial^{2}\xi}{\partial x^{2}}+\frac{\partial\xi^{*}}{\partial y}\frac{\partial^{2}\xi}{\partial x\partial y}\right)\right]e^{+2k_{z}^{m}(z-a/2)}\\
+\left(\begin{array}{l}+B_{+}^{*}B_{-}\\+B_{-}^{*}B_{+}\end{array}\right)\left[\frac{\tilde{\varepsilon}^{m}}{\varepsilon^{m}\varepsilon^{m}}\xi^{*}\frac{\partial\xi}{\partial x}+\left(\frac{\omega^{2}\tilde{\mu}^{m}}{\beta^{4}}-\frac{\tilde{\varepsilon}^{m}}{\varepsilon^{m}\varepsilon^{m}}\frac{k_{z}^{m}k_{z}^{m}}{\beta^{4}}\right)\left(\frac{\partial\xi^{*}}{\partial x}\frac{\partial^{2}\xi}{\partial x^{2}}+\frac{\partial\xi^{*}}{\partial y}\frac{\partial^{2}\xi}{\partial x\partial y}\right)\right]e^{-k_{z}^{m}a}\\
+B_{-}^{*}B_{-}\left[\frac{\tilde{\varepsilon}^{m}}{\varepsilon^{m}\varepsilon^{m}}\xi^{*}\frac{\partial\xi}{\partial x}+\left(\frac{\tilde{\varepsilon}^{m}}{\varepsilon^{m}\varepsilon^{m}}\frac{k_{z}^{m}k_{z}^{m}}{\beta^{4}}+\frac{\omega^{2}\tilde{\mu}^{m}}{\beta^{4}}\right)\left(\frac{\partial\xi^{*}}{\partial x}\frac{\partial^{2}\xi}{\partial x^{2}}+\frac{\partial\xi^{*}}{\partial y}\frac{\partial^{2}\xi}{\partial x\partial y}\right)\right]e^{-2k_{z}^{m}(z+a/2)}
\end{array}\right\}\\
A_{-}^{*}A_{-}\left[\frac{\tilde{\varepsilon}^{-}}{\varepsilon^{-}\varepsilon^{-}}\xi^{*}\frac{\partial\xi}{\partial x}+\left(\frac{\tilde{\varepsilon}^{-}}{\varepsilon^{-}\varepsilon^{-}}\frac{k_{z}^{-}k_{z}^{-}}{\beta^{4}}+\frac{\omega^{2}\tilde{\mu}^{-}}{\beta^{4}}\right)\left(\frac{\partial\xi^{*}}{\partial x}\frac{\partial^{2}\xi}{\partial x^{2}}+\frac{\partial\xi^{*}}{\partial y}\frac{\partial^{2}\xi}{\partial x\partial y}\right)\right]e^{+2k_{z}^{-}(z+a/2)}
\end{array}\right\}, \tag{S52a}$$

$$\tilde{p}_{oy}=\frac{1}{4\omega}\mathrm{Im}\left\{\begin{array}{l}
A_{+}^{*}A_{+}\left[\frac{\tilde{\varepsilon}^{+}}{\varepsilon^{+}\varepsilon^{+}}\xi^{*}\frac{\partial\xi}{\partial y}+\left(\frac{\tilde{\varepsilon}^{+}}{\varepsilon^{+}\varepsilon^{+}}\frac{k_{z}^{+}k_{z}^{+}}{\beta^{4}}+\frac{\omega^{2}\tilde{\mu}^{+}}{\beta^{4}}\right)\left(\frac{\partial\xi^{*}}{\partial x}\frac{\partial^{2}\xi}{\partial y\partial x}+\frac{\partial\xi^{*}}{\partial y}\frac{\partial^{2}\xi}{\partial y^{2}}\right)\right]e^{-2k_{z}^{+}(z-a/2)}\\
\left\{\begin{array}{l}
+B_{+}^{*}B_{+}\left[\frac{\tilde{\varepsilon}^{m}}{\varepsilon^{m}\varepsilon^{m}}\xi^{*}\frac{\partial\xi}{\partial y}+\left(\frac{\tilde{\varepsilon}^{m}}{\varepsilon^{m}\varepsilon^{m}}\frac{k_{z}^{m}k_{z}^{m}}{\beta^{4}}+\frac{\omega^{2}\tilde{\mu}^{m}}{\beta^{4}}\right)\left(\frac{\partial\xi^{*}}{\partial x}\frac{\partial^{2}\xi}{\partial y\partial x}+\frac{\partial\xi^{*}}{\partial y}\frac{\partial^{2}\xi}{\partial y^{2}}\right)\right]e^{+2k_{z}^{m}(z-a/2)}\\
+\left(\begin{array}{l}+B_{+}^{*}B_{-}\\+B_{-}^{*}B_{+}\end{array}\right)\left[\frac{\tilde{\varepsilon}^{m}}{\varepsilon^{m}\varepsilon^{m}}\xi^{*}\frac{\partial\xi}{\partial y}+\left(\frac{\omega^{2}\tilde{\mu}^{m}}{\beta^{4}}-\frac{\tilde{\varepsilon}^{m}}{\varepsilon^{m}\varepsilon^{m}}\frac{k_{z}^{m}k_{z}^{m}}{\beta^{4}}\right)\left(\frac{\partial\xi^{*}}{\partial x}\frac{\partial^{2}\xi}{\partial y\partial x}+\frac{\partial\xi^{*}}{\partial y}\frac{\partial^{2}\xi}{\partial y^{2}}\right)\right]e^{-k_{z}^{m}a}\\
+B_{-}^{*}B_{-}\left[\frac{\tilde{\varepsilon}^{m}}{\varepsilon^{m}\varepsilon^{m}}\xi^{*}\frac{\partial\xi}{\partial y}+\left(\frac{\tilde{\varepsilon}^{m}}{\varepsilon^{m}\varepsilon^{m}}\frac{k_{z}^{m}k_{z}^{m}}{\beta^{4}}+\frac{\omega^{2}\tilde{\mu}^{m}}{\beta^{4}}\right)\left(\frac{\partial\xi^{*}}{\partial x}\frac{\partial^{2}\xi}{\partial y\partial x}+\frac{\partial\xi^{*}}{\partial y}\frac{\partial^{2}\xi}{\partial y^{2}}\right)\right]e^{-2k_{z}^{m}(z+a/2)}
\end{array}\right\}\\
A_{-}^{*}A_{-}\left[\frac{\tilde{\varepsilon}^{-}}{\varepsilon^{-}\varepsilon^{-}}\xi^{*}\frac{\partial\xi}{\partial y}+\left(\frac{\tilde{\varepsilon}^{-}}{\varepsilon^{-}\varepsilon^{-}}\frac{k_{z}^{-}k_{z}^{-}}{\beta^{4}}+\frac{\omega^{2}\tilde{\mu}^{-}}{\beta^{4}}\right)\left(\frac{\partial\xi^{*}}{\partial x}\frac{\partial^{2}\xi}{\partial y\partial x}+\frac{\partial\xi^{*}}{\partial y}\frac{\partial^{2}\xi}{\partial y^{2}}\right)\right]e^{+2k_{z}^{-}(z+a/2)}
\end{array}\right\}, \tag{S52b}$$

and

$$\tilde{p}_{oz}=\begin{bmatrix}0 & z>+\frac{a}{2}\\ 0 & -\frac{a}{2}<z<\frac{a}{2}\\ 0 & z<-\frac{a}{2}\end{bmatrix}. \tag{S52c}$$

From the equations (S44) and (S52), one can reach two separated dispersive spin-momentum equations:

$$\alpha_{\parallel}\left(\nabla\times\tilde{\mathbf{p}}_o\right)_{\parallel}=\nabla^2\tilde{S}_{\parallel}+2k^2\gamma_t\tilde{S}_{\parallel}\,, \tag{S53a}$$

and

$$2\left(\nabla\times\tilde{\mathbf{p}}_o\right)_n=\nabla^2\tilde{S}_n+2k^2\gamma_t\tilde{S}_n-2k^2\gamma_l\tilde{S}_l\,. \tag{S53b}$$

Here, the symbol || indicates the horizontal components (for example: $x$ and $y$) and $n$ denotes the normal component. The parameters are

$$\alpha=\begin{cases}\alpha_{\parallel}=\left[2+\dfrac{2\left(1-\tilde{\eta}/\eta\right)}{n_{eff}^2-\left(1-\tilde{\eta}/\eta\right)}\right] & \text{horizontal}\\ \alpha_n=2 & \text{Normal}\end{cases}, \tag{S54a}$$

and

$$\gamma=\begin{cases}\gamma_t=2+\dfrac{n_{eff}^2\left(1-\tilde{\eta}/\eta\right)}{n_{eff}^2-\left(1-\tilde{\eta}/\eta\right)}\\ \gamma_l=\left[\left(3-2\dfrac{1}{n_{eff}^2}\right)+\left(1-2\dfrac{1}{n_{eff}^2}\right)\dfrac{\tilde{\eta}}{\eta}+\left(1-\dfrac{\tilde{\eta}}{\eta}\right)\dfrac{1-\left(1+\dfrac{\tilde{\eta}}{\eta}\right)\dfrac{1}{n_{eff}^2}}{1-\left(1-\dfrac{\tilde{\eta}}{\eta}\right)\dfrac{1}{n_{eff}^2}}\right]\Bigg/\left[2-2\dfrac{1}{n_{eff}^2}\right]\end{cases}. \tag{S54b}$$

To study the spin-momentum locking properties for the horizontal and normal components of surface EM modes in the dispersive medium, we first investigate the parameters in Eq. (S48) and Eq. (S54). We show the wavelength dependent character of parameters in Fig. S8.

In the dispersive noble metals, there is $-1<\tilde{\eta}/\eta=\left(1-\omega_p^2/\omega^2\right)/\left(1+\omega_p^2/\omega^2\right)<0$ since $\omega<<\omega_p$ at optical frequencies, as shown in Fig. S8(a). Thus, $0<1+\tilde{\eta}/\eta<1$, $1<1-\tilde{\eta}/\eta<2$ and $2<\chi_{\parallel}=2/\left[1+\tilde{\eta}/\eta\right]$, as shown in Fig. S8(d). For the air-metal-air configuration, the $n_{eff}^2$ in the air region is approximatively 1, the $n_{eff}^2$ in the metal region is less than 0 ($n_{eff}^2\approx 0$). Thus, $\chi_n\approx -1/n_{eff}^2>>1$, as shown in Fig. S8(d). On the other hand, since $n_{eff}^2\approx 0$, there are $\alpha_{\parallel}\approx 0$ and also $\gamma_t\approx 2-n_{eff}^2$, as shown in Fig. S8(e) and Fig. S8(f). In the metal-air-metal configuration, the $\tilde{\eta}/\eta=1$ in the layer, and thus the spin-momentum locking in the layer is consistent with the expression (S26).

From the expression (S51), we can understand this spin-momentum equation in three aspects. First, the spin-momentum locking in the dispersive medium is totally different from that of dispersionless medium as exhibited in equation (S26). In the dispersionless dielectric medium, the transverse spin is locked with the kinetic momentum/Poynting vector and the locking property satisfies the right-hand rule. Whereas in the dispersive noble materials, the whole transverse spin is still locked with the dispersive momentum (S37). However, this dispersion-dependent locking property between the SAM and dispersive momentum satisfies the left-hand rule as we consider the dispersive group permittivity in calculating the SAM, no matter whether the horizontal component or normal component of structured SAMs, as shown in Fig. S9 and Fig. S10. Remarkably, although there is an expression to connect the dispersive SAM and canonical momentum, there is no locking relation between these physical quantities (as shown in Fig. S9(e) and Fig. S9 (j), one may recognize that the horizontal dispersive SAM component is locked with the dispersive canonical momentum and satisfies the right-hand rule. However, in Fig. S10(f) and Fig. S10 (l), it can be observed that the normal dispersive SAM component is not locked with the dispersive canonical momentum.). Second, in the expressions of SAM (Eq. (S32)) and dispersive momentum (Eq. (S37)), one can find that these quantities are only affected by the dispersion but not the structural property. However, from our derivation in Eq. (S51), the spin-momentum locking property is definitely determined by the effective index $n_{eff}^2$, which is relative to the structural property. Thus, one can conclude that the complex structure and dispersion can engineer the spin-momentum locking property of EM field simultaneously.

From the equation (51), one can manipulate the spin-momentum locking in two ways: (1) the spin-momentum locking is proportional $1/k^2=1/\omega^2\varepsilon\mu$, which is relative to the permittivity $\varepsilon$ and the permeability $\mu$ of bulk materials (natural materials or artificial metamaterials). Thus, by tuning the permittivity and the permeability of the metamaterials or natural materials (such as Sb2Te3 [NPG Asia Materials **9**, e425(2017)]) from the metal characteristics to the negative refractive characteristics or from the dielectric characteristics to the metal characteristics, the spin-momentum locking properties would be inverted. (2) the spin-momentum locking is also proportional $2/[1+\tilde{\eta}/\eta]$. In the Drude-Lorentz model [62], there is

$$1+\frac{\tilde{\eta}}{\eta}=1+\frac{\varepsilon\tilde{\mu}}{\tilde{\varepsilon}\mu}=\frac{2-2\frac{\omega_{ep}^2}{\omega^2}\frac{\omega_{mp}^2}{\omega^2}}{\left(1+\frac{\omega_{ep}^2}{\omega^2}\right)\left(1-\frac{\omega_{mp}^2}{\omega^2}\right)},$$

where $\omega_{ep}$ and $\omega_{mp}$ are the electric and magnetic plasma frequencies, respectively. In the case that $\omega^2>\omega_{mp}^2$ and $\omega^2<\omega_{ep}^2$ to make $1-\omega_{mp}^2\omega_{ep}^2/\omega^4<0$, there is $1+\tilde{\eta}/\eta<0$. Whereas in the case that $\omega^2>\omega_{mp}^2$ and $\omega^2<\omega_{ep}^2$ to make $1-\omega_{mp}^2\omega_{ep}^2/\omega^4>0$, there is $1+\tilde{\eta}/\eta>0$. Since the $\omega_{ep}$ and $\omega_{mp}$ can be engineered by the structural design (spatial dispersion) in the metamaterials, the intrinsic spin-momentum locking can be manipulated flexibly.

Third, these asymmetry between the horizontal and normal components in expression (S47) is originated from the dispersion induced dual symmetry between the electric and magnetic properties breaking [63] since

$$-1<\frac{\tilde{\eta}}{\eta}=\frac{\tilde{\mu}\varepsilon}{\tilde{\varepsilon}\mu}=\left(1-\frac{\omega_p^2}{\omega^2}\right)\Big/\left(1+\frac{\omega_p^2}{\omega^2}\right)<0 \text{ and } \frac{1}{n_{eff}^2}\left(1-\tilde{\eta}/\eta\right)=2\frac{1}{n_{eff}^2}\frac{\omega_p^2}{\omega^2}\Big/\left(1+\frac{\omega_p^2}{\omega^2}\right)<0 \tag{S55}$$

in the dispersive noble metal. Assuming a specific case that

$$\varepsilon=\varepsilon_0\left(1-\omega_p^2/\omega^2\right) \text{ and } \mu=\mu_0\left(1-\omega_p^2/\omega^2\right), \tag{S56}$$

where the dispersion induced dual symmetry is protected (the permittivity and permeability have the equivalent dispersive properties or contain the same variation tendency), one can reach that

$$\frac{\tilde{\eta}}{\eta}=\frac{\tilde{\mu}\varepsilon}{\tilde{\varepsilon}\mu}=1 \text{ and } \frac{1}{n_{eff}^2}\left(1-\tilde{\eta}/\eta\right)=0. \tag{S57}$$

However, for the ordinary dispersive medium (the dispersion induced dual symmetry is broken consequentially), this asymmetry between the horizontal and normal components is inevitable.

Finally, we also investigate the longitudinal and transverse spins in the $z$-component SAMs for the symmetric mode and anti-symmetric mode in air-metal-air configuration and the symmetric mode in meal-air-metal configuration in Fig. S11.

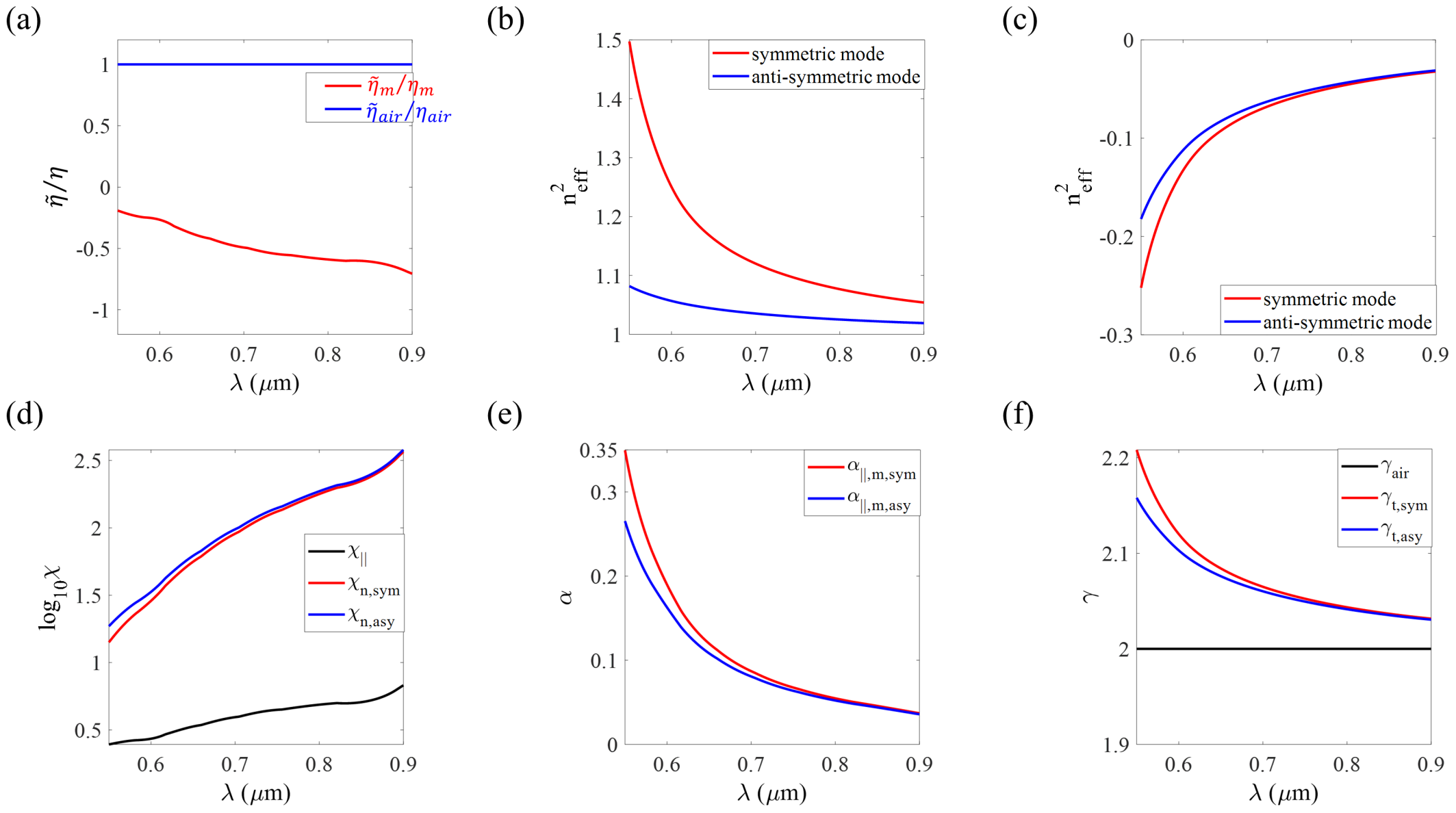

**Fig. S8.** Wavelength dependent (a) dispersive wave impedance $\tilde{\eta}/\eta$ for the silver material and air; wavelength dependent (b) effective index $n_{eff}^2$ for the symmetric mode (red) and anti-symmetric mode (blue) in the air space; (c) the effective index $n_{eff}^2$ for the symmetric mode (red) and anti-symmetric mode (blue) in the metal material; the parameters (d) χ, (e) horizontal $\alpha_{||}$, and (f) γ in the metal material for the symmetric mode (red) and anti-symmetric mode (blue). The black line is the horizontal $\chi_{||}$ in (e) and $\alpha_{air}$ in air space in (f). The multilayered configuration is air-silver-air with the thickness of metal layer is 50nm. The wavelength is 632.8nm.

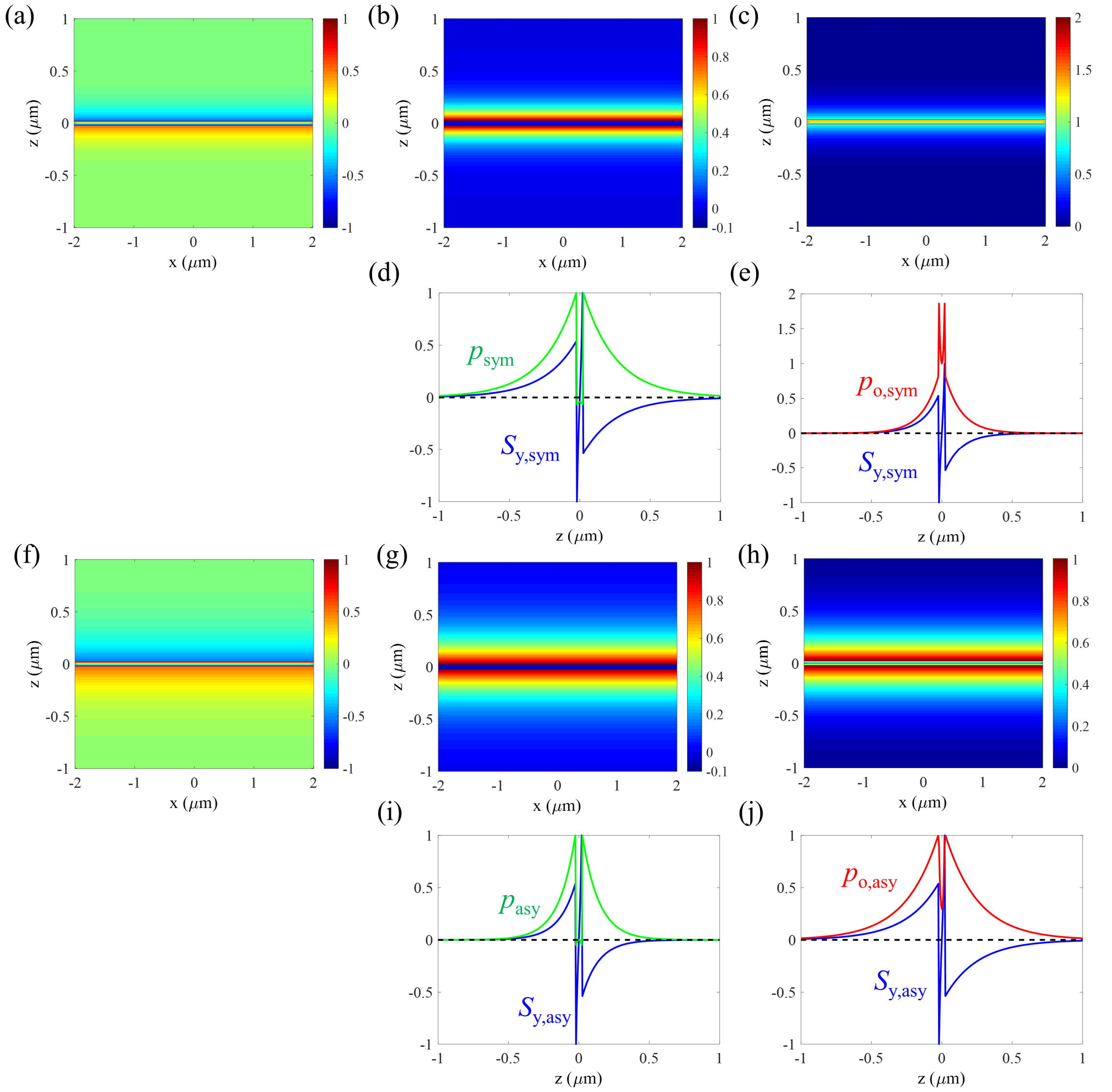

**Fig. S9.** The (a) $\tilde{S}_y$, (b) $\tilde{p}_x$, and (c) $\tilde{p}_{o,x}$ for the symmetric plane wave mode in the *xz*-plane ($y$=0), and the corresponding 1D contours of (e) $\tilde{S}_y$ and $\tilde{p}_x$, (d) $\tilde{S}_y$ and $\tilde{p}_{o,x}$ indicate the direction of dispersive canonical momentum are inverted to that of the dispersive momentum. The (f) $\tilde{S}_y$, (g) $\tilde{p}_x$, and (h) $\tilde{p}_{o,x}$ for the anti-symmetric plane wave mode in the *xz*-plane ($y$=0), and the corresponding 1D contours of (i) $\tilde{S}_y$ and $\tilde{p}_x$, (j) $\tilde{S}_y$ and $\tilde{p}_{o,x}$ indicate the direction of dispersive canonical momentum are also inverted to that of the dispersive momentum. In the plane wave case, if one utilizes the dispersive momentum to evaluate the spin-momentum locking in the layer, the spin vector and the dispersive momentum satisfy the left-hand rule. Whereas the spin vector and the dispersive canonical momentum satisfy the right-hand rule. The wavelength is 632.8nm and the thickness of layer is 50nm.

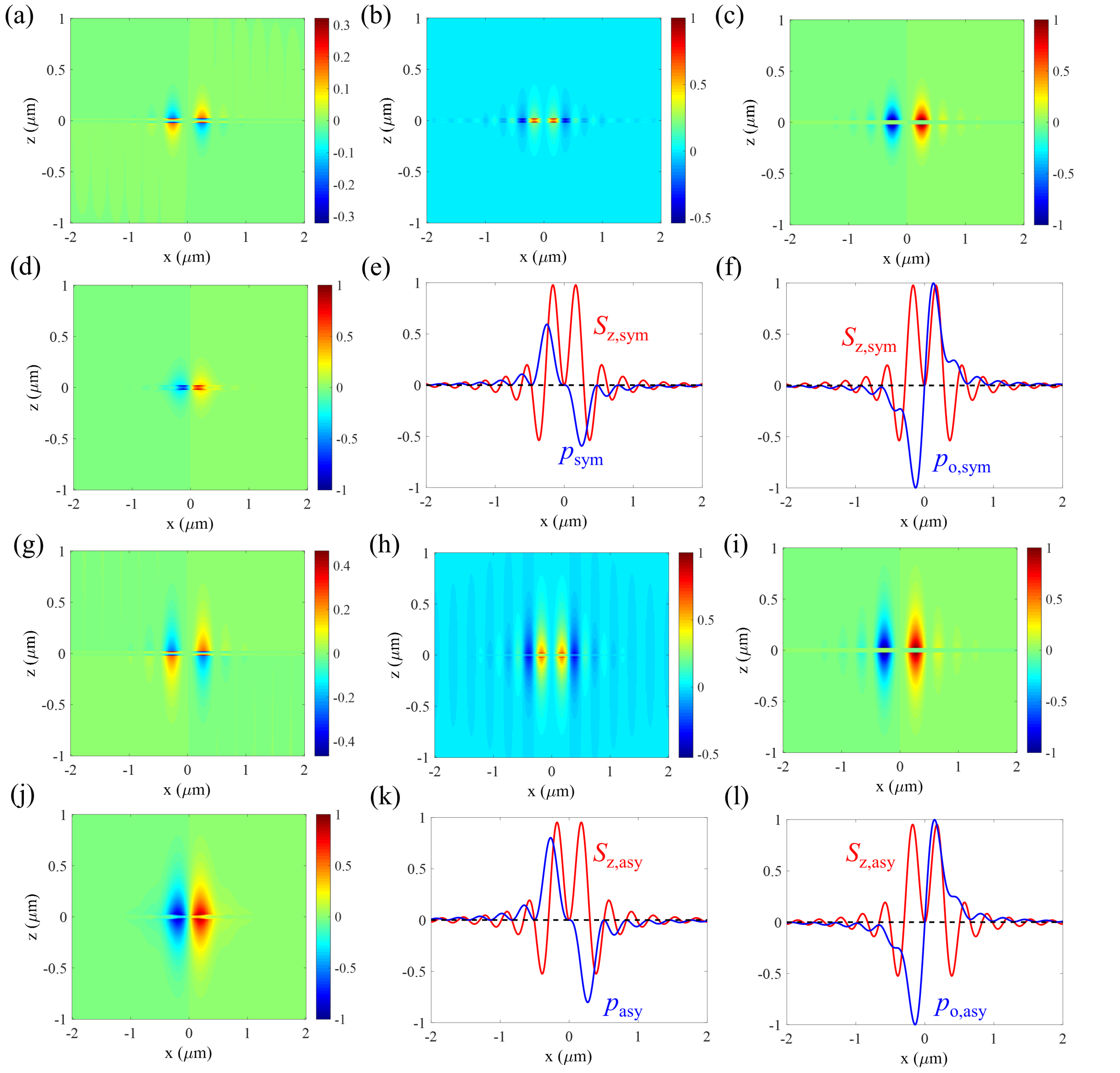

**Fig. S10.** The (a) $\tilde{S}_x$, (b) $\tilde{S}_z$, (c) $\tilde{p}_y$, and (d) $\tilde{p}_{o,y}$ for the symmetric Bessel function mode in the *xz*-plane (*y*=0), and the corresponding 1D contours of (e) $\tilde{S}_z$ and $\tilde{p}_y$, (f) $\tilde{S}_z$ and $\tilde{p}_{o,y}$ indicate the direction of dispersive canonical momentum are inverted to that of the dispersive momentum. The (g) $\tilde{S}_x$, (h) $\tilde{S}_z$, (i) $\tilde{p}_y$, and (j) $\tilde{p}_{o,y}$ for the symmetric Bessel function mode in the *xz*-plane (*y*=0), and the corresponding 1D contours of (k) $\tilde{S}_z$ and $\tilde{p}_y$, (l) $\tilde{S}_z$ and $\tilde{p}_{o,y}$ indicate the direction of dispersive canonical momentum are also inverted to that of the dispersive momentum. Remarkably, in the case, if one utilizes the dispersive momentum to evaluate the spin-momentum locking in the layer, the spin vector and the dispersive momentum satisfy the left-hand rule as shown in (e) and (k). However, the spin vector and the dispersive canonical momentum do not satisfy the spin-momentum locking, as shown in (f) and (l). The symmetric and anti-symmetric modes are constructed by +2-order Bessel function. The wavelength is 632.8nm and the thickness of layer is 50nm.

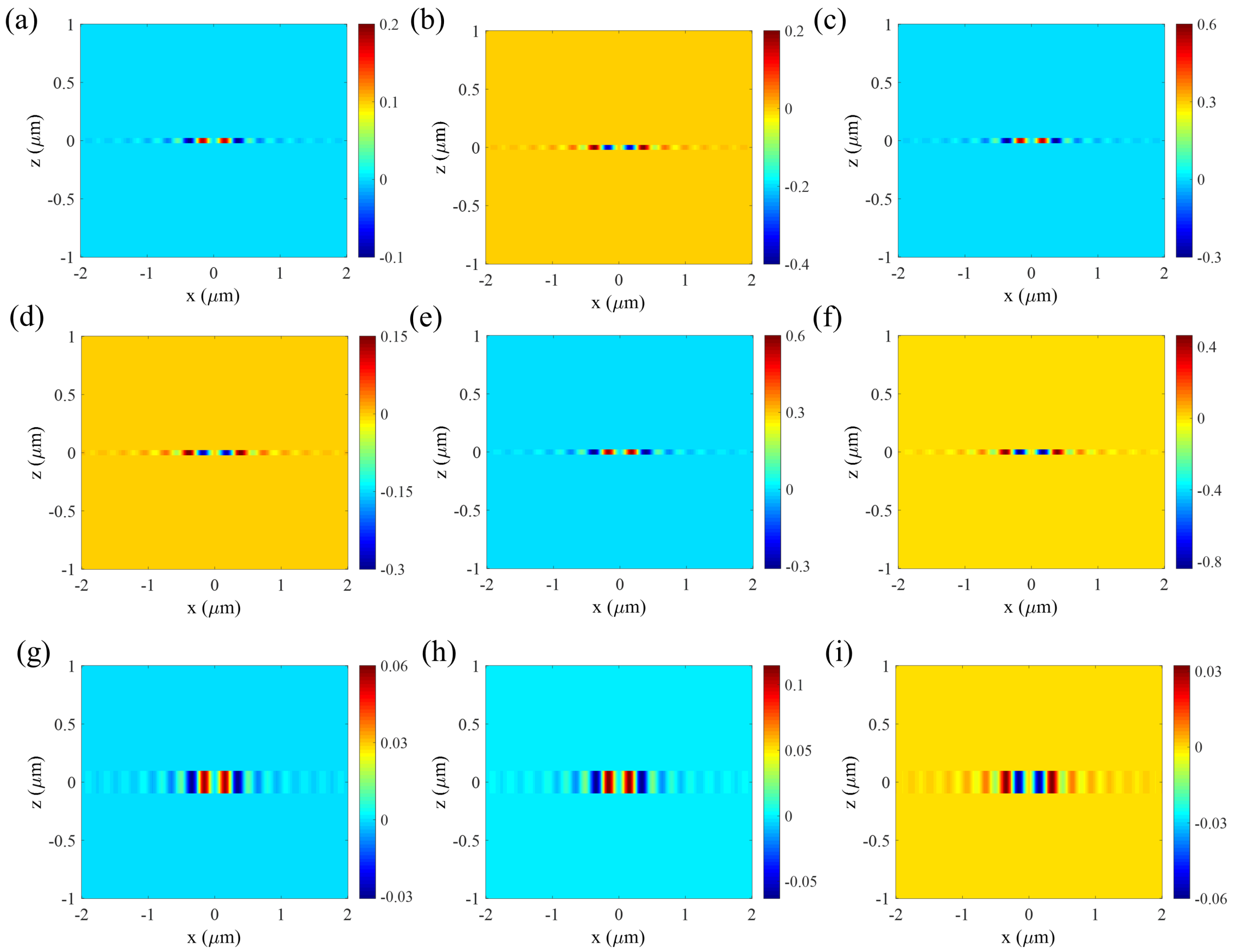

**Fig. S11.** In the *xz*-plane (*y*=0), the (a) coupling *z*-component SAM, (b) coupling *z*-component transverse spin and (c) coupling longitudinal spin for the symmetric Bessel function mode in the air-metal-air configuration; the (d) coupling *z*-component SAM, (e) coupling *z*-component transverse spin and (f) coupling longitudinal spin for the anti-symmetric Bessel function mode in the air-metal-air configuration; the (a) coupling *z*-component SAM, (b) coupling *z*-component transverse spin and (c) coupling longitudinal spin for the symmetric Bessel function mode in the metal-air-metal configuration. One can definitely observe that the coupling spin components in (c), (f), (i) are always inverted to the transverse spin components (b), (e), (h), which make the spin-momentum locking between the dispersive SAM and dispersive momentum consistently in diverse multilayered structures. The symmetric and anti-symmetric modes are constructed by +2-order Bessel function. The wavelength is 632.8nm; the thickness of metal layer is 50nm in the air-metal-air configuration; the thickness of dielectric layer is 200nm in the metal-air-metal configuration.

## IV. Photonic meron and skyrmion lattices in various rotational symmetric systems

The former spin-momentum locking properties can be utilized to investigate the spin-momentum dynamics of photonic topological solitons. Here, we first give the details of derivations of field distributions for the photonic skyrmion and meron lattices in C6 and C4 symmetries, respectively. The SAMs and dispersive momenta of the photonic skyrmion and meron lattices can be calculated by the equations (S43) and (S44). Then, we give several examples to indicate the relationship between the spin-momentum locking and photonic spin topological solitons.

As given in Eq. (S8), for the *p*-polarized surface wave in the one-layer configuration considered here, the normal electric field component $E_z = A/\varepsilon\beta^2\xi$ should fulfill the Helmholtz equation

$$\nabla^2\xi(x,y,z)+\beta^2\xi(x,y,z)=0\,, \tag{S58}$$

where the trial solution can be expressed as

$$\xi(x,y,z)=X(x)Y(y)e^{-k_z z}\,. \tag{S59}$$

By substituting equation (S59) into equation (S58), it can be obtained that

$$\frac{1}{X}\frac{\partial^2 X}{\partial x^2}+\frac{1}{Y}\frac{\partial^2 Y}{\partial y^2}+\beta^2=0\,, \tag{S60}$$

which can be separated into

$$\begin{cases}\dfrac{1}{X}\dfrac{\partial^2 X}{\partial x^2}+u\beta^2=0\\[2ex]\dfrac{1}{Y}\dfrac{\partial^2 Y}{\partial y^2}+v\beta^2=0\end{cases} \tag{S61}$$

with $u + v$=1. The nontrivial solution of expression (S61) is

$$\xi(x,y,z)=\begin{Bmatrix}A\sin\left(\sqrt{u}\beta x\right)\sin\left(\sqrt{v}\beta y\right)+B\cos\left(\sqrt{u}\beta x\right)\cos\left(\sqrt{v}\beta y\right)\\+C\sin\left(\sqrt{u}\beta x\right)\cos\left(\sqrt{v}\beta y\right)+D\cos\left(\sqrt{u}\beta x\right)\sin\left(\sqrt{v}\beta y\right)\end{Bmatrix}e^{-k_z z}\,. \tag{S62}$$

If we set

$$\begin{aligned}L_x&=\frac{2n\pi}{\sqrt{u}\beta}=2\lambda_{sp}\qquad n=\text{integer}\\L_y&=\frac{2m\pi}{\sqrt{v}\beta}=2\lambda_{sp}\qquad m=\text{integer}\end{aligned} \tag{S63}$$

with $\lambda_{sp}$=2π/$\beta$, one can get

$$\mu+v=\left(\frac{2n\pi}{L_x\beta}\right)^2+\left(\frac{2m\pi}{L_y\beta}\right)^2=\lambda_{sp}^2\left[\left(\frac{n}{L_x}\right)^2+\left(\frac{m}{L_y}\right)^2\right]=\frac{1}{4}\left[n^2+m^2\right]=1\,, \tag{S64}$$

and then one can obtain that two groups of solutions

$$\begin{aligned}n=0\qquad m=\pm 2\\n=\pm 2\qquad m=0\end{aligned}\,. \tag{S65}$$

Therefore, the nontrivial solution of expression (S62) is converted into

$$\begin{aligned}\xi(x,y,z)&=\left\{A\cos(\beta x)+B\cos(\beta y)+C\sin(\pm\beta x)+D\sin(\pm\beta y)\right\}e^{-k_z z}\\&=\left\{A'\cos(\beta x)+B'\cos(\beta y)+C'\sin(\beta x)+D'\sin(\beta y)\right\}e^{-k_z z}\end{aligned}\,. \tag{S66}$$

The parameters in expression (S66) can be calculated further with rotating symmetry with rotating matrix

$$R_z(\varphi)=\begin{pmatrix}\cos\varphi & \sin\varphi & 0\\ -\sin\varphi & \cos\varphi & 0\\ 0 & 0 & 1\end{pmatrix} \tag{S67}$$

as the solid-state physics [S10]. The rotating symmetry operator can be expressed as

$$R_z(\varphi)\left\{E_z\left[R_z(-\varphi)\vec{\mathbf{r}}\right]\hat{\mathbf{z}}\right\}=e^{il\varphi}\left\{E_z(\vec{\mathbf{r}})\hat{\mathbf{z}}\right\}. \tag{S68}$$

Note here that there is always $R_z(\varphi)\{E_z\hat{\mathbf{z}}\}=E_z\hat{\mathbf{z}}$ for the normal electric field component.

First, for the C4 rotational symmetry and $l = 0$, the calculated field components are

$$\begin{aligned}
E_x &= -\frac{k_z}{\beta^2}\frac{\partial E_z}{\partial x}=\frac{A_0}{\varepsilon}\frac{k_z}{\beta}\sin(\beta x)e^{-k_z z} & H_x &= -\frac{i\omega\varepsilon}{\beta^2}\frac{\partial E_z}{\partial y}=i\frac{A_0}{\varepsilon}\frac{\omega\varepsilon}{\beta}\sin(\beta y)e^{-k_z z}\\
E_y &= -\frac{k_z}{\beta^2}\frac{\partial E_z}{\partial y}=\frac{A_0}{\varepsilon}\frac{k_z}{\beta}\sin(\beta y)e^{-k_z z} & H_y &= \frac{i\omega\varepsilon}{\beta^2}\frac{\partial E_z}{\partial x}=-i\frac{A_0}{\varepsilon}\frac{\omega\varepsilon}{\beta}\sin(\beta x)e^{-k_z z}\ .\\
E_z &= \frac{A_0}{\varepsilon}\xi=\frac{A_0}{\varepsilon}\{\cos(\beta x)+\cos(\beta y)\}e^{-k_z z} & H_z &= 0
\end{aligned} \tag{S69}$$

For the C4 rotational symmetry and $l = 1$, the calculated field components are

$$\begin{aligned}
E_x &= -\frac{k_z}{\beta^2}\frac{\partial E_z}{\partial x}=-i\frac{A_1}{\varepsilon}\frac{k_z}{\beta}\cos(\beta x)e^{-k_z z} & H_x &= -\frac{i\omega\varepsilon}{\beta^2}\frac{\partial E_z}{\partial y}=-i\frac{A_1}{\varepsilon}\frac{\omega\varepsilon}{\beta}\cos(\beta y)e^{-k_z z}\\
E_y &= -\frac{k_z}{\beta^2}\frac{\partial E_z}{\partial y}=-\frac{A_1}{\varepsilon}\frac{k_z}{\beta}\cos(\beta y)e^{-k_z z} & H_y &= \frac{i\omega\varepsilon}{\beta^2}\frac{\partial E_z}{\partial x}=-\frac{A_1}{\varepsilon}\frac{\omega\varepsilon}{\beta}\cos(\beta x)e^{-k_z z}\ .\\
E_z &= \frac{A_1}{\varepsilon}\xi=\frac{A_1}{\varepsilon}\{\sin(\beta y)+i\sin(\beta x)\}e^{-k_z z} & H_z &= 0
\end{aligned} \tag{S70}$$

For the C4 rotational symmetry and $l = 2$, the calculated field components are

$$\begin{aligned}
E_x &= -\frac{k_z}{\beta^2}\frac{\partial E_z}{\partial x}=-\frac{A_2}{\varepsilon}\frac{k_z}{\beta}\sin(\beta x)e^{-k_z z} & H_x &= -\frac{i\omega\varepsilon}{\beta^2}\frac{\partial E_z}{\partial y}=i\frac{A_2}{\varepsilon}\frac{\omega\varepsilon}{\beta}\sin(\beta y)e^{-k_z z}\\
E_y &= -\frac{k_z}{\beta^2}\frac{\partial E_z}{\partial y}=\frac{A_2}{\varepsilon}\frac{k_z}{\beta}\sin(\beta y)e^{-k_z z} & H_y &= \frac{i\omega\varepsilon}{\beta^2}\frac{\partial E_z}{\partial x}=i\frac{A_2}{\varepsilon}\frac{\omega\varepsilon}{\beta}\sin(\beta x)e^{-k_z z}\ .\\
E_z &= \frac{A_2}{\varepsilon}\xi=\frac{A_2}{\varepsilon}\{\cos(\beta y)-\cos(\beta x)\}e^{-k_z z} & H_z &= 0
\end{aligned} \tag{S71}$$

For the C4 rotational symmetry and $l = 3$, the calculated field components are

$$\begin{aligned}
E_x &= -\frac{k_z}{\beta^2}\frac{\partial E_z}{\partial x}=i\frac{A_3}{\varepsilon}\frac{k_z}{\beta}\cos(\beta x)e^{-k_z z} & H_x &= -\frac{i\omega\varepsilon}{\beta^2}\frac{\partial E_z}{\partial y}=-i\frac{A_3}{\varepsilon}\frac{\omega\varepsilon}{\beta}\cos(\beta y)e^{-k_z z}\\
E_y &= -\frac{k_z}{\beta^2}\frac{\partial E_z}{\partial y}=-\frac{A_3}{\varepsilon}\frac{k_z}{\beta}\cos(\beta y)e^{-k_z z} & H_y &= \frac{i\omega\varepsilon}{\beta^2}\frac{\partial E_z}{\partial x}=\frac{A_3}{\varepsilon}\frac{\omega\varepsilon}{\beta}\cos(\beta x)e^{-k_z z}\ .\\
E_z &= \frac{A_3}{\varepsilon}\xi=\frac{A_3}{\varepsilon}\{\sin(\beta y)-i\sin(\beta x)\}e^{-k_z z} & H_z &= 0
\end{aligned} \tag{S72}$$

Using the equations (S8) and (S9), one can obtain the $z$-component electric field for meron lattices with $l = 3$ as

$$
\begin{aligned}
&z>+\frac{a}{2} && E_z^{+}=+\frac{A_+}{\varepsilon^{+}}\{\sin(\beta y)-i\sin(\beta x)\}e^{-k_z^{+}(z-a/2)}\\
&-\frac{a}{2}<z<+\frac{a}{2} && E_z^{m}=\begin{Bmatrix} +\frac{B_+}{\varepsilon^{m}}\{\sin(\beta y)-i\sin(\beta x)\}e^{+k_z^{m}(z-a/2)}\\ +\frac{B_-}{\varepsilon^{m}}\{\sin(\beta y)-i\sin(\beta x)\}e^{-k_z^{m}(z+a/2)}\end{Bmatrix}.\\
&z<-\frac{a}{2} && E_z^{-}=+\frac{A_-}{\varepsilon^{-}}\{\sin(\beta y)-i\sin(\beta x)\}e^{+k_z^{-}(z+a/2)}
\end{aligned}
\tag{S73}
$$

The horizontal electromagnetic field components are

$$
\begin{aligned}
&z>+\frac{a}{2} && \begin{aligned} E_x^{+}&=+\frac{A_+}{\varepsilon^{+}}\frac{ik_z^{+}}{\beta}\cos(\beta x)e^{-k_z^{+}(z-a/2)} & E_y^{+}&=-\frac{A_+}{\varepsilon^{+}}\frac{k_z^{+}}{\beta}\cos(\beta y)e^{-k_z^{+}(z-a/2)}\\ H_x^{+}&=-\frac{A_+}{\varepsilon^{+}}\frac{i\omega\varepsilon^{+}}{\beta}\cos(\beta y)e^{-k_z^{+}(z-a/2)} & H_y^{+}&=+\frac{A_+}{\varepsilon^{+}}\frac{\omega\varepsilon^{+}}{\beta}\cos(\beta x)e^{-k_z^{+}(z-a/2)}\end{aligned}\\
&-\frac{a}{2}<z<+\frac{a}{2} && \begin{aligned} E_x^{m}&=\begin{Bmatrix}-\frac{B_+}{\varepsilon^{m}}\frac{ik_z^{m}}{\beta}\cos(\beta x)e^{+k_z^{m}(z-a/2)}\\ +\frac{B_-}{\varepsilon^{m}}\frac{ik_z^{m}}{\beta^{2}}\cos(\beta x)e^{-k_z^{m}(z+a/2)}\end{Bmatrix} & E_y^{m}&=\begin{Bmatrix}+\frac{B_+}{\varepsilon^{m}}\frac{k_z^{m}}{\beta}\cos(\beta y)e^{+k_z^{m}(z-a/2)}\\ -\frac{B_-}{\varepsilon^{m}}\frac{k_z^{m}}{\beta}\cos(\beta y)e^{-k_z^{m}(z+a/2)}\end{Bmatrix}\\ H_x^{m}&=\begin{Bmatrix}-\frac{B_+}{\varepsilon^{m}}\frac{i\omega\varepsilon^{m}}{\beta}\cos(\beta y)e^{+k_z^{m}(z-a/2)}\\ -\frac{B_-}{\varepsilon^{m}}\frac{i\omega\varepsilon^{m}}{\beta}\cos(\beta y)e^{-k_z^{m}(z+a/2)}\end{Bmatrix} & H_y^{m}&=\begin{Bmatrix}+\frac{B_+}{\varepsilon^{m}}\frac{\omega\varepsilon^{m}}{\beta}\cos(\beta x)e^{+k_z^{m}(z-a/2)}\\ +\frac{B_-}{\varepsilon^{m}}\frac{\omega\varepsilon^{m}}{\beta}\cos(\beta x)e^{-k_z^{m}(z+a/2)}\end{Bmatrix}\end{aligned}.\\
&z<-\frac{a}{2} && \begin{aligned} E_x^{-}&=-\frac{A_-}{\varepsilon^{-}}\frac{ik_z^{-}}{\beta}\cos(\beta x)e^{+k_z^{-}(z+a/2)} & E_y^{-}&=+\frac{A_-}{\varepsilon^{-}}\frac{k_z^{-}}{\beta}\cos(\beta y)e^{+k_z^{-}(z+a/2)}\\ H_x^{-}&=-\frac{A_-}{\varepsilon^{-}}\frac{i\omega\varepsilon^{-}}{\beta}\cos(\beta y)e^{+k_z^{-}(z+a/2)} & H_y^{-}&=+\frac{A_-}{\varepsilon^{-}}\frac{\omega\varepsilon^{-}}{\beta}\cos(\beta x)e^{+k_z^{-}(z+a/2)}\end{aligned}
\end{aligned}
\tag{S74}
$$

The amplitude coefficients $A_+$, $B_+$, $B_-$ and $A_-$ have the relations:

$$
\frac{B_+}{B_-}=\frac{\frac{k_z^{m}}{\mu^{m}}-\frac{k_z^{+}}{\mu^{+}}}{\frac{k_z^{m}}{\mu^{m}}+\frac{k_z^{+}}{\mu^{+}}}e^{-k_z^{m}a}\quad \frac{B_-}{B_+}=\frac{\frac{k_z^{m}}{\mu^{m}}-\frac{k_z^{-}}{\mu^{-}}}{\frac{k_z^{m}}{\mu^{m}}+\frac{k_z^{-}}{\mu^{-}}}e^{-k_z^{m}a}\quad \begin{aligned}+A_+&=+B_++B_-e^{-k_z^{m}a}\\ +B_+e^{-k_z^{m}a}+B_-&=+A_-\end{aligned}.
\tag{S75}
$$

Remarkably, the solutions have a periodicity of 4 if we only consider the spin-momentum properties of photonic meron lattices here. Moreover, as the $l$=0 and $l$=2, the SAMs and dispersive momenta vanish simultaneously (Noteworthily, as $l$=0 or $l$=2, the electric field distributions can be regarded as the photonic meron lattices [39]. However, the spin-orbit interaction is absence in the cases, and hence these cases are outside the range of our study.). We only consider the properties of spin angular momenta and momenta for $l$=1 (in Fig. S12) and $l$=3 (in Fig. S14) here. As $l$=1, the photonic meron lattice in the air half-space has been researched in Ref. [39]. Here, we first investigated the spin-momentum locking and photonic meron lattices in the dispersive medium: 1. from the vector diagrams of dispersive momentum and SAMs in Fig. S12 and Fig. S14, it can be observed that the spin textures are locked with the dispersive momentum and satisfies the left-hand rule universally; 2. From the vector diagrams of SAMs, one can recognize that the skyrmion number of photonic meron lattices is ±1/2. We also give the corresponding abnormal results in Fig. S13 and Fig. S15 by ignoring the dispersion in the metal materials, which show the normal components of dispersionless SAMs are locked with the kinetic momentum and satisfies the right-hand rule for the symmetric modes, whereas for the anti-symmetric modes, the normal components of dispersionless SAMs are locked with the kinetic momentum and satisfies the left-hand rule.

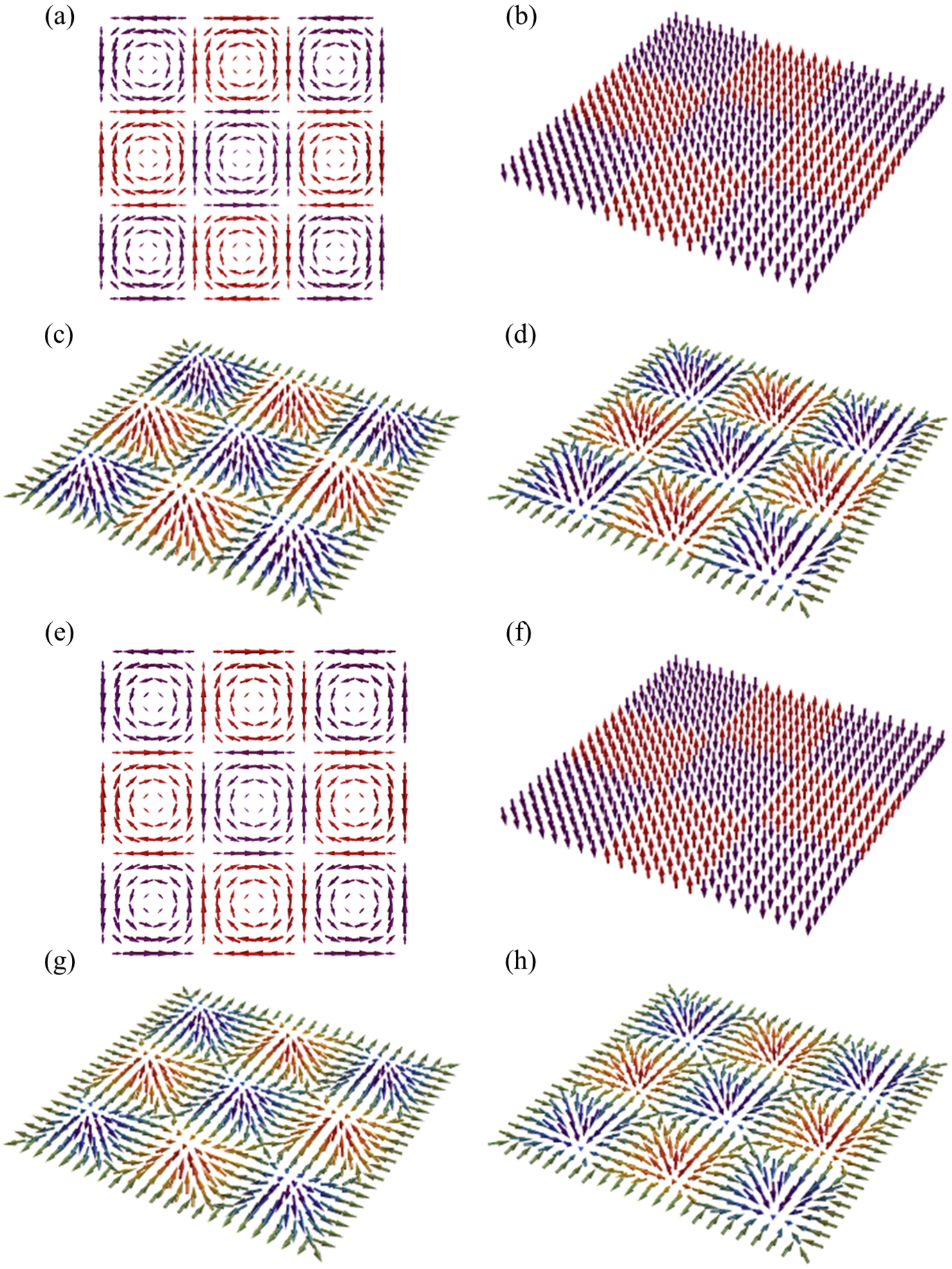

**Fig. S12.** As $l$=1 and C4 symmetry, the vector diagram of (a) dispersive momentum and the spin textures in the (b) z=0, (c) z=+12.5nm, (d) z=−12.5nm for the meron lattices in the layer of air-metal-air structure constructed by the symmetric modes and the vector diagram of (e) dispersive momentum and the spin textures in the (f) z=0, (g) z=+12.5nm, (h) z=−12.5nm for the meron lattices in the layer of air-metal-air structure constructed by the anti-symmetric modes. From the vector diagrams of dispersive momentum and SAMs, it can be observed that the spin textures are locked with the dispersive momentum and satisfies the left-hand rule universally. Noteworthily, the directions of whirling are opposite between the spin textures in the planes z=12.5nm and z=−12.5nm. This is because the horizontal SAM components are opposite in the planes z=12.5nm and z=−12.5nm. In addition, from the vector diagrams of SAMs, one can recognize that the skyrmion number of photonic meron lattices is ±1/2.

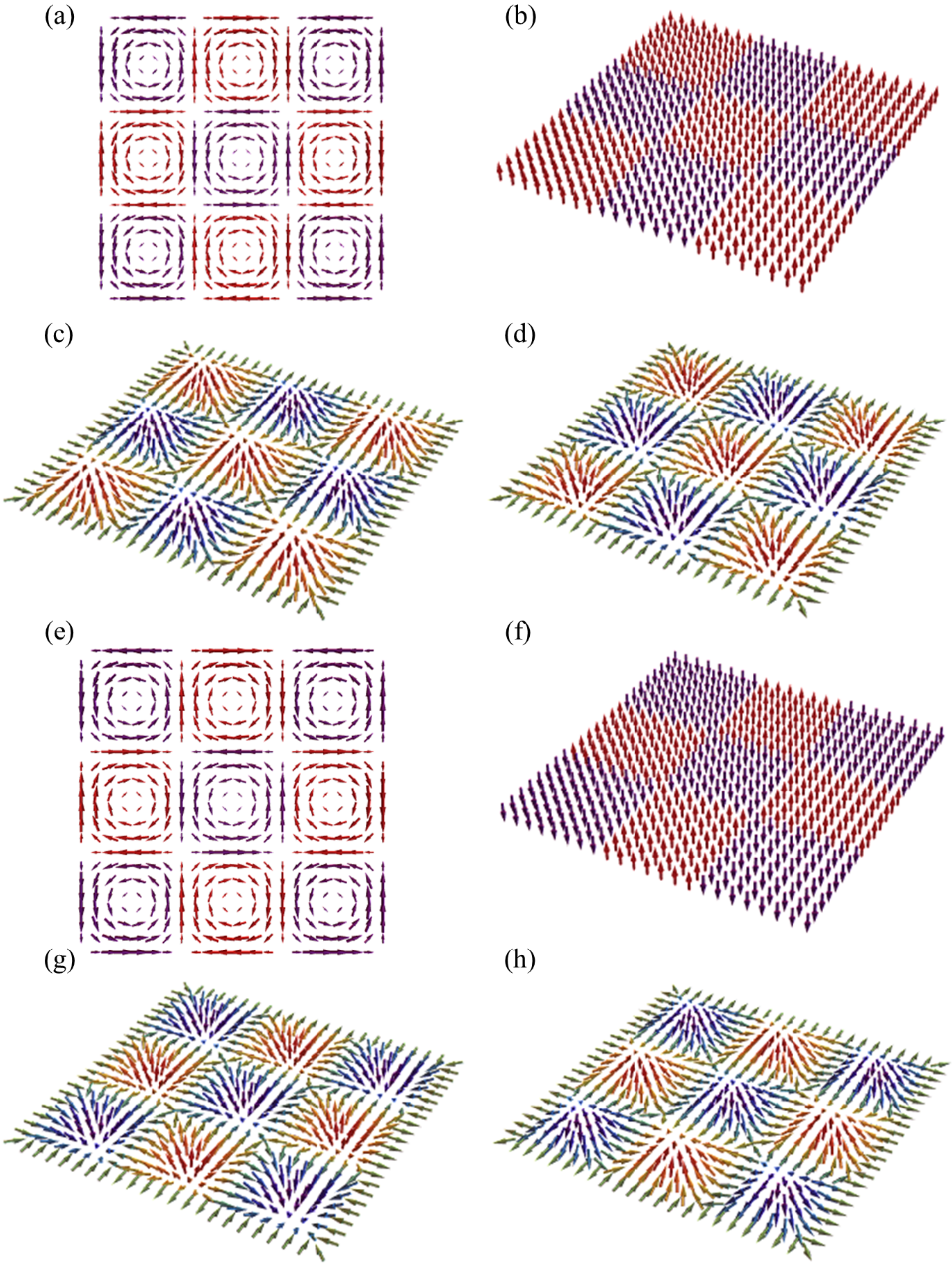

**Fig. S13.** As $l$=1 and C4 symmetry, the abnormal vector diagram of (a) kinetic momentum and the dispersiveless spin textures in the (b) z=0, (c) z=+12.5nm, (d) z=−12.5nm for the meron lattices in the layer of air-metal-air structure constructed by the symmetric modes and the abnormal vector diagram of (e) kinetic momentum and the dispersiveless spin textures in the (f) z=0, (g) z=+12.5nm, (h) z=−12.5nm for the meron lattices in the layer of air-metal-air structure constructed by the anti-symmetric modes. From the vector diagrams of kinetic momentum and SAMs, it can be observed that the normal components of spin textures are locked with the kinetic momentum and satisfies the right-hand rule for the symmetric modes, whereas for the anti-symmetric modes, the normal components of spin textures are locked with the kinetic momentum and satisfies the left-hand rule. The configuration is air-metal-air, and the thickness of metal layer is 50 nm. The wavelength is 632.8nm.

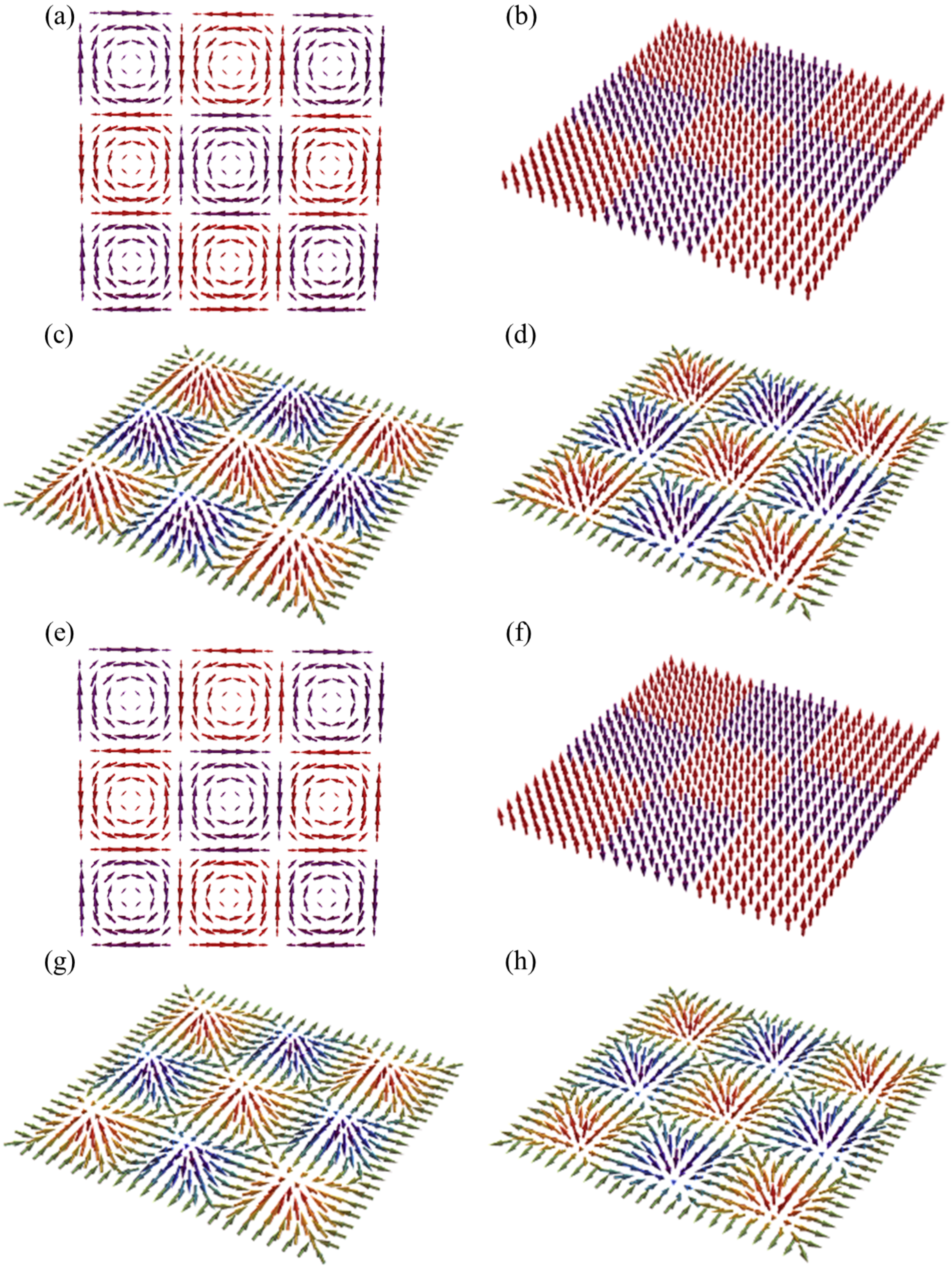

**Fig. S14.** As $l$=3 and C4 symmetry, the vector diagram of (a) dispersive momentum and the spin textures in the (b) z=0, (c) z=+12.5nm, (d) z=−12.5nm for the meron lattices in the layer of air-metal-air structure constructed by the symmetric modes and the vector diagram of (e) dispersive momentum and the spin textures in the (f) z=0, (g) z=+12.5nm, (h) z=−12.5nm for the meron lattices in the layer of air-metal-air structure constructed by the anti-symmetric modes. From the vector diagrams of dispersive momentum and SAMs, it can be observed that the spin textures are locked with the dispersive momentum and satisfies the left-hand rule universally. Noteworthily, the directions of whirling are opposite between the spin textures in the planes z=12.5nm and z=−12.5nm. This is because the horizontal SAM components are opposite in the planes z=12.5nm and z=−12.5nm. In addition, from the vector diagrams of SAMs, one can recognize that the skyrmion number of photonic meron lattices is ±1/2. The configuration is air-metal-air, and the thickness of metal layer is 50 nm. The wavelength is 632.8nm.

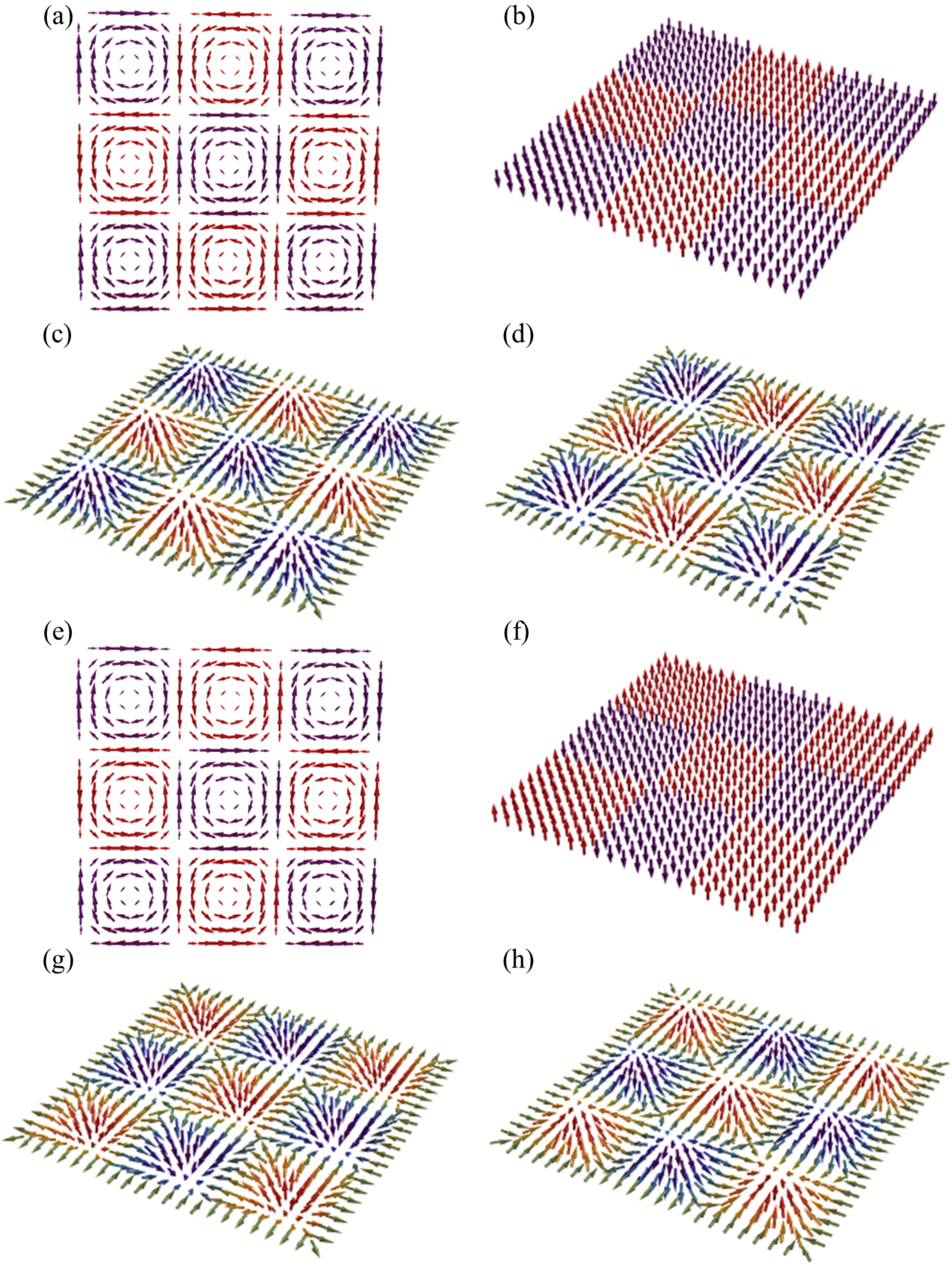

**Fig. S15.** As $l$=3 and C4 symmetry, the abnormal vector diagram of (a) kinetic momentum and the dispersiveless spin textures in the (b) z=0, (c) z=+12.5nm, (d) z=−12.5nm for the meron lattices in the layer of air-metal-air structure constructed by the symmetric modes and the abnormal vector diagram of (e) kinetic momentum and the dispersiveless spin textures in the (f) z=0, (g) z=+12.5nm, (h) z=−12.5nm for the meron lattices in the layer of air-metal-air structure constructed by the anti-symmetric modes. From the vector diagrams of kinetic momentum and SAMs, it can be observed that the normal components of spin textures are locked with the kinetic momentum and satisfies the right-hand rule for the symmetric modes, whereas for the anti-symmetric modes, the normal components of spin textures are locked with the kinetic momentum and satisfies the left-hand rule. The configuration is air-metal-air, and the thickness of metal layer is 50 nm. The wavelength is 632.8nm.

Then, for the C6 rotational symmetry and $l = 0$, the calculated field components are

$$\begin{aligned}
E_x &= -\frac{k_z}{\beta^2}\frac{\partial E_z}{\partial x} = -\frac{B_0}{\varepsilon}\frac{k_z}{2\beta}\left\{-2\sin(\beta x)-\sin\left(\frac{1}{2}\beta x+\frac{\sqrt{3}}{2}\beta y\right)-\sin\left(\frac{1}{2}\beta x-\frac{\sqrt{3}}{2}\beta y\right)\right\}e^{-k_z z} \\
H_x &= -\frac{i\omega\varepsilon}{\beta^2}\frac{\partial E_z}{\partial y} = -\frac{B_0}{\varepsilon}\frac{i\omega\varepsilon}{2\beta}\left\{-\sqrt{3}\sin\left(\frac{1}{2}\beta x+\frac{\sqrt{3}}{2}\beta y\right)+\sqrt{3}\sin\left(\frac{1}{2}\beta x-\frac{\sqrt{3}}{2}\beta y\right)\right\}e^{-k_z z} \\
E_y &= -\frac{k_z}{\beta^2}\frac{\partial E_z}{\partial y} = -\frac{B_0}{\varepsilon}\frac{k_z}{2\beta}\left\{-\sqrt{3}\sin\left(\frac{1}{2}\beta x+\frac{\sqrt{3}}{2}\beta y\right)+\sqrt{3}\sin\left(\frac{1}{2}\beta x-\frac{\sqrt{3}}{2}\beta y\right)\right\}e^{-k_z z} \\
H_y &= +\frac{i\omega\varepsilon}{\beta^2}\frac{\partial E_z}{\partial x} = \frac{B_0}{\varepsilon}\frac{i\omega\varepsilon}{2\beta}\left\{-2\sin(\beta x)-\sin\left(\frac{1}{2}\beta x+\frac{\sqrt{3}}{2}\beta y\right)-\sin\left(\frac{1}{2}\beta x-\frac{\sqrt{3}}{2}\beta y\right)\right\}e^{-k_z z} \\
E_z &= \frac{B_0}{\varepsilon}\xi = \frac{B_0}{\varepsilon}\left\{\cos(\beta x)+\cos\left(\frac{1}{2}\beta x+\frac{\sqrt{3}}{2}\beta y\right)+\cos\left(\frac{1}{2}\beta x-\frac{\sqrt{3}}{2}\beta y\right)\right\}e^{-k_z z} \\
H_z &= 0
\end{aligned} \quad . \quad \text{(S76)}$$

For the C6 rotational symmetry and $l = 1$, the calculated field components are

$$\begin{aligned}
E_x &= -\frac{k_z}{\beta^2}\frac{\partial E_z}{\partial x} = -\frac{B_1}{\varepsilon}\frac{k_z}{\beta}\left\{\frac{1}{\sqrt{3}}\cos\left(\frac{1}{2}\beta x\right)\cos\left(\frac{\sqrt{3}}{2}\beta y\right)-i\sin\left(\frac{1}{2}\beta x\right)\sin\left(\frac{\sqrt{3}}{2}\beta y\right)+\frac{2}{\sqrt{3}}\cos(\beta x)\right\}e^{-k_z z} \\
H_x &= -\frac{i\omega\varepsilon}{\beta^2}\frac{\partial E_z}{\partial y} = -\frac{B_1}{\varepsilon}\frac{i\omega\varepsilon}{\beta}\left\{-\sin\left(\frac{1}{2}\beta x\right)\sin\left(\frac{\sqrt{3}}{2}\beta y\right)+\sqrt{3}i\cos\left(\frac{1}{2}\beta x\right)\cos\left(\frac{\sqrt{3}}{2}\beta y\right)\right\}e^{-k_z z} \\
E_y &= -\frac{k_z}{\beta^2}\frac{\partial E_z}{\partial y} = -\frac{B_1}{\varepsilon}\frac{k_z}{\beta}\left\{-\sin\left(\frac{1}{2}\beta x\right)\sin\left(\frac{\sqrt{3}}{2}\beta y\right)+\sqrt{3}i\cos\left(\frac{1}{2}\beta x\right)\cos\left(\frac{\sqrt{3}}{2}\beta y\right)\right\}e^{-k_z z} \\
H_y &= +\frac{i\omega\varepsilon}{\beta^2}\frac{\partial E_z}{\partial x} = \frac{B_1}{\varepsilon}\frac{i\omega\varepsilon}{\beta}\left\{\frac{1}{\sqrt{3}}\cos\left(\frac{1}{2}\beta x\right)\cos\left(\frac{\sqrt{3}}{2}\beta y\right)-i\sin\left(\frac{1}{2}\beta x\right)\sin\left(\frac{\sqrt{3}}{2}\beta y\right)+\frac{2}{\sqrt{3}}\cos(\beta x)\right\}e^{-k_z z} \\
E_z &= \frac{B_1}{\varepsilon}\xi = \frac{B_1}{\varepsilon}\left\{\frac{2}{\sqrt{3}}\sin\left(\frac{1}{2}\beta x\right)\cos\left(\frac{\sqrt{3}}{2}\beta y\right)+2i\cos\left(\frac{1}{2}\beta x\right)\sin\left(\frac{\sqrt{3}}{2}\beta y\right)+\frac{2}{\sqrt{3}}\sin(\beta x)\right\}e^{-k_z z} \\
H_z &= 0
\end{aligned} \quad . \quad \text{(S77)}$$

For the C6 rotational symmetry and $l = 2$, the calculated field components are

$$\begin{aligned}
E_x &= -\frac{k_z}{\beta^2}\frac{\partial E_z}{\partial x} = -\frac{B_2}{\varepsilon}\frac{k_z}{\beta}\left\{-i\cos\left(\frac{1}{2}\beta x\right)\sin\left(\frac{\sqrt{3}}{2}\beta y\right)+\frac{1}{\sqrt{3}}\sin\left(\frac{1}{2}\beta x\right)\cos\left(\frac{\sqrt{3}}{2}\beta y\right)-\frac{2}{\sqrt{3}}\sin(\beta x)\right\}e^{-k_z z} \\
H_x &= -\frac{i\omega\varepsilon}{\beta^2}\frac{\partial E_z}{\partial y} = -\frac{B_2}{\varepsilon}\frac{i\omega\varepsilon}{\beta}\left\{-\sqrt{3}i\sin\left(\frac{1}{2}\beta x\right)\cos\left(\frac{\sqrt{3}}{2}\beta y\right)+\cos\left(\frac{1}{2}\beta x\right)\sin\left(\frac{\sqrt{3}}{2}\beta y\right)\right\}e^{-k_z z} \\
E_y &= -\frac{k_z}{\beta^2}\frac{\partial E_z}{\partial y} = -\frac{B_2}{\varepsilon}\frac{k_z}{\beta}\left\{-\sqrt{3}i\sin\left(\frac{1}{2}\beta x\right)\cos\left(\frac{\sqrt{3}}{2}\beta y\right)+\cos\left(\frac{1}{2}\beta x\right)\sin\left(\frac{\sqrt{3}}{2}\beta y\right)\right\}e^{-k_z z} \\
H_y &= +\frac{i\omega\varepsilon}{\beta^2}\frac{\partial E_z}{\partial x} = \frac{B_2}{\varepsilon}\frac{i\omega\varepsilon}{\beta}\left\{-i\cos\left(\frac{1}{2}\beta x\right)\sin\left(\frac{\sqrt{3}}{2}\beta y\right)+\frac{1}{\sqrt{3}}\sin\left(\frac{1}{2}\beta x\right)\cos\left(\frac{\sqrt{3}}{2}\beta y\right)-\frac{2}{\sqrt{3}}\sin(\beta x)\right\}e^{-k_z z} \\
E_z &= \frac{B_2}{\varepsilon}\xi = \frac{B_2}{\varepsilon}\left\{-2i\sin\left(\frac{1}{2}\beta x\right)\sin\left(\frac{\sqrt{3}}{2}\beta y\right)-\frac{2}{\sqrt{3}}\cos\left(\frac{1}{2}\beta x\right)\cos\left(\frac{\sqrt{3}}{2}\beta y\right)+\frac{2}{\sqrt{3}}\cos(\beta x)\right\}e^{-k_z z} \\
H_z &= 0
\end{aligned} \quad . \quad \text{(S78)}$$

For the C6 rotational symmetry and $l = 3$, the calculated field components are

$$
\begin{aligned}
E_x &= -\frac{k_z}{\beta^2}\frac{\partial E_z}{\partial x} = -\frac{B_3}{\varepsilon}\frac{k_z}{2\beta}\left\{2\cos(\beta x)-\cos\left(\frac{1}{2}\beta x+\frac{\sqrt{3}}{2}\beta y\right)-\cos\left(\frac{1}{2}\beta x-\frac{\sqrt{3}}{2}\beta y\right)\right\}e^{-k_z z} \\
H_x &= -\frac{i\omega\varepsilon}{\beta^2}\frac{\partial E_z}{\partial y} = -\frac{B_3}{\varepsilon}\frac{i\omega\varepsilon}{2\beta}\left\{-\sqrt{3}\cos\left(\frac{1}{2}\beta x+\frac{\sqrt{3}}{2}\beta y\right)+\sqrt{3}\cos\left(\frac{1}{2}\beta x-\frac{\sqrt{3}}{2}\beta y\right)\right\}e^{-k_z z} \\
E_y &= -\frac{k_z}{\beta^2}\frac{\partial E_z}{\partial y} = -\frac{B_3}{\varepsilon}\frac{k_z}{2\beta}\left\{-\sqrt{3}\cos\left(\frac{1}{2}\beta x+\frac{\sqrt{3}}{2}\beta y\right)+\sqrt{3}\cos\left(\frac{1}{2}\beta x-\frac{\sqrt{3}}{2}\beta y\right)\right\}e^{-k_z z} \\
H_y &= +\frac{i\omega\varepsilon}{\beta^2}\frac{\partial E_z}{\partial x} = \frac{B_3}{\varepsilon}\frac{i\omega\varepsilon}{2\beta}\left\{2\cos(\beta x)-\cos\left(\frac{1}{2}\beta x+\frac{\sqrt{3}}{2}\beta y\right)-\cos\left(\frac{1}{2}\beta x-\frac{\sqrt{3}}{2}\beta y\right)\right\}e^{-k_z z} \\
E_z &= \frac{B_3}{\varepsilon}\xi = \frac{B_3}{\varepsilon}\left\{\sin(\beta x)-\sin\left(\frac{1}{2}\beta x+\frac{\sqrt{3}}{2}\beta y\right)-\sin\left(\frac{1}{2}\beta x-\frac{\sqrt{3}}{2}\beta y\right)\right\}e^{-k_z z} \\
H_z &= 0
\end{aligned} \quad . \quad (S79)
$$

For the C6 rotational symmetry and $l = 4$, the calculated field components are

$$
\begin{aligned}
E_x &= -\frac{k_z}{\beta^2}\frac{\partial E_z}{\partial x} = -\frac{B_4}{\varepsilon}\frac{k_z}{\beta}\left\{i\cos\left(\frac{1}{2}\beta x\right)\sin\left(\frac{\sqrt{3}}{2}\beta y\right)+\frac{1}{\sqrt{3}}\sin\left(\frac{1}{2}\beta x\right)\cos\left(\frac{\sqrt{3}}{2}\beta y\right)-\frac{2}{\sqrt{3}}\sin(\beta x)\right\}e^{-k_z z} \\
H_x &= -\frac{i\omega\varepsilon}{\beta^2}\frac{\partial E_z}{\partial y} = -\frac{B_4}{\varepsilon}\frac{i\omega\varepsilon}{\beta}\left\{\sqrt{3}i\sin\left(\frac{1}{2}\beta x\right)\cos\left(\frac{\sqrt{3}}{2}\beta y\right)+\cos\left(\frac{1}{2}\beta x\right)\sin\left(\frac{\sqrt{3}}{2}\beta y\right)\right\}e^{-k_z z} \\
E_y &= -\frac{k_z}{\beta^2}\frac{\partial E_z}{\partial y} = -\frac{B_4}{\varepsilon}\frac{k_z}{\beta}\left\{\sqrt{3}i\sin\left(\frac{1}{2}\beta x\right)\cos\left(\frac{\sqrt{3}}{2}\beta y\right)+\cos\left(\frac{1}{2}\beta x\right)\sin\left(\frac{\sqrt{3}}{2}\beta y\right)\right\}e^{-k_z z} \\
H_y &= +\frac{i\omega\varepsilon}{\beta^2}\frac{\partial E_z}{\partial x} = \frac{B_4}{\varepsilon}\frac{i\omega\varepsilon}{\beta}\left\{i\cos\left(\frac{1}{2}\beta x\right)\sin\left(\frac{\sqrt{3}}{2}\beta y\right)+\frac{1}{\sqrt{3}}\sin\left(\frac{1}{2}\beta x\right)\cos\left(\frac{\sqrt{3}}{2}\beta y\right)-\frac{2}{\sqrt{3}}\sin(\beta x)\right\}e^{-k_z z} \\
E_z &= \frac{B_4}{\varepsilon}\xi = \frac{B_4}{\varepsilon}\left\{2i\sin\left(\frac{1}{2}\beta x\right)\sin\left(\frac{\sqrt{3}}{2}\beta y\right)-\frac{2}{\sqrt{3}}\cos\left(\frac{1}{2}\beta x\right)\cos\left(\frac{\sqrt{3}}{2}\beta y\right)+\frac{2}{\sqrt{3}}\cos(\beta x)\right\}e^{-k_z z} \\
H_z &= 0
\end{aligned} \quad . \quad (S80)
$$

For the C6 rotational symmetry and $l = 5$, the calculated field components are

$$
\begin{aligned}
E_x &= -\frac{k_z}{\beta^2}\frac{\partial E_z}{\partial x} = -\frac{B_5}{\varepsilon}\frac{k_z}{\beta}\left\{\frac{1}{\sqrt{3}}\cos\left(\frac{1}{2}\beta x\right)\cos\left(\frac{\sqrt{3}}{2}\beta y\right)+i\sin\left(\frac{1}{2}\beta x\right)\sin\left(\frac{\sqrt{3}}{2}\beta y\right)+\frac{2}{\sqrt{3}}\cos(\beta x)\right\}e^{-k_z z} \\
H_x &= -\frac{i\omega\varepsilon}{\beta^2}\frac{\partial E_z}{\partial y} = -\frac{B_5}{\varepsilon}\frac{i\omega\varepsilon}{\beta}\left\{-\sin\left(\frac{1}{2}\beta x\right)\sin\left(\frac{\sqrt{3}}{2}\beta y\right)-\sqrt{3}i\cos\left(\frac{1}{2}\beta x\right)\cos\left(\frac{\sqrt{3}}{2}\beta y\right)\right\}e^{-k_z z} \\
E_y &= -\frac{k_z}{\beta^2}\frac{\partial E_z}{\partial y} = -\frac{B_5}{\varepsilon}\frac{k_z}{\beta}\left\{-\sin\left(\frac{1}{2}\beta x\right)\sin\left(\frac{\sqrt{3}}{2}\beta y\right)-\sqrt{3}i\cos\left(\frac{1}{2}\beta x\right)\cos\left(\frac{\sqrt{3}}{2}\beta y\right)\right\}e^{-k_z z} \\
H_y &= +\frac{i\omega\varepsilon}{\beta^2}\frac{\partial E_z}{\partial x} = \frac{B_5}{\varepsilon}\frac{i\omega\varepsilon}{\beta}\left\{\frac{1}{\sqrt{3}}\cos\left(\frac{1}{2}\beta x\right)\cos\left(\frac{\sqrt{3}}{2}\beta y\right)+i\sin\left(\frac{1}{2}\beta x\right)\sin\left(\frac{\sqrt{3}}{2}\beta y\right)+\frac{2}{\sqrt{3}}\cos(\beta x)\right\}e^{-k_z z} \\
E_z &= \frac{B_5}{\varepsilon}\xi = \frac{B_5}{\varepsilon}\left\{\frac{2}{\sqrt{3}}\sin\left(\frac{1}{2}\beta x\right)\cos\left(\frac{\sqrt{3}}{2}\beta y\right)-2i\cos\left(\frac{1}{2}\beta x\right)\sin\left(\frac{\sqrt{3}}{2}\beta y\right)+\frac{2}{\sqrt{3}}\sin(\beta x)\right\}e^{-k_z z} \\
H_z &= 0
\end{aligned} \quad . \quad (S81)
$$

The electric/magnetic field components can be calculated as the processes in expressions (S73)-(S75).

Remarkably, the solutions have a periodicity of 6 if we only consider the spin-momentum properties of photonic Skyrmion lattices here. Moreover, as the $l$=0 and $l$=3, the spin angular momenta and momenta vanish simultaneously (Noteworthily, as $l$=0 or $l$=3, the electric field distributions can be regarded as the photonic skyrmion lattices [S11, S12]. However, the spin-orbit interaction is absence in the cases, and hence these cases are outside the range of our study.).